\documentclass[a4paper,11pt]{article}
\pdfoutput=1 % if your are submitting a pdflatex (i.e. if you have
\usepackage{jheppub} % for details on the use of the package, please
\usepackage{bm}

\usepackage[T1]{fontenc} % if needed

\newcommand{\bqa}{\begin{eqnarray}}
\newcommand{\eqa}{\end{eqnarray}}
\newcommand{\nn}{\nonumber \\}

\title{\boldmath
Entanglement transfer from quantum matter to classical geometry in an emergent holographic dual description of a scalar field theory}

\author[a,b]{Ki-Seok Kim}
\author[c]{and Shinsei Ryu}

\affiliation[a]{Department of Physics, POSTECH, Pohang, Gyeongbuk 37673, Korea}
\affiliation[b]{Asia Pacific Center for Theoretical Physics (APCTP), Pohang, Gyeongbuk 37673, Korea}
\affiliation[c]{Kadanoff Center for Theoretical Physics, University of Chicago, IL 60637, USA}

\abstract{
Applying recursive renormalization group transformations to a scalar field theory, we obtain an effective quantum gravity theory with an emergent extra dimension, described by a dual holographic Einstein-Klein-Gordon type action. Here, the dynamics of both the dual order-parameter field and the metric tensor field originate from density-density and energy-momentum tensor-tensor effective interactions, respectively, in the recursive renormalization group transformation, performed approximately in the Gaussian level. This linear approximation in the recursive renormalization group transformation for the gravity sector gives rise to a linearized quantum Einstein-scalar theory along the $z-$directional emergent space. In the large $N$ limit, where $N$ is the flavor number of the original scalar fields, quantum fluctuations of both dynamical metric and dual scalar fields are suppressed, leading to a classical field theory of the Einstein-scalar type in $(D+1)$-spacetime dimensions. We show that this emergent background gravity describes the renormalization group flows of coupling functions in the UV quantum field theory through the extra dimension. More precisely, the IR boundary conditions of the gravity equations correspond to the renormalization group $\beta$-functions of the quantum field theory, where the infinitesimal distance in the extra-dimensional space is identified with an energy scale for the renormalization group transformation. Finally, we also show that this dual holographic formulation describes quantum entanglement in a geometrical way, encoding the transfer of quantum entanglement from quantum matter to classical gravity in the large $N$ limit. We claim that this entanglement transfer serves as a microscopic foundation for the emergent holographic duality description.
}

\begin{document}
\maketitle
\flushbottom

\section{Introduction}

Renormalization group transformations lead us to construct an effective field theory of
renormalized low-energy fluctuations in terms of renormalized interaction vertices,
which result from quantum fluctuations of high-energy modes \cite{RG_Textbook}.
This renormalization group analysis requires that the resulting effective field
theory remains invariant in its mathematical form after the renormalization group transformation.
As a result, we can extract out how interaction vertices evolve as a function of the renormalization group energy scale,
referred to as renormalization group $\beta$-functions.
In addition, we obtain the so-called Callan-Symanzik equation that a correlation function has to satisfy
in the renormalization group transformation \cite{RG_Textbook}.
Solving this renormalization group equation for the correlation function,
we find a renormalized correlation function in terms of transfer momentum and energy,
which encodes anomalous scaling dimensions of renormalized low-energy
fluctuations
and renormalization group $\beta$-functions of interaction vertices.

One may suggest that the renormalization group energy scale
is introduced into an effective field theory
%as an extra dimensional space,
as an extra dimension,
manifesting a renormalization group flow of the effective field theory through
the extra dimensional space.
Renormalization group flows of interaction vertices are realized along the extra dimension,
serving as an effective curved spacetime for renormalized low-energy
fluctuations in this effective field theory.
Renormalization group flows of correlation functions can be described by
introducing external source fields,
coupled to conserved currents minimally, into the effective field theory,
regarded to be a dual description in this emergent spacetime with an extra dimension.
This novel construction for renormalization group transformations reminds us of the holographic duality conjecture \cite{Holographic_Duality_I,Holographic_Duality_II, Holographic_Duality_III,Holographic_Duality_IV,Holographic_Duality_V,Holographic_Duality_VI,Holographic_Duality_VII}.

In this study,
we follow this line of thoughts and
%we
suggest a microscopic foundation for the holographic duality conjecture \cite{RG_Holography_I,RG_Holography_II,RG_Holography_III,RG_Holography_IV,RG_Holography_V,RG_Holography_VI,RG_Holography_VII,RG_Holography_VIII,
RG_Holography_IX,RG_Holography_X,RG_Holography_XI,RG_Holography_XII,RG_Holography_XIII,RG_Holography_XIV,RG_Holography_XV,RG_Holography_XVI,
RG_Holography_XVII,RG_Holography_XVIII,RG_Holography_XIX,RG_Holography_XX,RG_Holography_XXI}.
%An
%{\color{red}To be concrete, the key}
The key idea is
the recursive applications of the renormalization group transformations.
% take {\color{red} perform} renormalization group transformations in a recursive way.
We follow the
  real space renormalization group scheme
  developed by Polchinski
  (although
  whether one works in real or momentum space
  should not matter ultimately).
%Although it does not matter to perform renormalization group transformations
%either in momentum space or in real space,
%we
%{\color{red}consider}
%% take
%the renormalization group transformations in real space a la Polchinski
%\cite{RG_realspace_Polchinski}.
  This renormalization group technique was further
  developed by Sung-Sik Lee,
  by applying the renormalization group transformations
  in a recursive way
%Sung-Sik Lee developed this renormalization group technique further,
%applying it in a recursive way
\cite{SungSik_Holography_I,SungSik_Holography_II,SungSik_Holography_III}.
Resorting to this renormalization group analysis,
we show that renormalization group flows of interaction vertices appear to form
an emergent curved spacetime with an extra dimension.
Here, the infinitesimal distance in this extra-dimensional space is identified
with the renormalization group energy scale.
To be concrete,
  we apply
 the Polchinski renormalization group technique a la
Sung-Sik Lee to a quantum field theory
of bosons with self-interactions in $D$-spacetime dimensions.
%More precisely, applying the Polchinski renormalization group technique a la
%Sung-Sik Lee to a quantum field theory
%of bosons with self-interactions in $D$-spacetime dimensions,
We derive an effective dual holographic description of the Einstein-Klein-Gordon type
theory in $(D+1)$-spacetime dimensions,
given by a quantum gravity theory of a modified Einstein-Hilbert action coupled to
fluctuating scalar fields dual to the density of the original bosons.
It turns out that quantum fluctuations of both dynamical metric and dual scalar
fields are suppressed in the large $N$ limit,
where $N$ is the number of boson flavors.
As a result, we obtain a classical field theory of the Einstein-scalar type in $(D+1)$-spacetime dimensions.

Since the quantum field theory is geometrized and described by a classical field
theory,
it is natural to ask how the quantum mechanical nature
of the original theory
can be encoded into the
classical and geometric description appropriately.
We thus study the entanglement entropy
%If the quantum mechanical nature means quantum entanglement, it is natural to
%consider the entanglement entropy
\cite{Entanglement_Entropy_Calabrese_Cardy_I,Entanglement_Entropy_Calabrese_Cardy_II,
  Entanglement_Entropy_Ryu_Takayanagi_I,Entanglement_Entropy_Ryu_Takayanagi_II,Entanglement_Entropy_Review_III,Entanglement_Entropy_Review_IV},
which measures
the quantum correlation between
  a given spatial region and its complement,
%how much a quantum state is entangled in itself.
by utilizing the classical holographic description.
As we will show, it
turns out that entanglement is transferred from quantum matter to classical
gravity through recursive renormalization group transformations.
We claim that the entanglement transfer serves as a microscopic foundation for the holographic duality conjecture.

\section{Overview} \label{Overview}

\subsection{Emergent gravity description for a scalar field theory}

%
%\subsubsection{Scalar field theory in a curved spacetime}
%

Since the present study
% shows
involves a lot of formulae,
discussing
our main results briefly with the introduction of the structure of this paper
would be helpful for readers to figure out the physics.
We start from an effective field theory of self-interacting bosons, where the partition function is
\begin{align}
  Z = \int D \phi_{\alpha}(x) \exp\Big\{ - S_{UV}[\phi_{\alpha}(x);g^{B}_{\mu\nu}(x)] \Big\}
\end{align}
and the effective action is
\begin{align}
  &
  S_{UV}[\phi_{\alpha}(x);g^{B}_{\mu\nu}(x)]
  \nonumber \\
  &\quad
    =
    \int d^{D} x \sqrt{g_{B}} \Big\{ g_{B}^{\mu\nu} (\partial_{\mu} \phi_{\alpha}) (\partial_{\nu} \phi_{\alpha}) + m^{2} \phi_{\alpha}^{2} + \xi R_{B} \phi_{\alpha}^{2} + \frac{u}{2N} \phi_{\alpha}^{2} \phi_{\beta}^{2} +
    \frac{\lambda}{2N} T_{\mu\nu} T^{\mu\nu}
    \Big\} . \label{UV_Effective_Action_with_Tensor_Interaction}
\end{align}
Here, $\phi_{\alpha}(x)$ is a real scalar field with a flavor index
$\alpha = 1, \ldots, N$
at a $D$-dimensional spacetime coordinate $x$.
We point out that the two flavor indices $\alpha$ and $\beta$ are summed independently.
$g_{B}^{\mu\nu}$ with $\mu ~ \& ~ \nu = 0, \ldots, D-1$ is a background metric
tensor to describe a $D$-dimensional curved spacetime manifold,
where the bosonic field lives. $g_{B}$ is the determinant of the metric tensor,
where $d^{D} x \sqrt{g_{B}}$ gives an invariant volume factor.
For example, we may consider the background metric as follows
\begin{align}
  d s^{2} = d r^{2} + r^{2} d \theta^{2}
  + \delta_{ij} d x_{\perp}^{i} d x_{\perp}^{j} ,
  \quad
  g_{rr}^{B} = 1 ,
  \quad
  g_{\theta\theta}^{B} = r^{2} ,
  \quad
  g_{ij}^{B \perp} = \delta_{ij}, ~~~~~ \sqrt{g_{B}} = r ,
\end{align}
which will be useful
% expected to be useful
for evaluation of the entanglement entropy
%in
using
the replica technique. Here,
$i$ and $j$ run from $2$ to $D-1$. $m$ denotes the mass of these scalar bosons
and $u$ represents their self-interactions.
$R_{B}$ is
the
Ricci scalar and $\xi$ is a coupling constant between scalar fields
and the background curvature \cite{Coupling_Scalarfields_Riccicurvature}.
Finally, the last term
% The last term
describes effective interactions between
energy-momentum tensor currents $T^{\mu\nu}$,
where $\lambda$ is the corresponding coupling constant.
Although these effective interactions are irrelevant at both Gaussian and
Wilson-Fisher fixed points in the perturbative regime \cite{RG_Textbook}, we
introduce this term explicitly,
which plays an important role in uplifting the background geometry into full dynamical degrees of freedom for the emergent bulk geometry \cite{TTbar_Deformation}.

%
%\subsubsection{Landau-Ginzburg effective field theory on an emergent curved spacetime with an extra dimension}
%

In Section \ref{Holography_Derivation},
by implementing the Polchinski real-space renormalization group technique
\cite{RG_realspace_Polchinski} to this effective field theory in a recursive way
a la Sung-Sik Lee
\cite{SungSik_Holography_I,SungSik_Holography_II,SungSik_Holography_III}, we
obtain a quantum gravity theory coupled to a dual scalar field in
one-dimensional higher spacetime.
It is given by the following partition function
\begin{align}
  Z &= Z_{\Lambda}
      \int D \varphi(x,z) D g_{\mu\nu}(x,z) \exp\Big\{ - S_{UV}[g_{\mu\nu}(x,0),\varphi(x,0)] - S_{IR}[g_{\mu\nu}(x,z_{f}),\varphi(x,z_{f})]
      \nonumber \\
    &\qquad  - S_{Bulk}[g_{\mu\nu}(x,z),\varphi(x,z)] \Big\} .
      \label{RG master part func}
\end{align}
Here, $z$ is the coordinate of an emergent extra dimension, identified with a
renormalization group scale as discussed in the introduction.
$g_{\mu\nu}(x,z)$ with $\mu ~ \& ~ \nu = 0, \ldots, D-1$ is an emergent
dynamical metric field, where metric tensors involved with this extra dimension
are gauge-fixed as $g_{DD}(x,z) = 1$ and $g_{\mu D}(x,z) = 0$.
$\varphi(x,z)$ is a dual scalar field, conventionally taken to be $\varphi(x) =
\frac{1}{N} \Big\langle \sum_{\alpha = 1}^{N} [\phi_{\alpha}(x)]^{2}
\Big\rangle$ in the large $N$ analysis of the quantum field theory
\cite{Large_N_phi4_Theory},
where $\langle \hat{O} \rangle$ is
% an ensemble average
  the expectation value (ensemble average)
of the operator $\hat{O}$ at the Gaussian fixed point.

The dynamics of both emergent metric and dual scalar fields are governed by the following bulk effective action
\begin{align}
  & S_{Bulk}[g_{\mu\nu}(x,z),\varphi(x,z)]
   = N \int_{0}^{z_{f}} d z \int d^{D} x \sqrt{g(x,z)} \Big\{
    \nonumber \\
  &\quad
    +\frac{1}{2u} [\partial_{z} \varphi(x,z)]^{2} + \frac{\mathcal{C}_{\varphi}}{2} g^{\mu\nu}(x,z) [\partial_{\mu} \varphi(x,z)] [\partial_{\nu} \varphi(x,z)] + \mathcal{C}_{\xi} R(x,z) [\varphi(x,z)]^{2}
    \nonumber \\
  & \quad - \frac{1}{2 \lambda} \Big(\partial_{z} g^{\mu\nu}(x,z) - g^{\mu\nu'}(x,z) \big(\partial_{\nu'} \partial_{\mu'} G_{xx'}[g_{\mu\nu}(x,z),\varphi(x,z)]\big)_{x' \rightarrow x} g^{\mu'\nu}(x,z) \Big)^{2}
    \nonumber \\
  &\quad
    + \frac{1}{2 \kappa} \Big( R(x,z) - 2 \Lambda \Big) \Big\} .
    \label{Einstein_Scalar_Theory_Holography}
\end{align}
Here,
the dynamics of $\varphi(x,z)$ is given by rather a conventional form of the
bosonic effective action on a curved spacetime manifold
\cite{Coupling_Scalarfields_Riccicurvature}.
%Here,
$\mathcal{C}_{\varphi}$ and $\mathcal{C}_{\xi}$ are positive constants,
which decrease as the mass of the original scalar fields increases.
The metric-tensor field is described
essentially by the Einstein-Hilbert type
action in a gauge-fixed version,
which originates from quantum fluctuations of matter fields.
This emergent Einstein-Hilbert action
is
% known to be the notion of
analogous to the notion of
induced gravity
\cite{Gradient_Expansion_Gravity_I,Gradient_Expansion_Gravity_II}.
%We point out that there also exists an important modification for the dynamics of the metric along the extra dimensional space.
%$G_{xx'}[g_{\mu\nu}(x,z),\varphi(x,z)]$ is the Green's function given by quantum fluctuations of heavy-mass scalar fields in the renormalization group transformation,
We also point out that
  the dynamics of the metric along the extra dimensional space
  is modified through
  $G_{xx'}[g_{\mu\nu}(x,z),\varphi(x,z)]$,
  which is
the Green's function given by quantum fluctuations of heavy-mass scalar fields in the renormalization group transformation,
\begin{align}
  & \Big\{- \frac{1}{\sqrt{g(x,z)}} \partial_{\mu} \Big( \sqrt{g(x,z)} g^{\mu\nu}(x,z) \partial_{\nu} \Big) + \frac{1}{2 d z} [m^{2} - i \varphi(x,z)] \Big\}
     G_{xx'}[g_{\mu\nu}(x,z),\varphi(x,z)]
     \nonumber\\
  &\quad
     = \frac{1}{\sqrt{g(x,z)}} \delta^{(D)}(x-x') ,
\end{align}
where $d z$ is an infinitesimal parameter of the renormalization group
transformation.
The evolution dynamics of both metric and dual scalar fields are purely
determined by this Green's function
in addition to the emergent local symmetry in the dynamical bulk geometry,
%here,
i.e.,
$(D+1)$-dimensional diffeomorphism invariance
\cite{SungSik_Holography_I,SungSik_Holography_II,SungSik_Holography_III},
%even if
although it
% this
is explicitly broken by the gauge fixing.
In section \ref{Discussion}, we argue that the dynamics of the metric tensor in the renormalization group flow along the extra-dimensional space may be interpreted as introduction of a higher curvature term into the Einstein-Hilbert action, combined with the Ricci flow and the holographic renormalization group.
As clearly
% shown
seen
in this expression, quantum fluctuations of both metric and
dual scalar fields are suppressed in the large $N$ limit.
As a result, both dynamics are described by coupled classical equations of motion in the presence of the emergent extra dimensional space.

Since these coupled equations of motion are
second order in derivatives
% in the second order for the derivative
with respect to the extra dimensional space, we need two boundary conditions.
The first boundary condition is described by an effective UV action
in Eq.\ \eqref{RG master part func}, given by,
\begin{align}
  &
    S_{UV}[g_{\mu\nu}(x,0),\varphi(x,0)]
    = N \int d^{D} x \sqrt{g(x,0)}
    \nonumber \\
  &\qquad \qquad
    \times
    \Big\{ \frac{1}{2u} \Big( \varphi(x,0) - i \xi R(x,0) \Big)^{2} - \frac{1}{2 \lambda} \Big(g^{\mu\nu}(x,0) - g_{B}^{\mu\nu}(x)\Big)^{2} \Big\} .
\end{align}
Of course, an actual equation corresponding to this UV boundary condition has to
take into account a linear $z$ derivative term at $z = 0$, which originates from
the above bulk effective action by the technique of integration-by-parts \cite{Gibbons_Hawking_York_I,Gibbons_Hawking_York_II}.
The other boundary condition is
%supported
provided by an effective IR action
\begin{align}
  &
  S_{IR}[g_{\mu\nu}(x,z_{f}),\varphi(x,z_{f})]
    \nonumber \\
  &\quad
    = N \int d^{D} x \sqrt{g(x,z_{f})}
    \Big\{
    \frac{\mathcal{C}_{\varphi}^{f}}{2} g^{\mu\nu}(x,z_{f}) [\partial_{\mu} \varphi(x,z_{f})] [\partial_{\nu} \varphi(x,z_{f})]
    + \mathcal{C}_{\xi}^{f} R(x,z_{f}) [\varphi(x,z_{f})]^{2}
    \nonumber \\
  &\qquad
    + \frac{1}{2 \kappa_{f}} \Big( R(x,z_{f}) - 2 \Lambda_{f} \Big) \Big\} ,
    \label{S IR}
\end{align}
where the original scalar fields have been integrated out, using the gradient expansion,
to give this effective action
%in the gradient expansion
for the dual scalar fields and metric tensors
\cite{Coupling_Scalarfields_Riccicurvature,Gradient_Expansion_Gravity_I,Gradient_Expansion_Gravity_II}
at $z = z_{f}$. Here, $\mathcal{C}_{\varphi}^{f}$ and $\mathcal{C}_{\xi}^{f}$
are positive coefficients, which also decrease as the mass $m$ increases.
The effective gravitational constant $\kappa_{f}$ and the cosmological parameter
$\Lambda_{f}$ are similarly determined from the gradient expansion.
Also, actual equations to describe the IR boundary condition have to be
supported by the linear $z$-derivative term at $z = z_{f}$.
All these boundary conditions including the Gibbons-Hawking-York term are self-consistently determined by these effective actions. We show all derivations for this emergent holographic construction in section \ref{Holography_Derivation}.

$z_{f}$
in Eqs.\ \eqref{Einstein_Scalar_Theory_Holography} and \eqref{S IR}
is an IR cutoff, which may be regarded to be inversely proportional to
temperature in the renormalization group transformation.
This implies
that the IR boundary conditions correspond to renormalization group equations.
More precisely,
as we discuss in Section \ref{RG_GR_Correspondence_Section}
and appendix \ref{RG_GR_Correspondence_Appendix},
the IR boundary equations for metric tensors correspond to
renormalization group $\beta$-functions of interaction vertices.
On the other hand,
the IR boundary conditions
%while that
of dual scalar fields does to a dual expression of the Callan-Symanzik equation.
%We discuss this aspect in section \ref{RG_GR_Correspondence_Section} and appendix \ref{RG_GR_Correspondence_Appendix} for more details.

Finally, we point out that the limit
% of
$z_{f} \rightarrow 0$
% gives rise
leads to
%a
the conventional large $N$ analysis as follows
\bqa && S_{UV}[g_{\mu\nu}(x,0),\varphi(x,0)] + \lim_{z_{f} \rightarrow 0} S_{IR}[g_{\mu\nu}(x,z_{f}),\varphi(x,z_{f})] = S_{UV}[g^{B}_{\mu\nu}(x),\varphi(x)] , \eqa
where the bulk effective action vanishes in the left-hand-side of the equality and the original scalar fields are integrated out to give an effective action in the large $N$ limit for the right-hand-side of the equality.

\subsection{Transferring quantum entanglement from matter to geometry}

%
%\subsubsection{Entanglement entropy of the IR effective field theory in the replica method}
%

To figure out how the quantum-entanglement structure can be translated into the
emergent classical geometry,
we calculate
in Section \ref{Entanglement_Entropy_Field_Theory},
the entanglement entropy
\cite{Entanglement_Entropy_Calabrese_Cardy_I,Entanglement_Entropy_Calabrese_Cardy_II,Entanglement_Entropy_Ryu_Takayanagi_I,Entanglement_Entropy_Ryu_Takayanagi_II,Entanglement_Entropy_Review_III,Entanglement_Entropy_Review_IV}.
Entanglement entropy may be regarded as an entropy of a subsystem,
given by the von-Neumann entropy of the reduced density matrix for the
subsystem.
Here, we perform a field-theory calculation of the entanglement entropy,
based on the replica technique in a Riemann surface with a conical singularity.
%where all details are shown in section \ref{Entanglement_Entropy_Field_Theory}.
It turns out that the entanglement entropy at the IR cutoff $z = z_{f}$ can be decomposed into $\mathcal{S}_{EE}^{\phi_{\alpha}}(z_{f})$ of the matter and $\mathcal{S}_{EE}^{GR}(z_{f})$ of the emergent gravity as follows
\bqa && \mathcal{S}_{EE}(z_{f}) = \mathcal{S}_{EE}^{\phi_{\alpha}}(z_{f}) + \mathcal{S}_{EE}^{GR}(z_{f}) . \eqa
%
%Here, we point out that all types of effective interactions are turned off for the discussion of the entanglement entropy.
%
Resorting to the heat-kernel method, one finds that the entanglement entropy of free bosons with the flavor degeneracy $N$ follows an area law, given by \cite{Entanglement_Entropy_Heat_Kernel}
\bqa && \mathcal{S}_{EE}^{\phi_{\alpha}}(z_{f}) = \frac{N}{6 (D - 2) (4\pi)^{\frac{D}{2} - 1}} \frac{\mathcal{A}[\Sigma(z_{f})]}{\epsilon^{D-2}} , \eqa
where
\bqa && \mathcal{A}[\Sigma(z_{f})] = \int_{\Sigma(z_{f})} d^{D-2} x_{\perp} \sqrt{\gamma(x,z_{f})} \eqa
is an area of the subsystem at $z = z_{f}$.
$\epsilon$ is a UV cutoff, here corresponding to a microscopic scale for mapping
from a discrete version of the holographic Einstein-Klein-Gordon type theory to the
above continuum field theory Eq.\ \eqref{Einstein_Scalar_Theory_Holography},
as we will discuss in detail in Section \ref{Holography_Derivation}.
% detailed discussed in section \ref{Holography_Derivation}.
On the other hand, the
% The
gravity contribution to the entanglement entropy is given by
\begin{align}
  & \mathcal{S}_{EE}^{GR}(z_{f})
    = N \int d^{D-2} x_{\perp} \int_{0}^{2 \pi} d \theta \int_{0}^{\infty} d r \int_{0}^{z_{f}} d z
    \sqrt{g_{n}(r,x_{\perp},z)} T_{\mu\nu, n}^{Bulk}(r,x_{\perp},z)
    \frac{\partial g_{n}^{\mu\nu}(r,x_{\perp},z)}{\partial n} \Big|_{n = 1}
%
% \nonumber \\   &\qquad     - N \int d^{D-2} x_{\perp} \int_{0}^{2 \pi} d \theta \int_{0}^{\infty} d r \int_{0}^{z_{f}} d z \sqrt{g(r,x_{\perp},z)}     \Big\{ \frac{1}{2u} [\partial_{z} \varphi(r,x_{\perp},z)]^{2}     \nonumber \\   &\qquad     + \frac{\mathcal{C}_{\varphi}}{2} g^{\mu\nu}(r,x_{\perp},z) [\partial_{\mu} \varphi(r,x_{\perp},z)] [\partial_{\nu} \varphi(r,x_{\perp},z)]     + \mathcal{C}_{\xi} R(r,x_{\perp},z) [\varphi(r,x_{\perp},z)]^{2}     \nonumber \\   &\qquad     - \frac{1}{2 \lambda}     \Big(\partial_{z} g^{\mu\nu}(r,x_{\perp},z) - \frac{1}{2 d z} g^{\mu\nu'}(r,x_{\perp},z) \big(\partial_{\nu'} \partial_{\mu'} G_{xx'}[g_{\mu\nu}(r,x_{\perp},z),\varphi(r,x_{\perp},z)]\big)_{x' \rightarrow x} g^{\mu'\nu}(r,x_{\perp},z) \Big)^{2}     \nonumber \\   &\qquad     + \frac{1}{2 \kappa} \Big( R(r,x_{\perp},z) - 2 \Lambda \Big) \Big\}
%
.
\end{align}
%{\color{blue}$g$ is replicated in the first line, but not in the other lines?
%Yes. This is essentially the same expression as that of the matter sector, given by Eqs. (83) and (85).
%For the field and $g$, they are classical?
%Yes, as far as only the large$-N$ limit is concerned.}
Here,
\bqa && T_{\mu\nu,n}^{Bulk}(r,x_{\perp},z) = \frac{1}{\sqrt{g_{n}(r,x_{\perp},z)}} \frac{\partial \sqrt{g_{n}(r,x_{\perp},z)} \mathcal{L}_{Bulk}[g_{\mu\nu,n}(r,x_{\perp},z),\varphi_{n}(r,x_{\perp},z)]}{\partial g^{\mu\nu}_{n}(r,x_{\perp},z)} \eqa
is an energy-momentum tensor of a $(D+1)$-dimensional modified Einstein gravity
in Eq.\ \eqref{Einstein_Scalar_Theory_Holography},
given by $S_{Bulk}[g_{\mu\nu,n}(x,z),\varphi_{n}(x,z)] = \int_{0}^{z_{f}} d z \int
d^{D} x \sqrt{g_{n}(x,z)} \mathcal{L}_{Bulk}[g_{\mu\nu,n}(x,z),\varphi_{n}(x,z)]$.
The quantum average of $\big\langle T_{\mu\nu,n}^{Bulk}(r,x_{\perp},z)
\big\rangle$ with respect to the quantum gravity action is reduced into its
classical value of $T_{\mu\nu,n}^{g}(r,x_{\perp},z)$ in the large $N$ limit. $n$
is the replica index to describe the conical singularity of the geometry
\cite{Entanglement_Entropy_Heat_Kernel}.
The detailed discussion is given in Section \ref{Entanglement_Entropy_Field_Theory}.

% First of all,
An important observation is that the entanglement entropy does not depend on the IR cutoff $z_{f}$ at all. In other words, it is a renormalization-group invariant, regardless of $z_{f}$. This originates from that fact that the partition function is a renormalization-group invariant. As a result, we find the renormalization-group flow equation for the entanglement entropy, given by
\bqa && \partial_{z_{f}} \mathcal{S}_{EE}(z_{f}) = \partial_{z_{f}} \mathcal{S}_{EE}^{\phi_{\alpha}}(z_{f}) + \partial_{z_{f}} \mathcal{S}_{EE}^{GR}(z_{f}) = 0 . \eqa
This observation leads us to conclude that the entanglement entropy of the
matter sector
at UV is transferred into that of the classical bulk gravity part at IR.
  (While we could not verify this relation
%  this renormalization group flow of the entanglement entropy
  based on explicit calculations for the general formula,
  we could check it
  % this relation
  after the first renormalization group transformation,
as discussed in detail
%detailed discussed
in appendix \ref{Entanglement_Entropy_RG_Invariant_General}.)
%This speculation is based on
Recalling
%the fact
that an emergent geometry of gapped
quantum fields is essentially a cap geometry, where the spacetime
does not exist
% after a scale
beyond some scale in the extra dimension
% of the extra dimensional space
\cite{SungSik_Holography_III,Holographic_Description_Kim,Horizon_critical_phenomenon},
%In other words,
the entanglement entropy of the matter sector vanishes after the
cap geometry in the extra dimensional space.
Our conclusion on the entanglement transfer is consistent with a recent study \cite{Entanglement_Entropy_SungSik}.

\section{Derivation of the holographic Landau-Ginzburg effective field theory} \label{Holography_Derivation}

\subsection{The first renormalization group transformation in real space a la Polchinski}

We start from an effective scalar field theory on a $D$-dimensional curved
spacetime described by a background metric tensor $g_{B}^{\mu\nu}$.
The partition function is given by \cite{Coupling_Scalarfields_Riccicurvature}
\begin{align}
  Z &=
      \int D \phi_{\alpha} D g_{\mu\nu}^{(0)} D T_{\mu\nu}^{(0)}
      \exp\Big[ - \int d^{D} x \sqrt{g^{(0)}}
      \Big\{
      \nonumber \\
  &\qquad
      g^{\mu\nu(0)} (\partial_{\mu} \phi_{\alpha}) (\partial_{\nu} \phi_{\alpha})
      + m^{2} \phi_{\alpha}^{2}
      + \xi R^{(0)} \phi_{\alpha}^{2}
      + \frac{u}{2N} \phi_{\alpha}^{2} \phi_{\beta}^{2}
    - N T_{\mu\nu}^{(0)} (g^{\mu\nu(0)} - g_{B}^{\mu\nu}) \Big\} \Big] .
\end{align}
Here, $T_{\mu\nu}^{(0)}$ is a Lagrange multiplier field to impose
the constraint
% equation of
$g^{\mu\nu(0)} = g_{B}^{\mu\nu}$.
For the time being, we do not
% take
consider the effective interaction between energy-momentum tensor
currents,
for simplicity in the presentation of derivations.
To deal with self-interactions between bosons, we introduce a dual scalar field $\varphi^{(0)}$ as
%
%\subsection{Introduction of an order parameter field for self-interactions}
%
\begin{align}
 Z &= \int D \phi_{\alpha} D \varphi^{(0)} D g_{\mu\nu}^{(0)} D
   T_{\mu\nu}^{(0)} \exp\Big[ - \int d^{D} x \sqrt{g^{(0)}} \Big\{
   \nonumber \\
  &\quad
   g^{\mu\nu(0)}
(\partial_{\mu} \phi_{\alpha}) (\partial_{\nu} \phi_{\alpha}) + m^{2}
   \phi_{\alpha}^{2} + \xi R^{(0)} \phi_{\alpha}^{2}
     - i \varphi^{(0)} \phi_{\alpha}^{2} + \frac{N}{2u} \varphi^{(0)2} - N T_{\mu\nu}^{(0)} (g^{\mu\nu(0)} - g_{B}^{\mu\nu}) \Big\} \Big]
\end{align}
by the Hubbard-Stratonovich transformation.

%
%\subsection{Real space renormalization group transformation}
%
%\subsubsection{Introduction of an auxiliary field for $\varphi^{(0)}$}
%

To implement the renormalization group transformation for the dynamics of dual scalar fields, we introduce an auxiliary field $\eta^{(0)}$ in the following way:
\begin{align}
Z &= \int D \phi_{\alpha} D \varphi^{(0)} D \eta^{(0)} D g_{\mu\nu}^{(0)}
    D T_{\mu\nu}^{(0)} \exp\Big[
    - \int d^{D} x \sqrt{g^{(0)}} \Big\{
    \nonumber \\
  &\quad
    g^{\mu\nu(0)}
(\partial_{\mu} \phi_{\alpha}) (\partial_{\nu} \phi_{\alpha}) + m^{2}
\phi_{\alpha}^{2} + \xi R^{(0)} \phi_{\alpha}^{2} - i \varphi^{(0)}
\phi_{\alpha}^{2}
       + \frac{N}{2u} \varphi^{(0)2} + \frac{N}{2u_{\eta}} \eta^{(0)2} - N T_{\mu\nu}^{(0)} (g^{\mu\nu(0)} - g_{B}^{\mu\nu}) \Big\} \Big] .
\end{align}
We emphasize that the introduction of this auxiliary field does not change
any physics except for the normalization constant of the partition function, omitted here for notational simplicity.

%
%\subsubsection{Introduction of low-energy and high-energy fields for $\varphi^{(0)}$}
%

Now, we separate low-energy and high-energy dual scalar fields, $\varphi^{(0)}$ and $\chi^{(0)}$, respectively, as follows
\bqa && \varphi^{(0)} \Longrightarrow \varphi^{(0)} + \chi^{(0)} , ~~~~~ \eta^{(0)} \Longrightarrow c_{\varphi}^{(0)} \varphi^{(0)} + c_{\chi}^{(0)} \chi^{(0)} . \eqa
Here, we determine two coefficients of $c_{\varphi}^{(0)}$ and $c_{\chi}^{(0)}$ as
\bqa && c_{\varphi}^{(0)} = \frac{u^{-1}}{\mu^{(0)} u_{\eta}^{-1/2}} ,
\qquad
c_{\chi}^{(0)} = - \frac{\mu^{(0)}}{u_{\eta}^{-1/2}} ,
\qquad
\mu^{(0)} = \frac{u^{-1/2}}{\sqrt{e^{2 \beta^{(0)} d z} - 1}},
\eqa
such that we
% which
do not generate mixing terms between such light and heavy dual scalar
fields in the mass sector.
% $\mu^{(0)} = \frac{u^{-1/2}}{\sqrt{e^{2 \beta^{(0)} d z} - 1}}$
$\mu^{(0)}$ is an effective mass for the heavy dual scalar field, clarified soon. $\beta^{(0)}$ represents a local speed of coarse graining for the dual scalar field, and $d z$ is an infinitesimal parameter for the renormalization group transformation, also to be clarified below. As a result, we rewrite the mass sector in terms of these light and heavy dual boson fields as follows
\bqa && \frac{N}{2u} \varphi^{(0)2} + \frac{N}{2u_{\eta}} \eta^{(0)2} \Longrightarrow
%
%\frac{N}{2u} (\varphi^{(0)} + \chi^{(0)})^{2} + \frac{N}{2u_{\eta}} (c_{\varphi}^{(0)} \varphi^{(0)} + c_{\chi}^{(0)} \chi^{(0)})^{2} =
%
\frac{N}{2u} e^{2 \beta^{(0)} d z} \varphi^{(0)2} + \frac{N}{2u} \frac{e^{2 \beta^{(0)} d z}}{e^{2 \beta^{(0)} d z} - 1} \chi^{(0)2} . \eqa
Now, it is clear why we call $\chi^{(0)}$ a heavy dual scalar field,
% where its
whose mass is given by $\mu^{(0)}$. Rescaling both dual scalar fields in the following way to return the mass sector to its original form,
\bqa && \varphi^{(0)} \Longrightarrow e^{- \beta^{(0)} d z} \varphi^{(0)} , ~~~~~ \chi^{(0)} \Longrightarrow e^{- \beta^{(0)} d z} \chi^{(0)} , \eqa
we rewrite the partition function in terms of these light and heavy dual scalar bosons as
\begin{align}
Z &= \int D \phi_{\alpha} D \varphi^{(0)} D \chi^{(0)} D g_{\mu\nu}^{(0)}
    D T_{\mu\nu}^{(0)} \exp\Big[ - \int d^{D} x \sqrt{g^{(0)}} \Big\{
    \nonumber \\
  &\qquad
    g^{\mu\nu(0)}
(\partial_{\mu} \phi_{\alpha}) (\partial_{\nu} \phi_{\alpha}) + m^{2}
\phi_{\alpha}^{2} + \xi R^{(0)} \phi_{\alpha}^{2}
    \nonumber \\
 &
       \qquad - i e^{- \beta^{(0)} d z} (\varphi^{(0)} + \chi^{(0)}) \phi_{\alpha}^{2} + \frac{N}{2u} \varphi^{(0)2} + \frac{N}{2u} \frac{1}{e^{2 \beta^{(0)} d z} - 1} \chi^{(0)2} - N T_{\mu\nu}^{(0)} (g^{\mu\nu(0)} - g_{B}^{\mu\nu}) \Big\} \Big] .
\end{align}

%
%\subsubsection{Integration of high-energy fields}
%

It is straightforward to perform the Gaussian integration for the heavy dual
scalar field,
%given by
resulting in the partition function
\begin{align}
  &
    Z = \int D \phi_{\alpha} D \varphi^{(0)} D g_{\mu\nu}^{(0)} D
T_{\mu\nu}^{(0)} \exp\Big[ - \frac{1}{2} \mbox{tr}_{xx'} \ln \frac{N}{2u}
\frac{1}{e^{2 \beta^{(0)} d z} - 1}
                \nn
  &
\qquad- \int d^{D} x \sqrt{g^{(0)}} \Big\{ g^{\mu\nu(0)} (\partial_{\mu}
\phi_{\alpha}) (\partial_{\nu} \phi_{\alpha}) + m^{2} \phi_{\alpha}^{2} + \xi
R^{(0)} \phi_{\alpha}^{2} - i e^{- \beta^{(0)} d z} \varphi^{(0)}
\phi_{\alpha}^{2} + \frac{N}{2u} \varphi^{(0)2}
    \nn
  &
    \qquad \qquad
    + \frac{u}{2N} (1 - e^{- 2\beta^{(0)} d z}) \phi_{\alpha}^{2} \phi_{\beta}^{2} - N T_{\mu\nu}^{(0)} (g^{\mu\nu(0)} - g_{B}^{\mu\nu}) \Big\} \Big] ,
\end{align}
%which generates
where self-interactions between the original scalar bosons
are generated, as expected.
To prepare for the second renormalization group transformation, we perform the
Hubbard-Stratonovich transformation once again for this newly generated
self-interaction term,
%given by
%
%\subsubsection{Introduction of an order parameter field for the resulting self-interaction}
%
\bqa && Z = \int D \phi_{\alpha} D \varphi^{(0)} D \varphi^{(1)} D
g_{\mu\nu}^{(0)} D T_{\mu\nu}^{(0)} \exp\Big[ - \frac{1}{2} \mbox{tr}_{xx'} \ln
\frac{N}{2u} \frac{1}{e^{2 \beta^{(0)} d z} - 1}
\nn &&\qquad  - \int d^{D} x \sqrt{g^{(0)}} \Big\{ g^{\mu\nu(0)} (\partial_{\mu}
\phi_{\alpha}) (\partial_{\nu} \phi_{\alpha}) + m^{2} \phi_{\alpha}^{2} + \xi
R^{(0)} \phi_{\alpha}^{2} - i e^{- \beta^{(0)} d z} \varphi^{(0)}
\phi_{\alpha}^{2} + \frac{N}{2u} \varphi^{(0)2}
\nn &&\qquad \qquad + \frac{N}{2u} \frac{e^{2\beta^{(0)} d z}}{e^{2\beta^{(0)} d z} - 1} \varphi^{(1)2} - i \varphi^{(1)} \phi_{\alpha}^{2} - N T_{\mu\nu}^{(0)} (g^{\mu\nu(0)} - g_{B}^{\mu\nu}) \Big\} \Big] . \eqa
%
%\subsubsection{Scaling and shifting of the new order parameter field}
%
Rescaling the dual scalar field as
\bqa && \varphi^{(1)} \Longrightarrow e^{- \beta^{(0)} d z} \varphi^{(1)} \eqa
%
%\bqa && Z_{1} = \int D \phi_{\alpha} D \varphi^{(0)} D \varphi^{(1)} D g_{\mu\nu}^{(0)} D T_{\mu\nu}^{(0)} \exp\Big[ - \frac{1}{2} \mbox{tr}_{xx'} \ln \frac{N}{2u} \frac{1}{e^{2 \beta^{(0)} d z} - 1} \nn && - \int d^{D} x \sqrt{g^{(0)}} \Big\{ g^{\mu\nu(0)} (\partial_{\mu} \phi_{\alpha}) (\partial_{\nu} \phi_{\alpha}) + m^{2} \phi_{\alpha}^{2} + \xi R^{(0)} \phi_{\alpha}^{2} - i e^{- \beta^{(0)} d z} \varphi^{(0)} \phi_{\alpha}^{2} + \frac{N}{2u} \varphi^{(0)2} \nn && + \frac{N}{2u} \frac{1}{e^{2\beta^{(0)} d z} - 1} \varphi^{(1)2} - i e^{- \beta^{(0)} d z} \varphi^{(1)} \phi_{\alpha}^{2} - N T_{\mu\nu}^{(0)} (g^{\mu\nu(0)} - g_{B}^{\mu\nu}) \Big\} \Big] \eqa
%
and
% taking the field shift of
shifting the field as
\bqa && \varphi^{(1)} \Longrightarrow \varphi^{(1)} - \varphi^{(0)} \eqa
%
%\bqa && Z = \int D \phi_{\alpha} D \varphi^{(0)} D \varphi^{(1)} D g_{\mu\nu}^{(0)} D T_{\mu\nu}^{(0)} \exp\Big[ - \frac{1}{2} \mbox{tr}_{xx'} \ln \frac{N}{2u} \frac{1}{e^{2 \beta^{(0)} d z} - 1} \nn && - \int d^{D} x \sqrt{g^{(0)}} \Big\{ g^{\mu\nu(0)} (\partial_{\mu} \phi_{\alpha}) (\partial_{\nu} \phi_{\alpha}) + m^{2} \phi_{\alpha}^{2} + \xi R^{(0)} \phi_{\alpha}^{2} - i e^{- \beta^{(0)} d z} \varphi^{(1)} \phi_{\alpha}^{2} \nn && + \frac{N}{2u} \varphi^{(0)2} + \frac{N}{2u} \frac{1}{e^{2\beta^{(0)} d z} - 1} (\varphi^{(1)} - \varphi^{(0)})^{2} - N T_{\mu\nu}^{(0)} (g^{\mu\nu(0)} - g_{B}^{\mu\nu}) \Big\} \Big] \eqa
%
%\subsubsection{Further shift of order parameter fields}
%
%\bqa && Z_{1} = \int D \phi_{\alpha} D \varphi^{(0)} D \varphi^{(1)} D g_{\mu\nu}^{(0)} D T_{\mu\nu}^{(0)} \exp\Big[ - \frac{1}{2} \mbox{tr}_{xx'} \ln \frac{N}{2u} \frac{1}{e^{2 \beta^{(0)} d z} - 1} \nn && - \int d^{D} x \sqrt{g^{(0)}} \Big\{ g^{\mu\nu(0)} (\partial_{\mu} \phi_{\alpha}) (\partial_{\nu} \phi_{\alpha}) - i e^{- \beta^{(0)} d z} \varphi^{(1)} \phi_{\alpha}^{2} \nn && + \frac{N}{2u} \varphi^{(0)2} + \frac{N}{2u} \frac{1}{e^{2\beta^{(0)} d z} - 1} \Big(\varphi^{(1)} - \varphi^{(0)} - i e^{\beta^{(0)} d z} m^{2} - i e^{- \beta^{(0)} d z} \xi R^{(0)}\Big)^{2} - N T_{\mu\nu}^{(0)} (g^{\mu\nu(0)} - g_{B}^{\mu\nu}) \Big\} \Big] \eqa
%
together with $\varphi^{(1)} \Longrightarrow \varphi^{(1)} - i e^{- \beta^{(0)} d z} \xi R^{(0)}$ and $\varphi^{(0)} \Longrightarrow \varphi^{(0)} - i e^{- \beta^{(0)} d z} \xi R^{(0)}$, we obtain the following expression of the partition function after the first renormalization group transformation for the dual scalar-field sector
\begin{align}
Z &= \int D \phi_{\alpha} D \varphi^{(0)} D \varphi^{(1)} D g_{\mu\nu}^{(0)} D
T_{\mu\nu}^{(0)} \exp\Big[ - \frac{1}{2} \mbox{tr}_{xx'} \ln \frac{N}{2u}
\frac{1}{e^{2 \beta^{(0)} d z} - 1}
    \nn
  &\qquad  - \int d^{D} x \sqrt{g^{(0)}} \Big\{ g^{\mu\nu(0)} (\partial_{\mu}
\phi_{\alpha}) (\partial_{\nu} \phi_{\alpha}) + m^{2} \phi_{\alpha}^{2} - i e^{-
  \beta^{(0)} d z} \varphi^{(1)} \phi_{\alpha}^{2}
    \nn
  &\qquad + \frac{N}{2u} \Big( \varphi^{(0)} - i e^{- \beta^{(0)} d z} \xi R^{(0)} \Big)^{2} + \frac{N}{2u} \frac{1}{e^{2\beta^{(0)} d z} - 1} (\varphi^{(1)} - \varphi^{(0)})^{2} - N T_{\mu\nu}^{(0)} (g^{\mu\nu(0)} - g_{B}^{\mu\nu}) \Big\} \Big] .
              \end{align}

%
%\subsubsection{Introduction of an auxiliary field for $\phi_{\alpha}$}
%

Now, we perform the first renormalization group transformation for the original boson sector.
As
% taken
in the renormalization group transformation for the dual scalar-field sector,
we introduce an auxiliary field $\psi_{\alpha}$ into the partition function
\begin{align}
& Z = \int D \phi_{\alpha} D \psi_{\alpha} D \varphi^{(0)} D \varphi^{(1)}
D g_{\mu\nu}^{(0)} D T_{\mu\nu}^{(0)} \exp\Big[ - \frac{1}{2} \mbox{tr}_{xx'}
\ln \frac{N}{2u} \frac{1}{e^{2 \beta^{(0)} d z} - 1}
\nn &\qquad  - \int d^{D} x \sqrt{g^{(0)}} \Big\{ g^{\mu\nu(0)} (\partial_{\mu}
\phi_{\alpha}) (\partial_{\nu} \phi_{\alpha}) + m^{2} \phi_{\alpha}^{2} - i e^{-
  \beta^{(0)} d z} \varphi^{(1)} \phi_{\alpha}^{2} + M^{2} \psi_{\alpha}^{2}
       \nn &\qquad  + \frac{N}{2u} \Big( \varphi^{(0)} - i e^{- \beta^{(0)} d z} \xi R^{(0)} \Big)^{2} + \frac{N}{2u} \frac{1}{e^{2\beta^{(0)} d z} - 1} (\varphi^{(1)} - \varphi^{(0)})^{2} - N T_{\mu\nu}^{(0)} (g^{\mu\nu(0)} - g_{B}^{\mu\nu}) \Big\} \Big] .
\end{align}
We point out again that the introduction of the auxiliary field $\psi_{\alpha}$ does not change any physics except for the normalization constant of the partition function, omitted here for notational simplicity.

%
%\subsubsection{Introduction of low-energy and high-energy fields for $\phi_{\alpha}$}
%

We separate low-energy and high-energy scalar fields, $\phi_{\alpha}$ and $\Phi_{\alpha}$, respectively,
\bqa && \phi_{\alpha} \Longrightarrow \phi_{\alpha} + \Phi_{\alpha} , ~~~~~ \psi_{\alpha} \Longrightarrow c_{\phi}^{(0)} \phi_{\alpha} + c_{\Phi}^{(0)} \Phi_{\alpha} , \eqa
where two coefficients of $c_{\phi}^{(0)}$ and $c_{\Phi}^{(0)}$ are given by
%
%\bqa (- i e^{- \beta^{(0)} d z} \varphi^{(1)}) \phi_{\alpha}^{2} + M^{2} \psi_{\alpha}^{2} &\Longrightarrow& (- i e^{- \beta^{(0)} d z} \varphi^{(1)}) (\phi_{\alpha} + \Phi_{\alpha})^{2} + M^{2} (c_{\phi}^{(0)} \phi_{\alpha} + c_{\Phi}^{(0)} \Phi_{\alpha})^{2} \nn &=& [(- i e^{- \beta^{(0)} d z} \varphi^{(1)}) + M^{2} c_{\phi}^{(0)2}] \phi_{\alpha}^{2} + [(- i e^{- \beta^{(0)} d z} \varphi^{(1)}) + M^{2} c_{\Phi}^{(0)2}] \Phi_{\alpha}^{2} \nn &+& 2 [(- i e^{- \beta^{(0)} d z} \varphi^{(1)}) + M^{2} c_{\phi}^{(0)} c_{\Phi}^{(0)}] \phi_{\alpha} \Phi_{\alpha} \eqa
%
%\bqa && (- i e^{- \beta^{(0)} d z} \varphi^{(1)}) + M^{2} c_{\phi}^{(0)} c_{\Phi}^{(0)} = 0 \eqa
%
\begin{align}
c_{\phi}^{(0)} = \frac{(m^{2} - i e^{- \beta^{(0)} d z} \varphi^{(1)})}{\mu^{(0)} M} , ~~~~~ c_{\Phi}^{(0)} = - \frac{\mu^{(0)}}{M} , ~~~~~ \mu^{(0)} = \frac{(m^{2} - i e^{- \beta^{(0)} d z} \varphi^{(1)})^{1/2}}{\sqrt{e^{2 \alpha^{(0)} d z} - 1}} .
\end{align}
Here, as $\beta^{(0)}$, $\alpha^{(0)}$ represents a local speed of coarse
graining for the dual scalar field.
Then, the mass sector of the original scalar fields reads
\begin{align}
  &
  (m^{2}
  - i e^{- \beta^{(0)} d z} \varphi^{(1)}) \phi_{\alpha}^{2} + M^{2}
  \psi_{\alpha}^{2}
%
%&=& \frac{(- i e^{- \beta^{(0)} d z} \varphi^{(1)})}{\mu^{(0)2}} [(- i e^{- \beta^{(0)} d z} \varphi^{(1)}) + \mu^{(0)2}] \phi_{\alpha}^{2} + [(- i e^{- \beta^{(0)} d z} \varphi^{(1)}) + \mu^{(0)2}] \Phi_{\alpha}^{2} \nn
%
     \nonumber \\
  &\Longrightarrow e^{2 \alpha^{(0)} d z} (m^{2} - i e^{- \beta^{(0)} d z} \varphi^{(1)}) \phi_{\alpha}^{2} + \frac{e^{2 \alpha^{(0)} d z}}{e^{2 \alpha^{(0)} d z} - 1} (m^{2} - i e^{- \beta^{(0)} d z} \varphi^{(1)}) \Phi_{\alpha}^{2} .
\end{align}
Rescaling both scalar fields as
\begin{align}
 \phi_{\alpha} \Longrightarrow e^{- \alpha^{(0)} d z} \phi_{\alpha} , ~~~~~ \Phi_{\alpha} \Longrightarrow e^{- \alpha^{(0)} d z} \Phi_{\alpha} ,
\end{align}
we rewrite the partition function in terms of light and heavy scalar fields as follows
\begin{align}
& Z = \int D \phi_{\alpha} D \Phi_{\alpha} D \varphi^{(0)} D \varphi^{(1)}
D g_{\mu\nu}^{(0)} D T_{\mu\nu}^{(0)} \exp\Big[ - \frac{1}{2} \mbox{tr}_{xx'}
\ln \frac{N}{2u} \frac{1}{e^{2 \beta^{(0)} d z} - 1}
                \nn
  &\qquad - \int d^{D} x \sqrt{g^{(0)}} \Big\{ g^{\mu\nu(0)} [\partial_{\mu}
e^{- \alpha^{(0)} d z} (\phi_{\alpha} + \Phi_{\alpha})] [\partial_{\nu} e^{-
  \alpha^{(0)} d z} (\phi_{\alpha} + \Phi_{\alpha})]
    \nn
  &\qquad + (m^{2} - i e^{- \beta^{(0)} d z} \varphi^{(1)})
\phi_{\alpha}^{2} + \frac{1}{e^{2 \alpha^{(0)} d z} - 1} (m^{2} - i e^{-
  \beta^{(0)} d z} \varphi^{(1)}) \Phi_{\alpha}^{2}
    \nn
  &\qquad + \frac{N}{2u} \Big( \varphi^{(0)} - i e^{- \beta^{(0)} d z} \xi R^{(0)} \Big)^{2} + \frac{N}{2u} \frac{1}{e^{2\beta^{(0)} d z} - 1} (\varphi^{(1)} - \varphi^{(0)})^{2} - N T_{\mu\nu}^{(0)} (g^{\mu\nu(0)} - g_{B}^{\mu\nu}) \Big\} \Big] .
\end{align}
Here, effective couplings between light and heavy scalar bosons arise in the kinetic-energy sector.

The next step is to perform the Gaussian integral for the heavy scalar fields. For our simple derivation, we fix a gauge for the local speed of coarse graining as
\bqa && \partial_{\mu} \alpha^{(0)} = 0 . \eqa
The physical meaning of this gauge fixing is that the renormalization group transformation is taken into account in a uniform way, where the coefficient in front of the infinitesimal parameter for the renormalization group transformation does not depend on the spacetime coordinate $x$.
%
%\bqa && Z_{1} = \int D \phi_{\alpha} D \Phi_{\alpha} D \varphi^{(0)} D \varphi^{(1)} D g_{\mu\nu}^{(0)} D T_{\mu\nu}^{(0)} \exp\Big[ - \frac{1}{2} \mbox{tr}_{xx'} \ln \frac{N}{2u} \frac{1}{e^{2 \beta^{(0)} d z} - 1} \nn && - \int d^{D} x \sqrt{g^{(0)}} \Big\{ (1 - 2 \alpha^{(0)} d z) g^{\mu\nu(0)} (\partial_{\mu} \phi_{\alpha}) (\partial_{\nu} \phi_{\alpha}) + (- i e^{- \beta^{(0)} d z} \varphi^{(1)}) \phi_{\alpha}^{2} \nn && + (1 - 2 \alpha^{(0)} d z) g^{\mu\nu(0)} (\partial_{\mu} \Phi_{\alpha}) (\partial_{\nu} \Phi_{\alpha}) + \frac{1}{e^{2 \alpha^{(0)} d z} - 1} (- i e^{- \beta^{(0)} d z} \varphi^{(1)}) \Phi_{\alpha}^{2} \nn && + (1 - 2 \alpha^{(0)} d z) g^{\mu\nu(0)} (\partial_{\mu} \phi_{\alpha}) (\partial_{\nu} \Phi_{\alpha}) + (1 - 2 \alpha^{(0)} d z) g^{\mu\nu(0)} (\partial_{\mu} \Phi_{\alpha}) (\partial_{\nu} \phi_{\alpha}) \nn && + \frac{N}{2u} \Big( \varphi^{(0)} - i e^{\beta^{(0)} d z} m^{2} - i e^{- \beta^{(0)} d z} \xi R^{(0)} \Big)^{2} + \frac{N}{2u} \frac{1}{e^{2\beta^{(0)} d z} - 1} (\varphi^{(1)} - \varphi^{(0)})^{2} - N T_{\mu\nu}^{(0)} (g^{\mu\nu(0)} - g_{B}^{\mu\nu}) \Big\} \Big] \eqa
%
Integrating over the heavy scalar fields with this gauge fixing,
%
%\bqa && Z = \int D \phi_{\alpha} D \Phi_{\alpha} D \varphi^{(0)} D \varphi^{(1)} D g_{\mu\nu}^{(0)} D T_{\mu\nu}^{(0)} \exp\Big[ - \frac{1}{2} \mbox{tr}_{xx'} \ln \frac{N}{2u} \frac{1}{e^{2 \beta^{(0)} d z} - 1} \nn && - \int d^{D} x \sqrt{g^{(0)}} \Big\{ e^{- 2 \alpha^{(0)} d z} g^{\mu\nu(0)} (\partial_{\mu} \phi_{\alpha}) (\partial_{\nu} \phi_{\alpha}) + (m^{2} - i e^{- \beta^{(0)} d z} \varphi^{(1)}) \phi_{\alpha}^{2} \nn && + e^{- 2 \alpha^{(0)} d z} g^{\mu\nu(0)} (\partial_{\mu} \Phi_{\alpha}) (\partial_{\nu} \Phi_{\alpha}) + \frac{(m^{2} - i e^{- \beta^{(0)} d z} \varphi^{(1)})}{e^{2 \alpha^{(0)} d z} - 1} \Phi_{\alpha}^{2} \nn && + e^{- 2 \alpha^{(0)} d z} g^{\mu\nu(0)} (\partial_{\mu} \phi_{\alpha}) (\partial_{\nu} \Phi_{\alpha}) + e^{- 2 \alpha^{(0)} d z} g^{\mu\nu(0)} (\partial_{\mu} \Phi_{\alpha}) (\partial_{\nu} \phi_{\alpha}) \nn && + \frac{N}{2u} \Big( \varphi^{(0)} - i e^{- \beta^{(0)} d z} \xi R^{(0)} \Big)^{2} + \frac{N}{2u} \frac{1}{e^{2\beta^{(0)} d z} - 1} (\varphi^{(1)} - \varphi^{(0)})^{2} - N T_{\mu\nu}^{(0)} (g^{\mu\nu(0)} - g_{B}^{\mu\nu}) \Big\} \Big] , \eqa
%
%\subsubsection{Integration of high-energy fields}
%
we obtain
%
%\bqa && Z = \int D \phi_{\alpha x} D \varphi_{x}^{(0)} D \varphi_{x}^{(1)} D g_{\mu\nu x}^{(0)} D T_{\mu\nu x}^{(0)} \nn && \exp\Big[ - \frac{N}{2} \mbox{tr}_{xx'} \ln \sqrt{g_{x}^{(0)}} \Big\{- \frac{e^{- 2 \alpha^{(0)} d z}}{\sqrt{g_{x}^{(0)}}} \partial_{\mu} \Big( \sqrt{g_{x}^{(0)}} g^{\mu\nu(0)}_{x} \partial_{\nu} \Big) + \frac{(m^{2} - i e^{- \beta^{(0)} d z} \varphi_{x}^{(1)})}{e^{2 \alpha^{(0)} d z} - 1} \Big\} \nn && - \frac{1}{2} \mbox{tr}_{xx'} \ln \frac{N}{2u} \frac{1}{e^{2 \beta^{(0)} d z} - 1} - \int d^{D} x \sqrt{g_{x}^{(0)}} \Big\{ e^{- 2 \alpha^{(0)} d z} g_{x}^{\mu\nu(0)} (\partial_{\mu} \phi_{\alpha x}) (\partial_{\nu} \phi_{\alpha x}) + (m^{2} - i e^{- \beta^{(0)} d z} \varphi_{x}^{(1)}) \phi_{\alpha x}^{2} \nn && - \int d^{D} x' \sqrt{g_{x'}^{(0)}} e^{- 4 \alpha^{(0)} d z} g^{\mu\nu(0)}_{x} g^{\mu'\nu'(0)}_{x'} (\partial_{\mu} \phi_{\alpha x}) (\partial_{\nu} \partial_{\mu'} G_{xx'}^{(0)}) (\partial_{\nu'} \phi_{\alpha x'}) \nn && + \frac{N}{2u} \Big( \varphi_{x}^{(0)} - i e^{- \beta^{(0)} d z} \xi R_{x}^{(0)} \Big)^{2} + \frac{N}{2u} \frac{1}{e^{2\beta^{(0)} d z} - 1} (\varphi_{x}^{(1)} - \varphi_{x}^{(0)})^{2} - N T_{\mu\nu x}^{(0)} (g_{x}^{\mu\nu(0)} - g_{B x}^{\mu\nu}) \Big\} \Big] . \eqa
%
%\bqa && \Big\{- \frac{e^{- 2 \alpha^{(0)} d z}}{\sqrt{g_{x}^{(0)}}} \partial_{\mu} \Big( \sqrt{g_{x}^{(0)}} g^{\mu\nu(0)}_{x} \partial_{\nu} \Big) + \frac{(m^{2} - i e^{- \beta^{(0)} d z} \varphi_{x}^{(1)})}{e^{2 \alpha^{(0)} d z} - 1} \Big\} G_{xx'}^{(0)} = \frac{1}{\sqrt{g_{x}^{(0)}}} \delta^{(D)}(x-x') \eqa
%
%\subsubsection{Redefinition of Green's function}
%
\begin{align}
  & Z = \int D \phi_{\alpha x} D \varphi_{x}^{(0)} D \varphi_{x}^{(1)} D
    g_{\mu\nu x}^{(0)} D T_{\mu\nu x}^{(0)} \exp\Big\{ - S_{\phi\phi}^{(0)} - \Delta
    S_{\phi\Phi}^{(0)} - S_{\Phi\Phi}^{(0)} - S_{\varphi\varphi}^{(1)}
    \nn
  &\qquad
    + N \int d^{D} x \sqrt{g_{x}^{(0)}} T_{\mu\nu x}^{(0)} (g_{x}^{\mu\nu(0)} - g_{B x}^{\mu\nu}) \Big\} .
    %
    % \nn && \exp\Big[ - \frac{N}{2} \mbox{tr}_{xx'} \ln e^{- 2 \alpha^{(0)} d z} \sqrt{g_{x}^{(0)}} \Big\{- \frac{1}{\sqrt{g_{x}^{(0)}}} \partial_{\mu} \Big( \sqrt{g_{x}^{(0)}} g^{\mu\nu(0)}_{x} \partial_{\nu} \Big) + \frac{e^{(2 \alpha^{(0)} - \beta^{(0)}) d z}}{e^{2 \alpha^{(0)} d z} - 1} (e^{- 2 \alpha^{(0)} d z} m^{2} - i \varphi_{x}^{(1)})\Big\} \nn && - \frac{1}{2} \mbox{tr}_{xx'} \ln \frac{N}{2u} \frac{1}{e^{2 \beta^{(0)} d z} - 1} - \int d^{D} x \sqrt{g_{x}^{(0)}} \Big\{ e^{- 2 \alpha^{(0)} d z} g_{x}^{\mu\nu(0)} (\partial_{\mu} \phi_{\alpha x}) (\partial_{\nu} \phi_{\alpha x}) + (m^{2} - i e^{- \beta^{(0)} d z} \varphi_{x}^{(1)}) \phi_{\alpha x}^{2} \nn && - \int d^{D} x' \sqrt{g_{x'}^{(0)}} e^{- 2 \alpha^{(0)} d z} g^{\mu\nu(0)}_{x} g^{\mu'\nu'(0)}_{x'} (\partial_{\mu} \phi_{\alpha x}) (\partial_{\nu} \partial_{\mu'} G_{xx'}^{(0)}) (\partial_{\nu'} \phi_{\alpha x'}) \nn && + \frac{N}{2u} \Big( \varphi_{x}^{(0)} - i e^{- \beta^{(0)} d z} \xi R_{x}^{(0)} \Big)^{2} + \frac{N}{2u} \frac{1}{e^{2\beta^{(0)} d z} - 1} (\varphi_{x}^{(1)} - \varphi_{x}^{(0)})^{2} - N T_{\mu\nu x}^{(0)} (g_{x}^{\mu\nu(0)} - g_{B x}^{\mu\nu}) \Big\} \Big] .
    %
\end{align}
Here, $S_{\phi\phi}^{(0)}$ is an effective action before the first renormalization group transformation, given by
\begin{align}
  S_{\phi\phi}^{(0)} = \int d^{D} x \sqrt{g_{x}^{(0)}} \Big\{ e^{- 2 \alpha^{(0)} d z} g_{x}^{\mu\nu(0)} (\partial_{\mu} \phi_{\alpha x}) (\partial_{\nu} \phi_{\alpha x}) + (m^{2} - i e^{- \beta^{(0)} d z} \varphi_{x}^{(1)}) \phi_{\alpha x}^{2} \Big\} .
\end{align}
$\Delta S_{\phi\Phi}^{(0)}$ is an effective action to result from the mixing terms between light and heavy scalar fields in the kinetic-energy part, given by
\begin{align}
\Delta S_{\phi\Phi}^{(0)} = - \int d^{D} x \sqrt{g_{x}^{(0)}} \int d^{D} x' \sqrt{g_{x'}^{(0)}} e^{- 2 \alpha^{(0)} d z} g^{\mu\nu(0)}_{x} g^{\mu'\nu'(0)}_{x'} (\partial_{\mu} \phi_{\alpha x}) (\partial_{\nu} \partial_{\mu'} G_{xx'}^{(0)}) (\partial_{\nu'} \phi_{\alpha x'}) ,
\end{align}
where $G_{xx'}^{(0)}$ is the Green's function of the heavy scalar bosons,
%given by
and obeys
\begin{align}
  & \Big\{- \frac{1}{\sqrt{g_{x}^{(0)}}} \partial_{\mu} \Big( \sqrt{g_{x}^{(0)}} g^{\mu\nu(0)}_{x} \partial_{\nu} \Big) + \frac{e^{(2 \alpha^{(0)} - \beta^{(0)}) d z}}{e^{2 \alpha^{(0)} d z} - 1} (e^{- 2 \alpha^{(0)} d z} m^{2} - i \varphi_{x}^{(1)}) \Big\} G_{xx'}^{(0)}
    \nonumber \\
  &
    = \frac{1}{\sqrt{g_{x}^{(0)}}} \delta^{(D)}(x-x') .
\end{align}
This Green's function plays a central role in the renormalization group transformation, regarded to be the only dynamic information. $S_{\Phi\Phi}^{(0)}$ serves as an effective vacuum action to originate from quantum fluctuations of the heavy scalar fields, given by
\bqa && S_{\Phi\Phi}^{(0)} = \frac{N}{2} \mbox{tr}_{xx'} \ln e^{- 2 \alpha^{(0)} d z} \sqrt{g_{x}^{(0)}} \Big\{- \frac{1}{\sqrt{g_{x}^{(0)}}} \partial_{\mu} \Big( \sqrt{g_{x}^{(0)}} g^{\mu\nu(0)}_{x} \partial_{\nu} \Big) + \frac{e^{(2 \alpha^{(0)} - \beta^{(0)}) d z}}{e^{2 \alpha^{(0)} d z} - 1} (e^{- 2 \alpha^{(0)} d z} m^{2} - i \varphi_{x}^{(1)})\Big\} . \nn \eqa
Finally, $S_{\varphi\varphi}^{(1)}$ is an effective action after the first renormalization group transformation for the dual scalar bosons, given by
\begin{align}
  & S_{\varphi\varphi}^{(1)} = \frac{1}{2} \mbox{tr}_{xx'} \ln \frac{N}{2u} \frac{1}{e^{2 \beta^{(0)} d z} - 1}
     \nonumber \\
  &\quad
     + \int d^{D} x \sqrt{g_{x}^{(0)}} \Big\{ \frac{N}{2u} \Big( \varphi_{x}^{(0)} - i e^{- \beta^{(0)} d z} \xi R_{x}^{(0)} \Big)^{2} + \frac{N}{2u} \frac{1}{e^{2\beta^{(0)} d z} - 1} (\varphi_{x}^{(1)} - \varphi_{x}^{(0)})^{2} \Big\} .
\end{align}

Taking the gradient expansion in the vacuum sector of $S_{\Phi\Phi}^{(0)}$ with respect to the mass of scalar bosons and keeping all terms up to the linear order in $d z$, we obtain
\begin{align}
  &
    Z = \int D \phi_{\alpha x} D \varphi_{x}^{(0)} D \varphi_{x}^{(1)} D
g_{\mu\nu x}^{(0)} D T_{\mu\nu x}^{(0)}
 \exp\Big[
    \nn
  &\qquad \frac{N}{2} \mbox{tr}_{xx'} \ln (2 \alpha^{(0)} d z) +
\frac{1}{2} \mbox{tr}_{xx'} \ln (2 \beta^{(0)} d z) - \frac{N}{2}
\mbox{tr}_{xx'} \ln \sqrt{g_{x}^{(0)}} m^{2} - \frac{1}{2} \mbox{tr}_{xx'} \ln
\frac{N}{2u}
    \nn
  &\qquad - \int d^{D} x \sqrt{g_{x}^{(0)}} \Big\{ (1 - 2 \alpha^{(0)} d z)
g_{x}^{\mu\nu(0)} (\partial_{\mu} \phi_{\alpha x}) (\partial_{\nu} \phi_{\alpha
  x}) + (m^{2} - i \varphi_{x}^{(1)}) \phi_{\alpha x}^{2}
    \nn
  &\qquad - \int d^{D} x' \sqrt{g_{x'}^{(0)}} g^{\mu\nu(0)}_{x}
g^{\mu'\nu'(0)}_{x'} (\partial_{\mu} \phi_{\alpha x}) (\partial_{\nu}
\partial_{\mu'} G_{xx'}^{(0)}) (\partial_{\nu'} \phi_{\alpha x'})
    \nn
  &\qquad + 2 \alpha^{(0)} d z N \Big( - \mathcal{C}_{\Lambda} +
\mathcal{C}_{R} R_{x}^{(0)} + \frac{\mathcal{C}_{\varphi}}{2} g_{x}^{\mu\nu(0)}
(\partial_{\mu} \varphi_{x}^{(1)}) (\partial_{\nu} \varphi_{x}^{(1)}) +
\mathcal{C}_{\xi} R_{x}^{(0)} \varphi_{x}^{(1) 2} \Big)
    \nn
  &\qquad + \frac{N}{2u} \Big( \varphi_{x}^{(0)} - i \xi R_{x}^{(0)} \Big)^{2} + \beta^{(0)} d z \frac{N}{4u} \Big(\frac{\varphi_{x}^{(1)} - \varphi_{x}^{(0)}}{\beta^{(0)} d z}\Big)^{2} - N T_{\mu\nu x}^{(0)} (g_{x}^{\mu\nu(0)} - g_{B x}^{\mu\nu}) \Big\} \Big] . \label{Non_Local_Quadratic_Term_Higher_Spin_Field}
\end{align}
We point out that the Einstein-Hilbert action appears to count vacuum
fluctuations of high-energy scalar bosons in the background geometry, known to
be the notion of induced gravity
\cite{Coupling_Scalarfields_Riccicurvature,Gradient_Expansion_Gravity_I,Gradient_Expansion_Gravity_II}.
In addition, dual scalar bosons acquire their kinetic energy, where
$\mathcal{C}_{\varphi}$ and $\mathcal{C}_{\xi}$ are positive constants.
% All coefficients of
All the coefficients
$\mathcal{C}_{\Lambda}$, $\mathcal{C}_{R}$, $\mathcal{C}_{\varphi}$, and
$\mathcal{C}_{\xi}$ decrease as the mass of scalar bosons increases. The Green's
function $G_{xx'}^{(0)}$ is determined
% given
%by the linear order in $d z$
to linear order in $d z$
%as follows
by the equation
\begin{align}
  \Big\{- \frac{1}{\sqrt{g_{x}^{(0)}}} \partial_{\mu} \Big( \sqrt{g_{x}^{(0)}} g^{\mu\nu(0)}_{x} \partial_{\nu} \Big) + \frac{1}{2 \alpha^{(0)} d z} (m^{2} - i \varphi_{x}^{(1)}) \Big\} G_{xx'}^{(0)} = \frac{1}{\sqrt{g_{x}^{(0)}}} \delta^{(D)}(x-x') .
\end{align}

\subsection{Recursive renormalization group transformations a la Sung-Sik Lee}

%
%\subsubsection{Approximation for locality}
%

%An
The idea for the second renormalization group transformation is that
$S_{\phi\phi}^{(0)} + \Delta S_{\phi\Phi}^{(0)}$ is reformulated as
$S_{\phi\phi}^{(1)}$, where the metric tensor is updated to be from
$g^{(0)\mu\nu}$ to $g^{(1)\mu\nu}$ appropriately.
%However, it turns out that this reformulation is not easy % to be performed
However, this reformulation turns out not to be straightforward
% because
since
$\Delta
S_{\phi\Phi}^{(0)}$ is a nonlocal action
% given by
because of
$G_{xx'}^{(0)}$, which differs
from the local kinetic energy term of $S_{\phi\phi}^{(0)}$. An important
observation is that this Green's function is given by an exponential form $\sim
\exp\Big(- (2 \alpha^{(0)} d z)^{-1/2} \sqrt{m^{2} - i \varphi_{x}^{(1)}} |x -
x'| \Big)$, where the inverse of the decay length is $(2 \alpha^{(0)} d
z)^{-1/2} \sqrt{m^{2} - i \varphi_{x}^{(1)}}$.
As a result, we may
only
keep
the local term
% only
in the gradient expansion as follows
\bqa && (\partial_{\nu} \partial_{\mu'} G_{xx'}^{(0)}) \approx (\partial_{\nu}
\partial_{\mu'} G_{xx'}^{(0)}) \frac{1}{\sqrt{g_{x'}}} \delta^{(D)}(x-x') +
\ldots , \label{Locality_Approximation_Green_Ft} \eqa
where higher gradient terms are in higher orders of $d z$.

%
%\subsubsection{Effective local field theory after the first renormalization group transformation}
%
%\bqa && Z = \int D \phi_{\alpha} D \varphi^{(0)} D \varphi^{(1)} D g_{\mu\nu}^{(0)} D T_{\mu\nu}^{(0)} \nn && \exp\Big[ \frac{N}{2} \mbox{tr}_{xx'} \ln \Big( {2 \alpha^{(0)} d z} \sqrt{g^{(0)}} \Big) + \frac{1}{2} \mbox{tr}_{xx'} \ln (2 \beta^{(0)} d z) - \frac{N}{2} \mbox{tr}_{xx'} \ln \sqrt{g^{(0)}} (- i \varphi^{(1)}) - \frac{1}{2} \mbox{tr}_{xx'} \ln \frac{N}{2u} \nn && - \int d^{D} x \sqrt{g^{(0)}} \Big\{ (1 - 2 \alpha^{(0)} d z) g^{\mu\nu(0)} (\partial_{\mu} \phi_{\alpha}) (\partial_{\nu} \phi_{\alpha}) + (- i \varphi^{(1)}) \phi_{\alpha}^{2} \nn && - g^{\mu\nu'(0)} (\partial_{\nu'} \partial_{\mu'} G_{xx'}^{(0)})_{x' \rightarrow x} g^{\mu'\nu(0)} (\partial_{\mu} \phi_{\alpha}) (\partial_{\nu} \phi_{\alpha}) + 2 \alpha^{(0)} d z N \Big( - \mathcal{C}_{\Lambda}[(- i \varphi^{(1)})] + \mathcal{C}_{R}[(- i \varphi^{(1)})] R^{(0)} \Big) \nn && + \frac{N}{2u} \Big( \varphi^{(0)} - i m^{2} - i \xi R^{(0)} \Big)^{2} + \beta^{(0)} d z \frac{N}{4u} \Big(\frac{\varphi^{(1)} - \varphi^{(0)}}{\beta^{(0)} d z}\Big)^{2} - N T_{\mu\nu}^{(0)} (g^{\mu\nu(0)} - g_{B}^{\mu\nu}) \Big\} \Big] \eqa
%

Taking into account this locality approximation
%justified in the linear order of $d z$,
to linear order in $dz$,
we obtain
\begin{align}
& Z = \int D \phi_{\alpha} D \varphi^{(0)} D \varphi^{(1)} D
g_{\mu\nu}^{(0)} D T_{\mu\nu}^{(0)}
\exp\Big[
                \nn
  &\qquad
\frac{N}{2} \mbox{tr}_{xx'} \ln (2 \alpha^{(0)} d z) +
\frac{1}{2} \mbox{tr}_{xx'} \ln (2 \beta^{(0)} d z) - \frac{N}{2}
\mbox{tr}_{xx'} \ln \sqrt{g^{(0)}} m^{2} - \frac{1}{2} \mbox{tr}_{xx'} \ln
\frac{N}{2u}
    \nn
  &\qquad - \int d^{D} x \sqrt{g^{(0)}} \Big\{ g^{\mu\nu(0)} (\partial_{\mu}
\phi_{\alpha}) (\partial_{\nu} \phi_{\alpha}) + (m^{2} - i \varphi^{(1)})
\phi_{\alpha}^{2}
    \nn
  &\qquad - 2 \alpha^{(0)} d z \Big( g^{\mu\nu(0)} +
  g^{\mu\nu'(0)} (\partial_{\nu'} \partial_{\mu'} G_{xx'}^{(0)})_{x'
  \rightarrow x} g^{\mu'\nu(0)} \Big) (\partial_{\mu} \phi_{\alpha})
(\partial_{\nu} \phi_{\alpha})
    \nn
  &\qquad + 2 \alpha^{(0)} d z N \Big( - \mathcal{C}_{\Lambda} +
\mathcal{C}_{R} R^{(0)} + \frac{\mathcal{C}_{\varphi}}{2} g^{\mu\nu(0)}
(\partial_{\mu} \varphi^{(1)}) (\partial_{\nu} \varphi^{(1)}) +
\mathcal{C}_{\xi} R^{(0)} \varphi^{(1) 2} \Big)
    \nn
  &\qquad + \frac{N}{2u} \Big( \varphi^{(0)} - i \xi R^{(0)} \Big)^{2} + \beta^{(0)} d z \frac{N}{4u} \Big(\frac{\varphi^{(1)} - \varphi^{(0)}}{\beta^{(0)} d z}\Big)^{2} - N T_{\mu\nu}^{(0)} (g^{\mu\nu(0)} - g_{B}^{\mu\nu}) \Big\} \Big]. \label{Absence_Higher_Spin_Fields}
\end{align}
%
%\subsubsection{Introduction of dynamical metric as an auxiliary field for the renormalization of the kinetic-energy term}
%

One may criticize this locality approximation for the Green's function because this approximation scheme does not take into account higher-spin fields \cite{Higher_Spin_Gauge_Theory_I,Higher_Spin_Gauge_Theory_II,Higher_Spin_Gauge_Theory_III,Higher_Spin_Gauge_Theory_IV} from the beginning. Such higher-spin fields are $O(N)$ singlets, and may arise due to the non-local reparameterization symmetry present in the quadratic action when the source for the quadratic term is promoted to a dynamical field in the bulk \cite{Holography_Higher_Spin_RG_I,Holography_Higher_Spin_RG_II,Holography_Higher_Spin_RG_III}. The higher-spin fields may be unavoidable because they are generated under the coarse graining even though the UV theory has only two derivatives. For example, $- \int d^{D} x \sqrt{g_{x}^{(0)}} \int d^{D} x' \sqrt{g_{x'}^{(0)}} g^{\mu\nu(0)}_{x} g^{\mu'\nu'(0)}_{x'} (\partial_{\mu} \phi_{\alpha x}) (\partial_{\nu} \partial_{\mu'} G_{xx'}^{(0)}) (\partial_{\nu'} \phi_{\alpha x'})$ in Eq. (\ref{Non_Local_Quadratic_Term_Higher_Spin_Field}) has a double-integration for a bi-local operator, and this can generate higher-spin operators if one reformulates this bi-local field in a local way. The derivatives in Eq. (\ref{Locality_Approximation_Green_Ft}) should act on the delta function, and the integration by part possibly give rise to higher-spin operators in Eq. (\ref{Absence_Higher_Spin_Fields}). Although this criticism does make sense in principle, we point out that our recursive renormalization group formulation keeps all the terms up to the spin-two field, neglecting the appearance of higher-spin fields. In Section \ref{Discussion}, we argue that this dual holographic effective field theory serves as a novel mean-field theory framework with non-perturbative quantum corrections, showing that the present renormalization group formulation serves as a physically meaningful truncation scheme in the absence of higher-spin fields, where the resulting classical field theory in the large $N$ limit takes into account !
 quantum corrections in the all-loop order. In particular, we show that the nonlocal diffeomorphism invariance for the origin of the appearance of higher-spin fields is explicitly broken by the existence of effective interactions.

Now, it is straightforward to update the metric tensor up to the linear order in $d z$ as follows
\begin{align}
& Z = \int D \phi_{\alpha} D \varphi^{(0)} D \varphi^{(1)} D g_{\mu\nu}^{(0)} D
T_{\mu\nu}^{(0)} D g_{\mu\nu}^{(1)} D T_{\mu\nu}^{(1)}
\exp\Big[
                \nn
  &\qquad
\frac{N}{2} \mbox{tr}_{xx'} \ln (2 \alpha^{(0)} d z) + \frac{1}{2}
\mbox{tr}_{xx'} \ln (2 \beta^{(0)} d z) - \frac{N}{2} \mbox{tr}_{xx'} \ln
\sqrt{g^{(0)}} m^{2} - \frac{1}{2} \mbox{tr}_{xx'} \ln \frac{N}{2u}
    \nn
 &\qquad - \int d^{D} x \sqrt{g^{(1)}} \Big\{ g^{(1)\mu\nu} (\partial_{\mu}
\phi_{\alpha}) (\partial_{\nu} \phi_{\alpha}) + (m^{2} - i \varphi^{(1)})
\phi_{\alpha}^{2} \Big\}
   \nn
  &\qquad - N \int d^{D} x \sqrt{g^{(0)}} \Big\{ \frac{1}{2u} \Big(
\varphi^{(0)} - i \xi R^{(0)} \Big)^{2} - T_{\mu\nu}^{(0)} (g^{(0)\mu\nu} -
g_{B}^{\mu\nu}) \Big\}
    \nn
  &\qquad - N \int d^{D} x \sqrt{g^{(0)}} \Big\{ 2 \alpha^{(0)} d z \Big( -
\mathcal{C}_{\Lambda} + \mathcal{C}_{R} R^{(0)} +
\frac{\mathcal{C}_{\varphi}}{2} g^{\mu\nu(0)} (\partial_{\mu} \varphi^{(1)})
(\partial_{\nu} \varphi^{(1)}) + \mathcal{C}_{\xi} R^{(0)} \varphi^{(1) 2} \Big)
    \nn
  &\qquad + \beta^{(0)} d z \frac{N}{4u} \Big(\frac{\varphi^{(1)} - \varphi^{(0)}}{\beta^{(0)} d z}\Big)^{2} - T_{\mu\nu}^{(1)} \Big(g^{(1)\mu\nu} - g^{(0)\mu\nu} - g^{(0)\mu\nu'} (\partial_{\nu'} \partial_{\mu'} G_{xx'}^{(0)})_{x' \rightarrow x} g^{(0)\mu'\nu}\Big) \Big\} \Big] .
\end{align}
Here, $T_{\mu\nu}^{(1)}$ is a Lagrange multiplier field to impose the update condition. This completes the first renormalization group transformation.

%
%\subsubsection{After the $(f-1)$th renormalization group transformation}
%

Based on the above expression, we can estimate the partition function after the $(f-1)$th renormalization group transformation, given by
\bqa && Z = \int D \phi_{\alpha} \Pi_{k = 0}^{f} D \varphi^{(k)} D g_{\mu\nu}^{(k)} D T_{\mu\nu}^{(k)} \exp\Big\{ - S_{\Lambda} - S_{UV} - S_{IR} - S_{Bulk} \Big\} .
%
%\nn && \exp\Big[ \frac{N}{2} \sum_{k = 1}^{f} \mbox{tr}_{xx'} \ln (2 d z) + \frac{1}{2} \sum_{k = 1}^{f} \mbox{tr}_{xx'} \ln (2 d z) \nn && - \frac{N}{2} \sum_{k = 1}^{f} \mbox{tr}_{xx'} \ln \sqrt{g^{(k-1)}} m^{2} - \frac{1}{2} \sum_{k = 1}^{f} \mbox{tr}_{xx'} \ln \frac{N}{2u} - \int d^{D} x \sqrt{g^{(f)}} \Big\{ g^{(f)\mu\nu} (\partial_{\mu} \phi_{\alpha}) (\partial_{\nu} \phi_{\alpha}) + (m^{2} + \varphi^{(f)}) \phi_{\alpha}^{2} \Big\} \nn && - N \int d^{D} x \Big\{ - \frac{1}{2u} \sqrt{g^{(0)}} \Big( \varphi^{(0)} - \xi R^{(0)} \Big)^{2} - \sqrt{g^{(0)}} T_{\mu\nu}^{(0)} (g^{(0)\mu\nu} - g_{B}^{\mu\nu}) - \frac{1}{2u} 2 d z \sum_{k = 1}^{f} \sqrt{g^{(k-1)}} \Big(\frac{\varphi^{(k)} - \varphi^{(k-1)}}{2 d z}\Big)^{2} \nn && - 2 d z \sum_{k = 1}^{f} \sqrt{g^{(k-1)}} T_{\mu\nu}^{(k)} \Big(\frac{g^{(k)\mu\nu} - g^{(k-1)\mu\nu}}{2 d z} - \frac{1}{2 d z} g^{(k-1)\mu\nu'} (\partial_{\nu'} \partial_{\mu'} G_{xx'}^{(k-1)})_{x' \rightarrow x} g^{(k-1)\mu'\nu}\Big) \nn && + 2 d z \sum_{k = 1}^{f} \sqrt{g^{(k-1)}} \Big( - \mathcal{C}_{\Lambda} + \mathcal{C}_{R} R^{(k-1)} - \frac{\mathcal{C}_{\varphi}}{2} g^{\mu\nu(k-1)} (\partial_{\mu} \varphi^{(k)}) (\partial_{\nu} \varphi^{(k)}) - \mathcal{C}_{\xi} R^{(k-1)} \varphi^{(k) 2} \Big) \Big\} \Big]
%
\eqa
Here,
$S_{\Lambda}$ is an effective action
% to be
which is
UV-divergent, and given by
\begin{align}
  S_{\Lambda}
  &= - \frac{N}{2} \sum_{k = 1}^{f} \mbox{tr}_{xx'} \ln (2 d z) - \frac{1}{2} \sum_{k = 1}^{f} \mbox{tr}_{xx'} \ln (2 d z)
    \nonumber \\
  &\quad
    + \frac{1}{2} \sum_{k = 1}^{f} \mbox{tr}_{xx'} \ln \frac{N}{2u} + \frac{N}{2} \sum_{k = 1}^{f} \mbox{tr}_{xx'} \ln \sqrt{g^{(k-1)}} m^{2} .
\end{align}
$S_{UV}$ is an effective action
imposing
% involved with
UV boundary conditions for both metric and dual scalar fields, given by
\bqa && S_{UV} = N \int d^{D} x \Big\{ - \frac{1}{2u} \sqrt{g^{(0)}} \Big( \varphi^{(0)} - \xi R^{(0)} \Big)^{2} - \sqrt{g^{(0)}} T_{\mu\nu}^{(0)} (g^{(0)\mu\nu} - g_{B}^{\mu\nu}) \Big\} . \eqa
$S_{IR}$ is an effective action
implementing
% involved with
IR boundary conditions of both metric and dual scalar fields, given by
\bqa && S_{IR} = \int d^{D} x \sqrt{g^{(f)}} \Big\{ g^{(f)\mu\nu} (\partial_{\mu} \phi_{\alpha}) (\partial_{\nu} \phi_{\alpha}) + (m^{2} + \varphi^{(f)}) \phi_{\alpha}^{2} \Big\} . \eqa
$S_{Bulk}$ is an effective action to govern the dynamics of both metric and dual scalar fields, given by
\begin{align}
 S_{Bulk} &= N (2 d z) \sum_{k = 1}^{f} \int d^{D} x \sqrt{g^{(k-1)}}
\Big\{ - \frac{1}{2u} \Big(\frac{\varphi^{(k)} - \varphi^{(k-1)}}{2 d
  z}\Big)^{2}
                \nn
  &\quad \quad - T_{\mu\nu}^{(k)} \Big(\frac{g^{(k)\mu\nu} - g^{(k-1)\mu\nu}}{2 d
  z} - g^{(k-1)\mu\nu'} (\partial_{\nu'} \partial_{\mu'}
G_{xx'}^{(k-1)})_{x' \rightarrow x} g^{(k-1)\mu'\nu}\Big)
    \nn
  &\quad \quad - \mathcal{C}_{\Lambda} + \mathcal{C}_{R} R^{(k-1)} - \frac{\mathcal{C}_{\varphi}}{2} g^{\mu\nu(k-1)} (\partial_{\mu} \varphi^{(k)}) (\partial_{\nu} \varphi^{(k)}) - \mathcal{C}_{\xi} R^{(k-1)} \varphi^{(k) 2} \Big\} .
\end{align}
$G_{xx'}^{(k-1)}$ is the Green's function for the $k$-th renormalization group transformation, given by
\begin{align}
  \Big\{- \frac{1}{\sqrt{g_{x}^{(k-1)}}} \partial_{\mu} \Big( \sqrt{g_{x}^{(k-1)}} g^{\mu\nu(k-1)}_{x} \partial_{\nu} \Big) + \frac{1}{2 d z} (m^{2} + \varphi_{x}^{(k)}) \Big\} G_{xx'}^{(k-1)} = \frac{1}{\sqrt{g_{x}^{(k-1)}}} \delta^{(D)}(x-x') .
\end{align}
Here, we
% took
replaced $\varphi^{(k)}$ with $i \varphi^{(k)}$
%$(- i \varphi^{(k)}) \Longrightarrow \varphi^{(k)}$
and adopt
gauge fixing for both uniform speeds of coarse graining as $\alpha^{(k)} =
\beta^{(k)} = 1$. More discussions on this gauge fixing
will be presented below.
%are shown below.

%
%\subsection{Emergent gravity description for a scalar field theory}
%

The last step is to rewrite the above partition function in the continuous coordinate representation instead of the discrete variable $k$. Considering
\begin{align}
  & (2 d z) \sum_{k = 1}^{f} \int d^{D} x \sqrt{g^{(k-1)}} ~ T_{\mu\nu}^{(k)} ~ \Big(\frac{g^{(k)\mu\nu} - g^{(k-1)\mu\nu}}{2 d z}\Big)
    \nonumber \\
  &\quad
    \Longrightarrow \int_{0}^{z_{f}} d z \int d^{D} x \sqrt{g(x,z)} ~ T_{\mu\nu}(x,z) ~ \partial_{z} g^{\mu\nu}(x,z)
\end{align}
with $(2 d z) \sum_{k = 1}^{f} \Longrightarrow \int_{0}^{z_{f}} d z$ and $\frac{g^{(k)\mu\nu} - g^{(k-1)\mu\nu}}{2 d z} \Longrightarrow \partial_{z} g^{\mu\nu}(x,z)$, we obtain
\begin{align}
  & Z = Z_{\Lambda} \int D \phi_{\alpha}(x) D \varphi(x,z) D g_{\mu\nu}(x,z)
    D T_{\mu\nu}(x,z)  \exp\Big[
    \nn
  &\qquad
    - \int d^{D} x \sqrt{g(x,z_{f})} \Big\{ g^{\mu\nu}(x,z_{f}) [\partial_{\mu}
    \phi_{\alpha}(x)] [\partial_{\nu} \phi_{\alpha}(x)] + [m^{2} + \varphi(x,z_{f})]
    \phi_{\alpha}^{2}(x) \Big\}
    \nn
  &\qquad - N \int d^{D} x \sqrt{g(x,0)} \Big\{- \frac{1}{2u} \Big(
    \varphi(x,0) - \xi R(x,0) \Big)^{2} - T_{\mu\nu}(x,0) \Big(g^{\mu\nu}(x,0) -
    g_{B}^{\mu\nu}(x)\Big) \Big\}
    \nn
  &\qquad - N \int_{0}^{z_{f}} d z \int d^{D} x \sqrt{g(x,z)}
    \Big\{
    \nonumber \\
  &\qquad
    -
    \frac{1}{2u} [\partial_{z} \varphi(x,z)]^{2} - \frac{\mathcal{C}_{\varphi}}{2}
    g^{\mu\nu}(x,z) [\partial_{\mu} \varphi(x,z)] [\partial_{\nu} \varphi(x,z)]
    \nonumber \\
  &\qquad
    -
    \mathcal{C}_{\xi} R(x,z) [\varphi(x,z)]^{2}
   + \frac{1}{2 \kappa} \Big( R(x,z) - 2 \Lambda \Big)
    \nonumber \\
  &\qquad
    - T_{\mu\nu}(x,z) \Big(\partial_{z} g^{\mu\nu}(x,z) - g^{\mu\nu'}(x,z) \big( \partial_{\nu'} \partial_{\mu'} G_{xx'}[g_{\mu\nu}(x,z),\varphi(x,z)]\big)_{x' \rightarrow x} g^{\mu'\nu}(x,z) \Big) \Big\} \Big] ,
\end{align}
where all UV divergent terms are absorbed into a normalization constant $Z_{\Lambda}$. For the Einstein-Hilbert action, we took the following replacements
\bqa && \mathcal{C}_{R} \equiv \frac{1}{2 \kappa} , ~~~~~ \frac{\mathcal{C}_{\Lambda}}{\mathcal{C}_{R}} \equiv 2 \Lambda . \eqa
All the
dynamical information is encoded into the Green's function, given by
\begin{align}
  &\Big\{- \frac{1}{\sqrt{g(x,z)}} \partial_{\mu} \Big( \sqrt{g(x,z)} g^{\mu\nu}(x,z) \partial_{\nu} \Big) + \frac{1}{2 d z} [m^{2} + \varphi(x,z)] \Big\} G_{xx'}[g_{\mu\nu}(x,z),\varphi(x,z)]
    \nonumber \\
  &\quad
    = \frac{1}{\sqrt{g(x,z)}} \delta^{(D)}(x-x') .
\end{align}
As discussed earlier in section \ref{Overview}, the physical description is
quite clear in this effective action.
The $D$-dimensional metric tensor evolves through the extra dimension,
%governed by
following the equation of motion
\bqa && \partial_{z} g^{\mu\nu}(x,z) = g^{\mu\nu'}(x,z) \big( \partial_{\nu'} \partial_{\mu'} G_{xx'}[g_{\mu\nu}(x,z),\varphi(x,z)] \big)_{x' \rightarrow x} g^{\mu'\nu}(x,z) . \eqa
As a result, a fully renormalized metric $g^{\mu\nu}(x,z_{f})$ appears
% into
in the IR effective action $\mathcal{S}_{IR} = \int d^{D} x \sqrt{g(x,z_{f})} \Big\{ g^{\mu\nu}(x,z_{f}) [\partial_{\mu} \phi_{\alpha}(x)] [\partial_{\nu} \phi_{\alpha}(x)] + [m^{2} + \varphi(x,z_{f})] \phi_{\alpha}^{2}(x) \Big\}$, describing all possible renormalizations such as field renormalization, mass renormalization, and interaction renormalization. This evolution equation plays essentially the same role as renormalization group $\beta$-functions.

\subsection{Remarks on the renormalization group transformation for the metric tensor}

Although the renormalization group flow of the metric tensor results from
quantum fluctuations of matter fields, the metric tensor is not fully dynamical
in contrast with the holographic duality conjecture
\cite{Holographic_Duality_I,Holographic_Duality_II,
 Holographic_Duality_III,Holographic_Duality_IV,Holographic_Duality_V,Holographic_Duality_VI,Holographic_Duality_VII}.
%On the other hand, the order-parameter field is promoted to be dynamical in the extra-dimensional space.
This should be contrasted with the order-parameter field which is promoted to be dynamical in the extra-dimensional space.
We recall that the dynamics of the order-parameter field originates
from effective self-interactions between matter fields.
In this respect we introduce effective tensor-field interactions into the partition function as follows
\begin{align}
Z &= \int D \phi_{\alpha} \exp\Big[ - \int d^{D} x \sqrt{g_{B}}
\Big\{ g_{B}^{\mu\nu}
(\partial_{\mu} \phi_{\alpha}) (\partial_{\nu} \phi_{\alpha})
+ m^{2} \phi_{\alpha}^{2} + \xi R_{B} \phi_{\alpha}^{2}
\nonumber \\
&\quad
+ \frac{u}{2N} \phi_{\alpha}^{2} \phi_{\beta}^{2}
+ \frac{\lambda}{2N} [(\partial^{\mu} \phi_{\alpha}) (\partial^{\nu} \phi_{\alpha})]
[(\partial_{\mu} \phi_{\beta}) (\partial_{\nu} \phi_{\beta})] \Big\} \Big] .
\end{align}

It is natural to expect that the effective interaction term $\frac{\lambda}{2N}
[(\partial^{\mu} \phi_{\alpha}) (\partial^{\nu} \phi_{\alpha})] [(\partial_{\mu}
\phi_{\beta}) (\partial_{\nu} \phi_{\beta})]$ is irrelevant in the
renormalization group sense as long as the coupling constant $\lambda$ remains
to be below a critical value. However, we point out that these tensor-type
quantum fluctuations promote the emergent metric tensor to be fully dynamical \cite{TTbar_Deformation}.
Although one can take into account an effective interaction term of the exact
energy-momentum tensor in the effective Lagrangian,
% we mention that
the above introduction of the effective interaction term is sufficient in discussing which approximation scheme has to be used for the metric renormalization-group transformation.

Performing the Hubbard-Stratonovich transformation for both effective interactions, we obtain
\bqa && Z = \int D \phi_{\alpha} D \varphi^{(0)} D g_{\mu\nu}^{(0)} D t_{\mu\nu}^{(0)} D g_{\mu\nu} \exp\Big[ - \int d^{D} x \sqrt{g^{(0)}} \Big\{ (g^{\mu\nu(0)} - i g^{\mu\nu}) (\partial_{\mu} \phi_{\alpha}) (\partial_{\nu} \phi_{\alpha}) \nn && + (m^{2} - i \varphi^{(0)}) \phi_{\alpha}^{2} + \xi R^{(0)} \phi_{\alpha}^{2} + \frac{N}{2 u} \varphi^{(0) 2} + \frac{N}{2 \lambda} g^{\mu\nu} g_{\mu\nu} - N t_{\mu\nu}^{(0)} (g^{\mu\nu(0)} - g_{B}^{\mu\nu}) \Big\} \Big] , \eqa
where $t_{\mu\nu}^{(0)}$ is a Lagrange multiplier field to impose the initial condition $g^{\mu\nu(0)} = g_{B}^{\mu\nu}$. Shifting the metric tensor in the following way
\bqa && g^{\mu\nu(0)} \Longrightarrow g^{\mu\nu(0)} + i g^{\mu\nu} , \eqa
we obtain
\begin{align}
  & Z = \int D \phi_{\alpha} D \varphi^{(0)} D g_{\mu\nu}^{(0)} D t_{\mu\nu}^{(0)} D g_{\mu\nu}
     \exp\Big[ - \int d^{D} x \sqrt{\mbox{det}[g_{\mu\nu}^{(0)} + i g_{\mu\nu}]} \Big\{ g^{\mu\nu(0)} (\partial_{\mu} \phi_{\alpha}) (\partial_{\nu} \phi_{\alpha})
     \nonumber \\
  &\quad
    + (m^{2} - i \varphi^{(0)}) \phi_{\alpha}^{2} + \xi R[g_{\mu\nu}^{(0)} + i g_{\mu\nu}] \phi_{\alpha}^{2} + \frac{N}{2 u} \varphi^{(0) 2} + \frac{N}{2 \lambda} g^{\mu\nu} g_{\mu\nu} - N t_{\mu\nu}^{(0)} (g^{\mu\nu(0)} + i g^{\mu\nu} - g_{B}^{\mu\nu}) \Big\} \Big] .
\end{align}
%Here, we
We now
perform the path integral $\int D t_{\mu\nu}^{(0)} D g_{\mu\nu}$. The approximation that we have to use is to keep quantum fluctuations of the metric tensor up to the linear order. In other words, we neglect the $i g_{\mu\nu}$ contribution in both the determinant and Ricci scalar. As a result, we obtain
\begin{align}
  & Z \approx \int D \phi_{\alpha} D \varphi^{(0)} D g_{\mu\nu}^{(0)} \exp\Big[ - \int d^{D} x \sqrt{g^{(0)}} \Big\{ g^{\mu\nu(0)} (\partial_{\mu} \phi_{\alpha}) (\partial_{\nu} \phi_{\alpha}) + (m^{2} - i \varphi^{(0)}) \phi_{\alpha}^{2} + \xi R^{(0)} \phi_{\alpha}^{2}
     \nonumber \\
  &\quad
     + \frac{N}{2 u} \varphi^{(0) 2} - \frac{N}{2 \lambda} (g^{\mu\nu(0)} - g_{B}^{\mu\nu}) (g_{\mu\nu}^{(0)} - g_{B \mu\nu}) \Big\} \Big] .
\end{align}

To perform the renormalization group transformation for the metric tensor, we introduce an auxiliary field in the following way
\bqa && Z = \int D \phi_{\alpha} D \varphi^{(0)} D \varphi^{(1)} D g_{\mu\nu}^{(0)} D G_{\mu\nu}^{(0)} \nn && \exp\Big[ - \int d^{D} x \sqrt{i^{-1} \mbox{det}[g_{\mu\nu}^{(0)} + g_{B \mu\nu}]} \Big\{ i [g^{\mu\nu(0)} + g_{B}^{\mu\nu}] (\partial_{\mu} \phi_{\alpha}) (\partial_{\nu} \phi_{\alpha}) + (m^{2} - i e^{- \beta^{(0)} d z} \varphi^{(1)}) \phi_{\alpha}^{2} \nn && + \frac{N}{2u} \Big( \varphi^{(0)} - i e^{- \beta^{(0)} d z} \xi R[i (g_{\mu\nu}^{(0)} + g_{B \mu\nu})] \Big)^{2} + \frac{N}{2u} \frac{1}{e^{2\beta^{(0)} d z} - 1} (\varphi^{(1)} - \varphi^{(0)})^{2} \nn && + \frac{N}{2 \lambda} g^{\mu\nu(0)} g_{\mu\nu}^{(0)} + \frac{N}{2 \lambda_{G}} G^{\mu\nu(0)} G_{\mu\nu}^{(0)} \Big\} \Big] , \eqa
where the renormalization group transformation for the order-parameter field has been performed and $g^{\mu\nu (0)} \Longrightarrow i (g^{\mu\nu (0)} + g_{B}^{\mu\nu})$ has been taken. Separating slow and fast degrees of freedom in the metric tensor as
\bqa && g^{\mu\nu (0)} \Longrightarrow g^{\mu\nu (0)} + \mathcal{G}^{\mu\nu (0)} , ~~~~~ G^{\mu\nu (0)} \Longrightarrow c_{g}^{(0)} g^{\mu\nu (0)} + c_{G}^{(0)} \mathcal{G}^{\mu\nu (0)} , \eqa
where both coefficients are given by
\bqa && c_{g}^{(0)} = \frac{\lambda^{-1}}{\mu^{(0)} \lambda_{G}^{-1/2}} , ~~~~~ c_{G}^{(0)} = - \frac{\mu^{(0)}}{\lambda_{G}^{-1/2}} , ~~~~~ \mu^{(0)} = \frac{\lambda^{-1/2}}{\sqrt{e^{2 \delta^{(0)} d z} - 1}} , \eqa
and rescaling both degrees of freedom as
\bqa && g^{\mu\nu (0)} \Longrightarrow e^{- \delta^{(0)} d z} g^{\mu\nu (0)} , ~~~~~ \mathcal{G}^{\mu\nu (0)} \Longrightarrow e^{- \delta^{(0)} d z} \mathcal{G}^{\mu\nu (0)} , \eqa
where $\delta^{(0)}$ is the local speed of coarse graining, we have
\bqa && Z = \int D \phi_{\alpha} D \varphi^{(0)} D \varphi^{(1)} D g_{\mu\nu}^{(0)} D \mathcal{G}_{\mu\nu}^{(0)} \exp\Big[ - \int d^{D} x \sqrt{i^{-1} \mbox{det}[e^{\delta^{(0)} d z} (g_{\mu\nu}^{(0)} + e^{- \delta^{(0)} d z} g_{B \mu\nu} + \mathcal{G}_{\mu\nu}^{(0)})]} \nn && \Big\{ i e^{- \delta^{(0)} d z} (g^{\mu\nu(0)} + e^{\delta^{(0)} d z} g_{B}^{\mu\nu} + \mathcal{G}^{\mu\nu(0)}) (\partial_{\mu} \phi_{\alpha}) (\partial_{\nu} \phi_{\alpha}) + (m^{2} - i e^{- \beta^{(0)} d z} \varphi^{(1)}) \phi_{\alpha}^{2} \nn && + \frac{N}{2u} \Big( \varphi^{(0)} - i e^{- \beta^{(0)} d z} \xi R[i e^{\delta^{(0)} d z} (g_{\mu\nu}^{(0)} + e^{- \delta^{(0)} d z} g_{B \mu\nu} + \mathcal{G}_{\mu\nu}^{(0)})] \Big)^{2} + \frac{N}{2u} \frac{1}{e^{2\beta^{(0)} d z} - 1} (\varphi^{(1)} - \varphi^{(0)})^{2} \nn && + \frac{N}{2 \lambda} g^{\mu\nu (0)} g_{\mu\nu}^{(0)} + \frac{N}{2 \lambda} \frac{1}{e^{2 \delta^{(0)} d z} - 1} \mathcal{G}^{\mu\nu (0)} \mathcal{G}_{\mu\nu}^{(0)} \Big\} \Big] . \eqa

To perform the renormalization group transformation for the metric tensor, we consider the linear approximation again for the $\mathcal{G}_{\mu\nu}^{(0)}$ path integral, where the $\mathcal{G}_{\mu\nu}^{(0)}$ contribution in both the determinant and the Ricci scalar is neglected. As a result, we obtain
\bqa && Z = \int D \phi_{\alpha} D \varphi^{(0)} D \varphi^{(1)} D g_{\mu\nu}^{(0)} \exp\Big[ - \int d^{D} x \sqrt{e^{\delta^{(0)} d z} g^{(0)}} \Big\{ e^{- \delta^{(0)} d z} g^{\mu\nu(0)} (\partial_{\mu} \phi_{\alpha}) (\partial_{\nu} \phi_{\alpha}) \nn && + (m^{2} - i e^{- \beta^{(0)} d z} \varphi^{(1)}) \phi_{\alpha}^{2} + \frac{N}{2u} \Big( \varphi^{(0)} - i e^{- \beta^{(0)} d z} \xi R^{(0)} \Big)^{2} + \frac{N}{2u} \frac{1}{e^{2\beta^{(0)} d z} - 1} (\varphi^{(1)} - \varphi^{(0)})^{2} \nn && - \frac{N}{2 \lambda} (g^{\mu\nu(0)} - e^{\delta^{(0)} d z} g_{B}^{\mu\nu}) (g_{\mu\nu}^{(0)} - e^{- \delta^{(0)} d z} g_{B \mu\nu}) + \frac{\lambda}{2 N} [1 - e^{- 2 \delta^{(0)} d z}] (\partial^{\mu} \phi_{\alpha}) (\partial^{\nu} \phi_{\alpha}) (\partial_{\mu} \phi_{\beta}) (\partial_{\nu} \phi_{\beta}) \Big\} \Big] , ~~~~~~~ \eqa
where an effective interaction term $\frac{\lambda}{2 N} [1 - e^{- 2 \delta^{(0)} d z}] (\partial^{\mu} \phi_{\alpha}) (\partial^{\nu} \phi_{\alpha}) (\partial_{\mu} \phi_{\beta}) (\partial_{\nu} \phi_{\beta})$ has been generated. Performing the Hubbard-Stratonovich transformation for the newly generated effective interaction term and updating the metric tensor in the boundary action, we obtain
\bqa && Z = \int D \phi_{\alpha} D \varphi^{(0)} D \varphi^{(1)} D g_{\mu\nu}^{(0)} D \tilde{g}_{\mu\nu}^{(0)} \exp\Big[ - \int d^{D} x \sqrt{e^{\delta^{(0)} d z} \tilde{g}^{(0)}} \Big\{ e^{- \delta^{(0)} d z} \tilde{g}^{\mu\nu (0)} (\partial_{\mu} \phi_{\alpha}) (\partial_{\nu} \phi_{\alpha}) \nn && + (m^{2} - i e^{- \beta^{(0)} d z} \varphi^{(1)}) \phi_{\alpha}^{2} + \frac{N}{2u} \frac{1}{e^{2\beta^{(0)} d z} - 1} (\varphi^{(1)} - \varphi^{(0)})^{2} \nn && - \frac{N}{2 \lambda} \frac{1}{e^{2 \delta^{(0)} d z} - 1} (\tilde{g}^{\mu\nu (0)} - g^{\mu\nu (0)}) (\tilde{g}_{\mu\nu}^{(0)} - g_{\mu\nu}^{(0)}) \Big\} - \int d^{D} x \sqrt{e^{\delta^{(0)} d z} g^{(0)}} \Big\{ \frac{N}{2u} \Big( \varphi^{(0)} - i e^{- \beta^{(0)} d z} \xi R^{(0)} \Big)^{2} \nn && - \frac{N}{2 \lambda} (g^{\mu\nu(0)} - e^{\delta^{(0)} d z} g_{B}^{\mu\nu}) (g_{\mu\nu}^{(0)} - e^{- \delta^{(0)} d z} g_{B \mu\nu}) \Big\} \Big] , \eqa
where $\tilde{g}^{\mu\nu (0)}$ is an updated metric tensor.

Finally, we perform the renormalization group transformation for the matter sector as discussed in the previous section. Taking the locality approximation for the Green's function and keeping all terms up to the linear order in $d z$, we find
\bqa && Z = \int D \phi_{\alpha} D \varphi^{(0)} D \varphi^{(1)} D g_{\mu\nu}^{(0)} D \tilde{g}_{\mu\nu}^{(0)} \nn && \exp\Big[ - \int d^{D} x \sqrt{\tilde{g}^{(0)}} \Big\{ \Big( \tilde{g}^{\mu\nu(0)} - \tilde{g}^{\mu\delta(0)} (\partial_{\delta} \partial_{\delta'} G_{xx'}^{(0)})_{x' \rightarrow x} \tilde{g}^{\delta'\nu(0)} \Big) (\partial_{\mu} \phi_{\alpha}) (\partial_{\nu} \phi_{\alpha}) + (m^{2} - i \varphi^{(1)}) \phi_{\alpha}^{2} \nn && + 2 \beta^{(0)} d z \frac{N}{2u} \Big(\frac{\varphi^{(1)} - \varphi^{(0)}}{2 \beta^{(0)} d z}\Big)^{2} - 2 \delta^{(0)} d z \frac{N}{2 \lambda} \Big(\frac{\tilde{g}^{\mu\nu (0)} - g^{\mu\nu (0)}}{2 \delta^{(0)} d z}\Big) \Big(\frac{\tilde{g}_{\mu\nu}^{(0)} - g_{\mu\nu}^{(0)}}{2 \delta^{(0)} d z}\Big) \nn && + 2 \alpha^{(0)} d z N \Big( - \mathcal{C}_{\Lambda} + \mathcal{C}_{R} \tilde{R}^{(0)} + \frac{\mathcal{C}_{\varphi}}{2} \tilde{g}^{\mu\nu (0)} (\partial_{\mu} \varphi^{(1)}) (\partial_{\nu} \varphi^{(1)}) + \mathcal{C}_{\xi} \tilde{R}^{(0)} \varphi^{(1) 2} \Big) \Big\} \nn && - \int d^{D} x \sqrt{g^{(0)}} \Big\{ \frac{N}{2u} \Big( \varphi^{(0)} - i \xi R^{(0)} \Big)^{2} - \frac{N}{2 \lambda} (g^{\mu\nu(0)} - g_{B}^{\mu\nu}) (g_{\mu\nu}^{(0)} - g_{B \mu\nu}) \Big\} \Big] . \eqa
Upgrading the metric tensor $\tilde{g}_{\mu\nu}^{(0)}$ to make the kinetic-energy term be invariant, we rewrite the above expression as follows
\bqa && Z = \int D \phi_{\alpha} D \varphi^{(0)} D \varphi^{(1)} D g_{\mu\nu}^{(0)} D g_{\mu\nu}^{(1)} \nn && \exp\Big[ - \int d^{D} x \sqrt{g^{(1)}} \Big\{ g^{\mu\nu (1)} (\partial_{\mu} \phi_{\alpha}) (\partial_{\nu} \phi_{\alpha}) + (m^{2} - i \varphi^{(1)}) \phi_{\alpha}^{2} + 2 \beta^{(0)} d z \frac{N}{2u} \Big(\frac{\varphi^{(1)} - \varphi^{(0)}}{2 \beta^{(0)} d z}\Big)^{2} \nn && - 2 \delta^{(0)} d z \frac{N}{2 \lambda} \Big(\frac{g^{\mu\nu(1)} - g^{\mu\nu (0)}}{2 \delta^{(0)} d z} - g^{\mu\delta(0)} (\partial_{\delta} \partial_{\delta'} G_{xx'}^{(0)})_{x' \rightarrow x} g^{\delta'\nu(0)}\Big) \Big(\frac{g_{\mu\nu}^{(1)} - g_{\mu\nu}^{(0)}}{2 \delta^{(0)} d z} - g_{\mu\delta}^{(0)} (\partial^{\delta} \partial^{\delta'} G_{xx'}^{(0)})_{x' \rightarrow x} g_{\delta'\nu}^{(0)}\Big) \nn && + 2 \alpha^{(0)} d z N \Big( - \mathcal{C}_{\Lambda} + \mathcal{C}_{R} R^{(1)} + \frac{\mathcal{C}_{\varphi}}{2} g^{\mu\nu (1)} (\partial_{\mu} \varphi^{(1)}) (\partial_{\nu} \varphi^{(1)}) + \mathcal{C}_{\xi} R^{(1)} \varphi^{(1) 2} \Big) \Big\} \nn && - \int d^{D} x \sqrt{g^{(0)}} \Big\{ \frac{N}{2u} \Big( \varphi^{(0)} - i \xi R^{(0)} \Big)^{2} - \frac{N}{2 \lambda} (g^{\mu\nu(0)} - g_{B}^{\mu\nu}) (g_{\mu\nu}^{(0)} - g_{B \mu\nu}) \Big\} \Big] . \eqa

Repeating all these renormalization group transformations within the recursion framework and rewriting the recursive expression in the continuous variable $z$, we obtain \begin{align}
  & Z = Z_{\Lambda} \int D \phi_{\alpha}(x) D \varphi(x,z) D g_{\mu\nu}(x,z)
    \exp\Big[
    \nn
  &\quad
    - \int d^{D} x \sqrt{g(x,z_{f})} \Big\{ g^{\mu\nu}(x,z_{f}) [\partial_{\mu} \phi_{\alpha}(x)] [\partial_{\nu} \phi_{\alpha}(x)]
    + [m^{2} + \varphi(x,z_{f})] \phi_{\alpha}^{2}(x)
    %
    %- i g^{\mu\nu}(x,z_{f}) T_{\mu\nu}(x)
    %
    \Big\}
    \nn
  &\quad - N \int d^{D} x \sqrt{g(x,0)} \Big\{
    - \frac{1}{2u} \Big(
    \varphi(x,0) - \xi R(x,0) \Big)^{2} - \frac{1}{2 \lambda} \Big(g^{\mu\nu}(x,0) -
    g_{B}^{\mu\nu}(x)\Big)^{2} \Big\}
    \nn
  &\quad - N \int_{0}^{z_{f}} d z \int d^{D} x \sqrt{g(x,z)} \Big\{
    \nonumber \\
  &\qquad
    -
    \frac{1}{2u} [\partial_{z} \varphi(x,z)]^{2} - \frac{\mathcal{C}_{\varphi}}{2}
    g^{\mu\nu}(x,z) [\partial_{\mu} \varphi(x,z)] [\partial_{\nu} \varphi(x,z)] -
    \mathcal{C}_{\xi} R(x,z) [\varphi(x,z)]^{2}
    \nn
  &\qquad - \frac{1}{2 \lambda} \Big(\partial_{z} g^{\mu\nu}(x,z) - g^{\mu\nu'}(x,z) \big( \partial_{\nu'} \partial_{\mu'} G_{xx'}[g_{\mu\nu}(x,z),\varphi(x,z)] \big)_{x' \rightarrow x} g^{\mu'\nu}(x,z) \Big)^{2}
    \nonumber \\
  &\qquad
    + \frac{1}{2 \kappa} \Big( R(x,z) - 2 \Lambda \Big) \Big\} \Big] . \label{Full_Holographic_Dual}
\end{align}
Here,
the second derivative with respect to $z$ coordinatizing the extra dimension
% the second-order derivative for the coordinate $z$ of the extra dimensional
%space
arises in the metric evolution, resulting from essentially the same procedure as
that in the order-parameter evolution $[\partial_{z} \varphi(x,z)]^{2}$, where the linear approximation for the metric tensor has been used. It is interesting to notice that the linear approximation for quantum fluctuations of the metric tensor gives rise to the highly nonlinear renormalization group flow, shown in Eq.\ (\ref{Full_Holographic_Dual}), where the metric-tensor renormalization is intertwined with the renormalization of the order-parameter field through the Green's function of the matter field. In this case renormalization group $\beta$-functions are given by IR boundary conditions, where the linear derivative in $z$ appears from the bulk effective action by the technique of integration-by-parts. In the next section we demonstrate that the evolution equations of the metric tensor along the extra dimensional space are nothing but the renormalization group $\beta$-functions.

Before going further, we point out that this holographic dual effective field
theory is not covariant but a gauge-fixed version.
In other words, the metric-tensor components involved with the extra dimension
are gauge-fixed as follows:
$g_{DD}(x,z) = 1$ and $g_{\mu D}(x,z) = 0$ with $\mu = 0, \ldots, D-1$.
It turns out that the gauge choice of $\alpha^{(k)} = 1$ gives rise to
$g_{DD}(x,z) = 1$.
On the other hand,
the gauge freedom
in $g_{\mu D}(x,z) = 0$ results from the invariance of the partition function
with respect to $D$-dimensional diffeomorphism after the renormalization group
transformation with $d z$
\cite{SungSik_Holography_I,SungSik_Holography_II,SungSik_Holography_III}.
%not discussed carefully in this study.
While we do not delve into this issue here,
a fully covariant formulation has been constructed in the absence of dual scalar
fields,
where the $D$-dimensional Einstein-Hilbert action is uplifted into the $(D+1)$-dimensional Einstein-Hilbert one via recursive renormalization group transformations \cite{SungSik_Holography_I,SungSik_Holography_II,SungSik_Holography_III}.

We also point out
that the order of the renormalization group transformations
for the dual and original boson fields is immaterial.
%One may ask the order of renormalization group transformations between the dual
%scalar field and the original boson field.
We recall that the $ d z \rightarrow 0$ limit with $z_{f} \equiv d z \sum_{k =
  1}^{f} = f d z$ controls our recursive renormalization group transformations.
The order of
the
renormalization group transformations does not matter in this controllable limit.

\section{Physical interpretation of the emergent dynamical metric tensor: From renormalization group equations of coupling parameters to the evolution equation of the metric tensor} \label{RG_GR_Correspondence_Section}

%
%\subsection{Emergent geometric description for a lattice field theory in one dimension}
%

In this section we demonstrate explicitly that renormalization group
$\beta$-functions in the quantum field theory are nothing but IR boundary
conditions in the large $N$ limit of the holographic dual field theory.
We do so by comparing our holographic theory and
  yet another real space renormalization group transformation (the Kadanoff block-spin transformation).
For simplicity in the presentation, we
%focus on the absence of
switch off the
self-interactions and consider a one-dimensional lattice field theory, given by
\begin{align}
  Z = \int D \Phi_{i} \exp\Big[ - \int_{0}^{\beta} d \tau \sum_{i = 1}^{M} \Big\{ (\partial_{\tau} \Phi_{i})^{2} - t (\Phi_{i} \Phi_{i+1} + \Phi_{i+1} \Phi_{i}) + m^{2} \Phi_{i}^{2} \Big\} \Big] .
\end{align}
Here, $\Phi_{i}$ is a real scalar field at site $i$. $t$ is a hopping integral
and $m$ is a mass parameter.
In the case of one dimensional lattice, it is
straightforward
% easy
to perform the Kadanoff block-spin transformation \cite{Kadanoff_RG}.
% , regarded to be a real-space renormalization group transformation.
Integrating out all dynamical fields at even lattice sites, we obtain an effective field theory for odd-site scalar fields with renormalized hopping and mass parameters. Implementing this renormalization group transformation in a recursive way \cite{Holographic_Description_Kim}, we obtain the following expression of the partition function
\begin{align}
  &
    Z = \int D \Phi(i,\tau) D t(z) D m^{2}(z)
     \nn
  &
     \qquad
     \times \delta \Big(t(0) - t\Big)
     \delta \Big(m^{2}(0) - m^{2}\Big) \delta \Big( \partial_{z} m^{2}(z) +
     \frac{[t(z)]^{2}}{a m^{2}(z)} \Big)
     \delta \Big( \partial_{z} t(z) + \frac{1}{a} t(z) -
     \frac{[t(z)]^{2}}{2 a m^{2}(z)} \Big)
    \nn
  &\qquad \times
            \exp\Bigg[ - \int_{0}^{\beta} d \tau
            \sum_{i = 1}^{M} \Bigg\{ \Big(\partial_{\tau} \Phi(i,\tau)\Big)^{2} +
            m^{2}(z_{f}) [\Phi(i,\tau)]^{2}
    \nn
  &\qquad
                   \quad - t(z_{f}) \Big( \Phi(i,\tau) \Phi(i+1,\tau) + \Phi(i+1,\tau) \Phi(i,\tau) \Big) \Bigg\} - \frac{M}{4 a}  \int_{0}^{z_{f}} d z \sum_{i \omega_{n}} \ln \Big( \omega_{n}^{2} + m^{2}(z) \Big) \Bigg] . \label{Emergent_Gravity_RG_Flow}
\end{align}
Here, $a$ is a scale for the renormalization group transformation, defined in Eqs. (\ref{a_definition_I}) and (\ref{a_definition_II}), which corresponds to $d z$ in the previous section.
All details are shown in appendix \ref{RG_GR_Correspondence_Appendix}, where
recursive Kadanoff block-spin transformations have been performed in the
presence of self-interactions. We point out that both the hopping and mass
parameters are renormalized to appear in the IR boundary action.
In this respect two $\delta$-function constraints serve as renormalization group $\beta$-functions for the hopping and mass parameters, respectively.

%
%\subsection{Emergent gravity description for a scalar field theory in one dimension}
%

To show the equivalence between the above renormalization group $\beta-$functions and the evolution equations for the metric-tensor fields, we recall the gravity formulation, given by
\begin{align}
  & Z = \int D \phi_{\alpha}(x) D g_{\mu\nu}(x,z)
     \delta\Big(g^{\mu\nu}(x,0)
     - g_{B}^{\mu\nu}\Big)
     \nonumber \\
  &\qquad
    \times
     \delta\Big(\partial_{z} g^{\mu\nu}(x,z) -
g^{\mu\nu'}(x,z) \big(\partial_{\nu'} \partial_{\mu'}
G_{xx'}[g_{\mu\nu}(x,z)]\big)_{x' \rightarrow x} g^{\mu'\nu}(x,z) \Big)
    \nn
  &\qquad
    \times
\exp\Big[ - \int d^{D} x \sqrt{g(x,z_{f})} \Big\{ g^{\mu\nu}(x,z_{f})
(\partial_{\mu} \phi_{\alpha}(x)) (\partial_{\nu} \phi_{\alpha}(x)) + m^{2}
\phi_{\alpha}^{2}(x) \Big\}
    \nn
  & \qquad - \frac{N}{2 \kappa} \int_{0}^{z_{f}} d z \int d^{D} x \sqrt{g(x,z)} \Big( R(x,z) - 2 \Lambda \Big) \Big] , \label{Emergent_Gravity_Ricci_Flow}
\end{align}
which results from the UV theory of
\bqa && Z = \int D \phi_{\alpha}(x) \exp\Big[ - \int d^{D} x \sqrt{g_{B}} \Big\{ g_{B}^{\mu\nu} (\partial_{\mu} \phi_{\alpha}) (\partial_{\nu} \phi_{\alpha}) + m^{2} \phi_{\alpha}^{2} \Big\} \Big] . \eqa

%
%\subsection{Identification between the lattice model and the continuum field theory}
%

Comparing each term between Eq.\ (\ref{Emergent_Gravity_RG_Flow}) and Eq.\ (\ref{Emergent_Gravity_Ricci_Flow}), we observe the following correspondences,
\bqa && \sqrt{g(x,z_{f})} m^{2} = m^{2}(z_{f}) - 2 t(z_{f}) \Longrightarrow \frac{1}{\sqrt{g(x,z_{f})}} = \frac{m^{2}}{m^{2}(z_{f}) - 2 t(z_{f})} \eqa
%{\color{blue} May be $-2a^2 t(z_f)$ instead(?).
%Previously, I made a mistake to identify $a$ with a lattice scale. I correct it now. $a$ is not a lattice scale but an RG scale, corresponding to $d z$ in the previous section. So, $-2 t(z_f)$ is correct.}
for the metric determinant,
\bqa && \sqrt{g(x,z_{f})} g^{xx}(x,z_{f}) = t(z_{f}) \Longrightarrow g^{xx}(x,z_{f}) = \frac{m^{2}}{m^{2}(z_{f}) - 2 t(z_{f})} t(z_{f}) \eqa
for the $g^{xx}(x,z_{f})$ component, and
\bqa && \sqrt{g(x,z_{f})} g^{\tau\tau}(x,z_{f}) = 1 \Longrightarrow g^{\tau\tau}(x,z_{f}) = \frac{m^{2}}{m^{2}(z_{f}) - 2 t(z_{f})} \eqa
for the $g^{\tau\tau}(x,z_{f})$ component. We recall the gauge choice
%of
\bqa && g^{zz}(x,z) = 1 . \eqa

%
%\subsection{Evolution equation for the metric tensor}
%

We also point out the correspondence in the Green's function, given by
\bqa && G[q_{0},q_{x};t(z),m(z)] = \frac{1}{q_{0}^{2} + t(z) q_{x}^{2} + \frac{m^{2}(z) - 2 t(z)}{2 d z}} , \eqa
where $q_0$ and $q_x$ are temporal and spatial
components of the momentum, respectively.
Introducing all these correspondences into the evolution equations for the metric tensors, given by
\bqa && \partial_{z} g^{\tau\tau}(z) = g^{\tau\tau}(z) \Big\{ \int d q_{0} d q_{x} \frac{q_{0}^{2}}{q_{0}^{2} + t(z) q_{x}^{2} + \frac{m^{2}(z) - 2 t(z)}{2 d z}} \Big\} g^{\tau\tau}(z) \eqa
and
\bqa && \partial_{z} g^{xx}(z) = g^{xx}(z) \Big\{ \int d q_{0} d q_{x} \frac{q_{x}^{2}}{q_{0}^{2} + t(z) q_{x}^{2} + \frac{m^{2}(z) - 2 t(z)}{2 d z}} \Big\} g^{xx}(z) , \eqa
%
%\subsection{Reformulation of the metric equation in terms of renormalized mass and hopping parameters}
%
we obtain the following equations for both mass and hopping parameters,
\begin{align}
\partial_{z} m^{2}(z) - 2 \partial_{z} t(z) = - \pi \frac{ m^{2} }{ 2 d z } \Big( \int d q_{0} \frac{ q_{0}^{2} }{\sqrt{ q_{0}^{2} + 1}} \Big) \frac{m^{2}(z) - 2 t(z)}{\sqrt{t(z)}}
\end{align}
and
\begin{align}
  m^{2}(z) \partial_{z} t(z) - t(z) \partial_{z} m^{2}(z) = \pi \frac{m^{2}}{ 2 d z } \Big( \int d q_{x} \frac{q_{x}^{2}}{\sqrt{q_{x}^{2} + 1}} \Big) \sqrt{t(z)} [m^{2}(z) - 2 t(z)] .
\end{align}

%
%\subsection{Equivalence between the lattice model and the continuum field theory}
%

These two coupled equations are reduced into
\bqa && \partial_{z} m^{2}(z) = - \pi \frac{ m^{2} }{ 2 d z } \Big( \int d q_{0} \frac{ q_{0}^{2} }{\sqrt{ q_{0}^{2} + 1}} \Big) \frac{m^{2}(z) - 2 t(z)}{\sqrt{t(z)}} \label{Gravity_Mass_RG_Flow} \eqa
and
\bqa &&
\partial_z t(z) = 0 . \eqa
Redefining the mass parameter as
\bqa && M^{2}(z) \equiv m^{2}(z) + 2 t(z) , \eqa
we rewrite Eq. (\ref{Gravity_Mass_RG_Flow}) as
\bqa && \partial_{z} M^{2}(z) = - \pi \frac{ m^{2} }{ 2 d z \sqrt{t}} \Big( \int d q_{0} \frac{ q_{0}^{2} }{\sqrt{ q_{0}^{2} + 1}} \Big) M^{2}(z) . \label{Gravity_RG_Flow_Final} \eqa

To show the equivalence between this equation
and the $\beta$-function of the mass parameter, we consider the following fixed point, given by
\bqa && \partial_{z} t(z) = - \frac{1}{a} t(z) + \frac{[t(z)]^{2}}{2 a m^{2}(z)} \Longrightarrow 0 \eqa
in the $\beta$-function of the hopping parameter. Then, the $\beta$-function of the mass parameter is
\bqa && \partial_{z} m^{2}(z) = - \frac{[t(z)]^{2}}{a m^{2}(z)} \Longrightarrow - \frac{4}{a} m^{2}(z) . \label{QFT_RG_Flow_Final} \eqa
Comparing Eq.\ (\ref{Gravity_RG_Flow_Final}) with Eq.\ (\ref{QFT_RG_Flow_Final}),
we obtain the correspondence between
the scale $d z$ of the Polchinski real-space renormalization group transformation and the scale $a$ of the Kadanoff block-spin transformation as follows
\bqa && \frac{2 d z}{a} = \pi \frac{ m^{2} }{ 4 \sqrt{t}} \Big( \int d q_{0} \frac{ q_{0}^{2} }{\sqrt{ q_{0}^{2} + 1}} \Big) . \eqa
This completes our demonstration
that the evolution equations of the metric tensor are nothing but the renormalization-group $\beta$-functions of the coupling constants.

%
%\section{Towards effective hydrodynamics in the emergent holographic description}
%

\section{Evaluation of entanglement entropy based on the heat kernel method} \label{Entanglement_Entropy_Field_Theory}

\subsection{Entanglement entropy}

As we have demonstrated that
% Since
the quantum field theory is geometrized in the large $N$ limit,
we
%need
now
%to
ask how quantum entanglement in the quantum field theory is encoded into the
classical geometry in the holographic dual reformulation.
As discussed before, we address this issue by calculating the entanglement entropy.

Entanglement entropy is a subsystem von Neumann entropy $\mathcal{S}_{EE} = - \mbox{tr}_{A} \rho_{A} \ln \rho_{A}$, given by a reduced density matrix of the subsystem $\rho_{A} = \mbox{tr}_{B} \rho_{A \cup B}$, where $\rho_{A \cup B}$ is the density matrix of a total system $A \cup B$. This von Neumann entropy can be obtained from the Renyi entropy $\mathcal{S}_{RE} = \frac{1}{1 - n} \mbox{tr}_{A} \rho_{A}^{n}$ as follows \cite{Entanglement_Entropy_Calabrese_Cardy_I,Entanglement_Entropy_Calabrese_Cardy_II,Entanglement_Entropy_Ryu_Takayanagi_I,Entanglement_Entropy_Ryu_Takayanagi_II,Entanglement_Entropy_Review_III,Entanglement_Entropy_Review_IV}
\begin{align}
  \mathcal{S}_{EE}(z_{f})
  &= \lim_{n \rightarrow 1} \frac{1}{1 - n} \mbox{tr}_{A} \rho_{A}^{n}
    = \lim_{n \rightarrow 1} \frac{\ln Z_{n}(z_{f}) - n \ln Z_{1}(z_{f})}{1 - n}
    \nonumber \\
  &
    = - [\partial_{n} \ln Z_{n}(z_{f})]_{n = 1} + \ln Z_{1}(z_{f}) .
\end{align}
Here, $Z_{1}(z_{f})$ is the partition function of the emergent holographic dual
description with an IR cutoff $z_{f}$, defined in the Riemann geometry without a
conical singularity and given by the $D$-dimensional metric of $d s^{2} = d
r^{2} + r^{2} d \theta^{2} + \delta_{ij} d x_{\perp}^{i} d x_{\perp}^{j}$
with $i ~ \& ~ j = 2, \ldots, D-1$.
% Here,
For simplicity,
we
do not consider the self-interactions in the channel of energy-momentum tensor-currents.
%consider the absence of self-interactions in the channel of energy-momentum tensor-currents.
Then, the partition function is given by
\begin{align}
  &
    Z_{1}(z_{f}) = \int D \phi_{\alpha}(x) D \varphi(x,z) D g_{\mu\nu}(x,z)
    \delta\Big(g^{\mu\nu}(x,0) - g_{B}^{\mu\nu}\Big)
    \nn
  &\quad
    \times \delta\Big(\partial_{z} g^{\mu\nu}(x,z) - g^{\mu\nu'}(x,z) \big(
    \partial_{\nu'} \partial_{\mu'} G_{xx'}[g_{\mu\nu}(x,z),\varphi(x,z)] \big)_{x'
    \rightarrow x} g^{\mu'\nu}(x,z) \Big)
    \nn
  &\quad
    \times \exp\Big[ - \int d^{D} x \sqrt{g(x,z_{f})} \Big\{
    g^{\mu\nu}(x,z_{f}) [\partial_{\mu} \phi_{\alpha}(x)] [\partial_{\nu}
    \phi_{\alpha}(x)] + [m^{2} - i \varphi(x,z_{f})] \phi_{\alpha}^{2}(x) \Big\}
    \nn
  &\qquad - N \int d^{D} x \sqrt{g(x,0)} \Big\{ \frac{1}{2u} \Big(
    \varphi(x,0) - i \xi R(x,0) \Big)^{2} \Big\}
    \nn
  &\qquad - N \int_{0}^{z_{f}} d z \int d^{D} x \sqrt{g(x,z)} \Big\{
    \nonumber \\
  &\qquad\quad
    +\frac{1}{2u} [\partial_{z} \varphi(x,z)]^{2} + \frac{\mathcal{C}_{\varphi}}{2}
    g^{\mu\nu}(x,z) [\partial_{\mu} \varphi(x,z)] [\partial_{\nu} \varphi(x,z)] +
    \mathcal{C}_{\xi} R(x,z) [\varphi(x,z)]^{2}
    \nn
  &\qquad \quad + \frac{1}{2 \kappa} \Big( R(x,z) - 2 \Lambda \Big) \Big\} \Big].
\end{align}
On the other hand,
$Z_{n}(z_{f})$ is the partition function of the emergent holographic dual
description with an IR cutoff $z_{f}$, defined in the Riemann geometry with a
conical singularity and given by the $D$-dimensional metric of $d s^{2} = d
r^{2} + n^{2} r^{2} d \theta^{2} + \delta_{ij} d x_{\perp}^{i} d x_{\perp}^{j}$
with $i ~ \& ~ j = 2, \ldots, D-1$ \cite{Entanglement_Entropy_Heat_Kernel}.
%Similarly, it is given by
It is given by
\begin{align}
& Z_{n}(z_{f}) = \int D \phi_{\alpha,n}(x) D \varphi_{n}(x,z) D
g_{\mu\nu,n}(x,z) \delta\Big(g^{\mu\nu}_{n}(x,0) - g_{B n}^{\mu\nu}\Big)
                \nn
 &\quad
\times \delta\Big(\partial_{z} g_{n}^{\mu\nu}(x,z) -
g_{n}^{\mu\nu'}(x,z) \big( \partial_{\nu'} \partial_{\mu'}
G_{xx'}[g_{\mu\nu,n}(x,z),\varphi_{n}(x,z)] \big)_{x' \rightarrow x}
g_{n}^{\mu'\nu}(x,z) \Big)
   \nn
  &\quad
\times \exp\Big[ - \int d^{D} x \sqrt{g_{n}(x,z_{f})} \Big\{
g_{n}^{\mu\nu}(x,z_{f}) [\partial_{\mu} \phi_{\alpha,n}(x)] [\partial_{\nu}
\phi_{\alpha,n}(x)] + [m^{2} - i \varphi_{n}(x,z_{f})] \phi_{\alpha,n}^{2}(x) \Big\}
    \nn
  &\qquad  - N \int d^{D} x \sqrt{g_{n}(x,0)} \Big\{ \frac{1}{2u} \Big(
\varphi_{n}(x,0) - i \xi R_{n}(x,0) \Big)^{2} \Big\}
    \nn
  &\qquad - N \int_{0}^{z_{f}} d z \int d^{D} x \sqrt{g_{n}(x,z)} \Big\{
    \nonumber \\
  &\qquad \quad
+\frac{1}{2u} [\partial_{z} \varphi_{n}(x,z)]^{2} +
\frac{\mathcal{C}_{\varphi}}{2} g_{n}^{\mu\nu}(x,z) [\partial_{\mu}
\varphi_{n}(x,z)] [\partial_{\nu} \varphi_{n}(x,z)] + \mathcal{C}_{\xi}
R_{n}(x,z) [\varphi_{n}(x,z)]^{2}
    \nn
  &\qquad \quad  + \frac{1}{2 \kappa} \Big( R_{n}(x,z) - 2 \Lambda \Big) \Big\} \Big] . \label{Dual_Holographic_Partition_Function_Conical_Singularity}
\end{align}
%
%\bqa && Z_{n}(z_{f}) = \int D \phi_{\alpha}(x) D g_{\mu\nu}^{n}(x,z) \delta\Big(g^{\mu\nu}_{n}(x,0) - g_{B n}^{\mu\nu}\Big) \delta\Big(\partial_{z} g_{n}^{\mu\nu}(x,z) - \frac{1}{2 d z} g_{n}^{\mu\nu'}(x,z) [\partial_{\nu'} \partial_{\mu'} G_{xx'}^{n}(z)]_{x' \rightarrow x} g_{n}^{\mu'\nu}(x,z) \Big) \nn && \exp\Big[ - \int d^{D} x \sqrt{g_{n}(x,z_{f})} \Big\{ g_{n}^{\mu\nu}(x,z_{f}) (\partial_{\mu} \phi_{\alpha}(x)) (\partial_{\nu} \phi_{\alpha}(x)) + m^{2} \phi_{\alpha}^{2}(x) \Big\} \nn && - \frac{N}{2 \kappa} \int_{0}^{z_{f}} d z \int d^{D} x \sqrt{g_{n}(x,z)} \Big( R_{n}(x,z) - 2 \Lambda \Big) \Big] . \label{Dual_Holographic_Partition_Function_Conical_Singularity} \eqa
%
%\subsubsection{Total entanglement entropy}
%

Following this prescription, we find that the entanglement entropy
is composed
% by
of
two pieces, given by
\bqa && \mathcal{S}_{EE}(z_{f}) = \mathcal{S}_{EE}^{\phi_{\alpha}}(z_{f}) + \mathcal{S}_{EE}^{GR}(z_{f}) , \eqa
where $\mathcal{S}_{EE}^{\phi_{\alpha}}(z_{f})$ is the entanglement entropy from the matter sector at the IR cutoff $z = z_{f}$ and $\mathcal{S}_{EE}^{GR}(z_{f})$ is that from the emergent geometry also at the IR cutoff $z = z_{f}$.

It is straightforward to see that the matter contribution is \cite{Generalized_Gravitational_Entropy}
\begin{align}
&
\mathcal{S}_{EE}^{\phi_{\alpha}}(z_{f}) = \int d^{D-2} x_{\perp} \int_{0}^{2
  \pi} d \theta \int_{0}^{\infty} d r \sqrt{g_{n}(r,x_{\perp},z_{f})} \Big\langle
T_{\mu\nu,n}^{\phi_{\alpha}}(r,x_{\perp},z_{f}) \Big\rangle \frac{\partial
  g_{n}^{\mu\nu}(r,x_{\perp},z_{f})}{\partial n} \Big|_{n = 1}
%
%                \nn   &\qquad  - \frac{N}{2} \mbox{tr}_{xx'} \ln \sqrt{g(x,z_{f})} \Big\{ - \frac{1}{\sqrt{g(x,z_{f})}} \partial_{\mu} \Big( \sqrt{g(x,z_{f})} g^{\mu\nu}(x,z_{f}) \partial_{\nu} \Big) + m^{2} - i \varphi(x,z_{f}) \Big\}
%
. \label{EE_Matter_Formula_I}
\end{align}
Here, the energy-momentum tensor is given by
\begin{align}
   T_{\mu\nu,n}^{\phi_{\alpha}}(x,z_{f})
  &\equiv \frac{1}{\sqrt{g_{n}(x,z_{f})}}
\frac{\partial \sqrt{g_{n}(x,z_{f})}
  \mathcal{L}_{\phi_{\alpha}}[g_{\mu\nu,n}(x,z_{f}),\varphi_{n}(x,z_{f})]}{\partial
  g^{\mu\nu}_{n}(x,z_{f})}
                \nn
  &
     = 2 (\partial_{\mu} \phi_{\alpha,n}(x)) (\partial_{\nu} \phi_{\alpha,n}(x))
    \nonumber \\
  &\quad
     + g_{\mu\nu,n}(x,z_{f}) \Big( g^{\mu'\nu'}_{n}(x,z_{f}) (\partial_{\mu'} \phi_{\alpha,n}(x)) (\partial_{\nu'} \phi_{\alpha,n}(x)) + [m^{2} - i \varphi_{n}(x,z_{f})] \phi_{\alpha,n}^{2}(x) \Big) ,
\end{align}
where the effective Lagrangian for the matter sector is
$\mathcal{L}_{\phi_{\alpha}}[g_{\mu\nu,n}(x,z_{f}),\varphi_{n}(x,z_{f})] = g^{\mu\nu}_{n}(x,z_{f}) [\partial_{\mu} \phi_{\alpha,n}(x)] [\partial_{\nu} \phi_{\alpha,n}(x)] + [m^{2} - i \varphi_{n}(x,z_{f})] \phi_{\alpha,n}^{2}(x)$. Inserting this energy-momentum tensor into the above expression, we obtain
\bqa
&& \mathcal{S}_{EE}^{\phi_{\alpha}}(z_{f}) =
\nonumber \\
&&\quad
N \int d^{D-2} x_{\perp}
\int_{0}^{2 \pi} d \theta \int_{0}^{\infty} d r \sqrt{g_{n}(r,x_{\perp},z_{f})}
\Big\{ 2 \big(\partial_{\mu} \partial_{\nu}
G[x,x;g_{\mu\nu,n}(x,z_{f}),\varphi_{n}(x,z_{f})]\big)
\nn &&
\quad\quad  +
g_{\mu\nu,n}(r,x_{\perp},z_{f}) \Big( g^{\mu'\nu'}_{n}(r,x_{\perp},z_{f})
\big(\partial_{\mu'} \partial_{\nu'}
G[x,x;g_{\mu\nu,n}(x,z_{f}),\varphi_{n}(x,z_{f})]\big)
\nn &&\qquad  + [m^{2} - i
\varphi_{n}(x,z_{f})] G[x,x;g_{\mu\nu,n}(x,z_{f}),\varphi_{n}(x,z_{f})] \Big) \Big\}
\frac{\partial g_{n}^{\mu\nu}(r,x_{\perp},z_{f})}{\partial n} \Big|_{n = 1}
%
%\nn && \quad - N \int d^{D-2} x_{\perp} \int_{0}^{2 \pi} d \theta \int_{0}^{\infty} d r \sqrt{g(r,x_{\perp},z_{f})} \Big\{ \frac{\mathcal{C}_{\varphi}^{f}}{2} g^{\mu\nu}(r,x_{\perp},z_{f}) [\partial_{\mu} \varphi(r,x_{\perp},z_{f})] [\partial_{\nu} \varphi(r,x_{\perp},z_{f})] \nn && \qquad + \mathcal{C}_{\xi}^{f} R(r,x_{\perp},z_{f}) [\varphi(r,x_{\perp},z_{f})]^{2} + \frac{1}{2 \kappa_{f}} \Big( R(r,x_{\perp},z_{f}) - 2 \Lambda_{f} \Big) \Big\}
%
, \eqa
where $G[x,x';g_{\mu\nu,n}(x,z_{f}),\varphi_{n}(x,z_{f})] = \Big\langle \frac{1}{N}
\sum_{\alpha = 1}^{N} \phi_{\alpha,n}(x) \phi_{\alpha,n}(x') \Big\rangle$ is the
Green's function of the original scalar fields at $z = z_{f}$ with the conical singularity.
Both the Klein-Gordon action and the induced Einstein-Hilbert gravity action at $z = z_{f}$ in the last two lines result from the gradient expansion of the last logarithmic term \cite{Gradient_Expansion_Gravity_I,Gradient_Expansion_Gravity_II} in Eq. (\ref{EE_Matter_Formula_I}).

The entanglement entropy from the classical geometry in the presence of dual scalar fields is given by
\bqa && \mathcal{S}_{EE}^{GR}(z_{f}) =
\nonumber\\
&&\quad
N \int d^{D-2} x_{\perp} \int_{0}^{2 \pi}
d \theta \int_{0}^{\infty} d r \int_{0}^{z_{f}} d z \sqrt{g_{n}(r,x_{\perp},z)} ~
T_{\mu\nu,n}^{GR}(r,x_{\perp},z) ~ \frac{\partial
  g_{n}^{\mu\nu}(r,x_{\perp},z)}{\partial n} \Big|_{n = 1}
%
%\nn &&\quad - N \int d^{D-2} x_{\perp} \int_{0}^{2 \pi} d \theta \int_{0}^{\infty} d r \int_{0}^{z_{f}} d z \sqrt{g(r,x_{\perp},z)} \Big\{ \frac{1}{2u} [\partial_{z} \varphi(r,x_{\perp},z)]^{2} \nn &&\qquad + \frac{\mathcal{C}_{\varphi}}{2} g^{\mu\nu}(r,x_{\perp},z) [\partial_{\mu} \varphi(r,x_{\perp},z)] [\partial_{\nu} \varphi(r,x_{\perp},z)] + \mathcal{C}_{\xi} R(r,x_{\perp},z) [\varphi(r,x_{\perp},z)]^{2} \nn &&\qquad + \frac{1}{2 \kappa} \Big( R(r,x_{\perp},z) - 2 \Lambda \Big) \Big\}
%
. \eqa
The energy-momentum tensor of the gravity sector is
\bqa
&& T_{\mu\nu,n}^{GR}(x,z) \equiv \frac{1}{\sqrt{g_{n}(x,z)}} \frac{\partial \sqrt{g_{n}(x,z)} \mathcal{L}_{GR}[g_{\mu\nu,n}(x,z),\varphi_{n}(x,z)]}{\partial g^{\mu\nu}_{n}(x,z)} \equiv T_{\mu\nu,n}^{g_{\mu\nu}}(x,z) + T_{\mu\nu,n}^{\varphi}(x,z) , \nn \eqa
where the bulk Lagrangian in terms of bulk metric tensor and dual scalar fields is given by
\begin{align}
  \mathcal{L}_{GR}[g_{\mu\nu,n}(x,z),\varphi_{n}(x,z)]
  &= \frac{1}{2u} [\partial_{z} \varphi_{n}(x,z)]^{2} + \frac{\mathcal{C}_{\varphi}}{2} g^{\mu\nu}_{n}(x,z) [\partial_{\mu} \varphi_{n}(x,z)] [\partial_{\nu} \varphi_{n}(x,z)] + \mathcal{C}_{\xi} R_{n}(x,z) [\varphi_{n}(x,z)]^{2}
    \nn
  &\quad + \frac{1}{2 \kappa} \Big( R_{n}(x,z) - 2 \Lambda \Big) .
\end{align}
As a result, the energy-momentum tensor of our classical gravity is \cite{Energy_Momentum_Tensor_Gravity}
\begin{align}
  &2 \kappa T_{\mu\nu,n}^{g_{\mu\nu}}(x,z)
  = 2 \Big( R_{\mu\nu,n}(x,z) - \frac{1}{2} g_{\mu\nu,n}(x,z) R_{n}(x,z) + \Lambda g_{\mu\nu,n}(x,z) \Big)
    \nn
  &\quad + \frac{1}{g_{n}(x,z)} g_{\mu\mu',n}(x,z) g_{\nu\nu',n}(x,z) \partial_{\mu''} \partial_{\nu''} \Big\{g_{n}(x,z) \Big( g^{\mu'\nu'}_{n}(x,z) g^{\mu''\nu''}_{n}(x,z) - g^{\mu'\mu''}_{n}(x,z) g^{\nu'\nu''}_{n}(x,z) \Big) \Big\} ,
\end{align}
and that of the dual scalar field is
\begin{align}
  T_{\mu\nu,n}^{\varphi}(x,z)
  &= \frac{2}{u} \delta_{\mu z} \delta_{\nu z} [\partial_{z} \varphi_{n}(x,z)]^{2} + 2 (1 - \delta_{\mu z} \delta_{\nu z}) \mathcal{C}_{\varphi} [\partial_{\mu} \varphi_{n}(x,z)] [\partial_{\nu} \varphi_{n}(x,z)]
    \nn
  &\quad + g_{\mu\nu,n}(x,z) \Big( \frac{1}{u} [\partial_{z} \varphi_{n}(x,z)]^{2} + \mathcal{C}_{\varphi} g^{\mu'\nu'}_{n}(x,z) [\partial_{\mu'} \varphi_{n}(x,z)] [\partial_{\nu'} \varphi_{n}(x,z)] \Big)
    \nn
  &\quad + 2\mathcal{C}_{\xi} \Big( R_{\mu\nu,n} - \frac{1}{2} g_{\mu\nu,n} R_{n} \Big) [\varphi_{n}(x,z)]^{2} + 2 \mathcal{C}_{\xi} \Big( g_{\mu\nu,n} \partial_{\rho} \partial^{\rho} - \partial_{\mu} \partial_{\nu} \Big) [\varphi_{n}(x,z)]^{2}.
\end{align}

\subsection{Heat kernel method for entanglement entropy} \label{Heat_kernel_method_for_entanglement_entropy}

%
%\subsubsection{Setup}
%

To show the entanglement transfer from matter to geometry through recursive
renormalization group transformations, we need to simplify the
entanglement-entropy formula of the matter sector. Here, we review the
entanglement-entropy formula based on the heat kernel method
\cite{Entanglement_Entropy_Heat_Kernel}. An effective free energy on
%the $n-$Riemann sheet
$n$-sheeted Riemann
with a conical singularity can be represented by
\bqa && - \ln Z_{n}^{\phi_{\alpha}}(z_{f}) = - \frac{N}{2} \int_{\epsilon^{2}}^{\infty} \frac{d s}{s} \mbox{tr}_{xx'} K(s,x,x';z_{f}) , \eqa
where $K(s,x,x';z_{f}) \equiv \langle x z_{f} | e^{- s \mathcal{D}} | x' z_{f} \rangle$ is the corresponding heat kernel with the differential operator $\mathcal{D} \equiv - \frac{1}{\sqrt{g(x,z_{f})}} \partial_{\mu} \Big( \sqrt{g(x,z_{f})} g^{\mu\nu}(x,z_{f}) \partial_{\nu} \Big) + m^{2}$. For the time being, we consider the case of a constant mass, given by $m^{2} - i \varphi(x,z_{f}) \Longrightarrow m^{2}$.
%
%\bqa && K(s,x,x';z_{f}) \equiv \langle x z_{f} | e^{- s \mathcal{D}} | x' z_{f} \rangle , ~~~~~ \mathcal{D} \equiv - \frac{1}{\sqrt{g(x,z_{f})}} \partial_{\mu} \Big( \sqrt{g(x,z_{f})} g^{\mu\nu}(x,z_{f}) \partial_{\nu} \Big) + m^{2} \eqa
%
This heat kernel satisfies
an effective heat-diffusion equation $(\partial_{s} + \mathcal{D})
K(s,x,x';z_{f}) = 0$ with an initial condition $K(s = 0, x, x';z_{f}) =
\frac{1}{\sqrt{g(x,z_{f})}} \delta^{(D)}(x-x')$
%which assigns the name of heat kernel to this propagator.
-- this is the origin of the name of the heat kernel.
%
%\bqa && (\partial_{s} + \mathcal{D}) K(s,x,x';z_{f}) = 0, ~~~~~ K(s = 0, x, x';z_{f}) = \frac{1}{\sqrt{g(x,z_{f})}} \delta^{(D)}(x-x') \eqa
%
%\subsubsection{The heat kernel expansion on a space with a conical singularity}
%
$\epsilon$ is a UV cutoff to cure the UV divergence of the entanglement entropy, which will be clarified below.

To deal with curvature integrals in the presence of a conical singularity, we
need to understand the curvature structure of this geometry
\cite{Entanglement_Entropy_Heat_Kernel}.
Let $E_{\alpha}$ be an $\alpha$-fold Riemann sheet, which covers a smooth manifold $E$ along the Killing vector $\partial_{\theta}$. A codimension two surface $\Sigma$ is a stationary point ($r = 0$) of this isometry. Then, the space $E_{\alpha}$ is a direct product near $\Sigma$ ($r = 0$), given by $\Sigma \times C_{\alpha}$, where $C_{\alpha}$ is a two-dimensional cone with angle deficit $\delta = 2 \pi (1 - \alpha)$. In other words, the metric of $E_{\alpha}$ is $d s_{E_{\alpha}}^{2} = d s_{C_{\alpha}}^{2} + d s_{\Sigma}^{2}$, where the metric of the two dimensional cone is $d s_{C_{\alpha}}^{2} = g_{rr}^{\alpha}(r,x_{\perp},z) d r^{2} + g_{\theta\theta}^{\alpha}(r,x_{\perp},z) d \theta^{2}$ and the metric of the codimension two surface is $d s_{\Sigma}^{2} = \gamma_{ij}(r,x_{\perp},z) d x_{\perp}^{i} d x_{\perp}^{j}$ at a fixed renormalization group scale $z$. Outside this singular surface $\Sigma$, $E_{\alpha}$ is reduced to the smooth manifold $E$, where their curvature tensors coincide. On the other hand, the conical singularity at the surface $\Sigma$ gives rise to a singular contribution to the curvatures. Here, the positive integer $n$ is analytically continued to a real value $\alpha$, which results from the abelian isometry generated by the Killing vector $\partial_{\theta}$.
%
%\bqa && E_{\alpha} \approx \Sigma \times C_{\alpha} \eqa
%
%\bqa && d s_{E_{\alpha}}^{2} \equiv g_{\mu\nu}^{\alpha}(x,z) dx^{\mu} d x^{\nu} = g_{rr}^{\alpha}(r,x_{\perp},z) d r^{2} + g_{\theta\theta}^{\alpha}(r,x_{\perp},z) d \theta^{2} + d s_{\Sigma}^{2} , ~~~~~ d s_{\Sigma}^{2} = \gamma_{ij}(r,x_{\perp},z) d x_{\perp}^{i} d x_{\perp}^{j} \eqa
%

One may take an expansion for small $s$ in the heat kernel, given by
\bqa && \mbox{tr}_{xx'} K_{E_{\alpha}}(s) = \frac{1}{(4\pi s)^{\frac{d}{2}}} \sum_{n = 0} a_{n} s^{n} . \label{Heat_Kernel_Expansion} \eqa
Then, we have two types of coefficients in this expansion,
\bqa && a_{n} = a_{n}^{reg} + a_{n}^{\Sigma} , \eqa
where $a_{n}^{reg}$ is a regular contribution resulting from the bulk and
$a_{n}^{\Sigma}$ is a singular one coming from the surface.
Regular bulk coefficients are given by
\begin{align}
   a_{0}^{reg} &= \int_{E} d^{D} x \sqrt{g(x,z_{f})} ,
    \nonumber \\
  a_{1}^{reg} &=
\int_{E} d^{D} x \sqrt{g(x,z_{f})} \Big( \frac{1}{6} R(x,z_{f}) - m^{2} \Big) ,
                \nonumber \\
 a_{2}^{reg} &= \int_{E} d^{D} x \sqrt{g(x,z_{f})} \Big\{ \frac{1}{180}
R^{2}_{\mu\nu\alpha\beta}(x,z_{f}) - \frac{1}{180} R_{\mu\nu}^{2}(x,z_{f})
                               \nonumber \\
&
                                              \quad + \frac{1}{6} \frac{1}{\sqrt{g(x,z_{f})}} \partial_{\mu} \Big( \sqrt{g(x,z_{f})} g^{\mu\nu}(x,z_{f}) \partial_{\nu} \Big) \Big( \frac{1}{5} R(x,z_{f}) - m^{2} \Big) + \frac{1}{2} \Big( \frac{1}{6} R(x,z_{f}) - m^{2} \Big)^{2} \Big\} ,
\end{align}
for general curvature tensors \cite{Entanglement_Entropy_Heat_Kernel}. On the other hand, singular surface coefficients are given by \cite{Entanglement_Entropy_Heat_Kernel}
\begin{align}
  a_{0}^{\Sigma}
  &= 0 ,
    \nonumber \\
  a_{1}^{\Sigma}
  &= \frac{\pi}{3}
\frac{(1-\alpha)(1+\alpha)}{\alpha} \int_{\Sigma(z_{f})} d^{D-2} x_{\perp}
    \sqrt{\gamma(x,z_{f})} ,
    \nonumber \\
  a_{2}^{\Sigma}
  &= \frac{\pi}{3}
\frac{(1-\alpha)(1+\alpha)}{\alpha} \int_{\Sigma(z_{f})} d^{D-2} x_{\perp}
    \sqrt{\gamma(x,z_{f})} \Big( \frac{1}{6} R(x,z_{f}) - m^{2} \Big)
    \nonumber \\
  &\quad
    - \frac{\pi}{180} \frac{(1-\alpha)(1+\alpha)(1+\alpha^{2})}{\alpha^{3}} \int_{\Sigma(z_{f})} d^{D-2} x_{\perp} \sqrt{\gamma(x,z_{f})} \Big( R_{ii}(x,z_{f}) - 2 R_{ijij}(x,z_{f}) \Big) .
\label{Heat_Kernel_Expansion_Singular_Coefficients}
                                                                         \end{align}
We recall $d s_{\Sigma}^{2} = \gamma_{ij}(r,x_{\perp},z) d x_{\perp}^{i} d x_{\perp}^{j}$.

Based on this general curvature formula from the heat kernel expansion, we obtain the following expression of the entanglement entropy for the matter sector
\begin{align}
 & \mathcal{S}_{EE}^{\phi_{\alpha}}(z_{f}) = (\alpha \partial_{\alpha} - 1)
[- \ln Z_{\phi_{\alpha}}(\alpha;z_{f})] \Big|_{\alpha = 1}
   \nn
  &
\qquad \approx - N \frac{\pi}{6} \frac{1}{(4\pi)^{\frac{D}{2}}} \Big\{ (\alpha \partial_{\alpha} - 1) \frac{(1-\alpha)(1+\alpha)}{\alpha} \Big\}_{\alpha = 1} \Bigg( \int_{\epsilon^{2}}^{\infty} \frac{d s}{s} s^{1 - \frac{D}{2}} \Bigg) \Big\{ \int_{\Sigma(z_{f})} d^{D-2} x_{\perp} \sqrt{\gamma(x,z_{f})} \Big\} .
\end{align}
This leading contribution results from the $a_{1}^{\Sigma}$ term in
Eq.\ (\ref{Heat_Kernel_Expansion_Singular_Coefficients}).
Considering that the area of the singular surface $\Sigma(z_{f})$ is given by
\bqa && \mathcal{A}[\Sigma(z_{f})] = \int_{\Sigma(z_{f})} d^{D-2} x_{\perp} \sqrt{\gamma(x,z_{f})} \eqa
at the renormalization group scale $z = z_{f}$, we obtain the area law for the entanglement entropy of the matter sector
%
%\bqa && \int_{\epsilon^{2}}^{\infty} \frac{d s}{s} s^{1 - \frac{D}{2}} = \frac{1}{- \frac{D}{2} + 1} s^{- \frac{D}{2} + 1} \Big|_{\epsilon^{2}}^{\infty} = \frac{1}{\frac{D}{2} - 1} \frac{1}{\epsilon^{D-2}} , ~~~~~ D > 2 \eqa
%
\bqa && \mathcal{S}_{EE}^{\phi_{\alpha}}(z_{f}) \approx \frac{N}{6 (D - 2) (4\pi)^{\frac{D}{2} - 1}} \frac{\mathcal{A}[\Sigma(z_{f})]}{\epsilon^{D-2}} . \eqa
%
%\subsubsection{$D = 2$}
%
It is straightforward to see that this formula is reduced
% into
to
the well known result in two spacetime dimensions, given by
\bqa && \mathcal{S}_{EE}^{\phi_{\alpha}}(z_{f}) = \frac{N}{6} \ln \frac{\xi}{\epsilon} . \label{EE_2D} \eqa
Here, $\xi$ is the correlation length proportional to the inverse of the mass parameter $m$, which appears from the introduction of an IR cutoff in the $s-$integral, given by $\int_{\epsilon^{2}}^{\xi^{2}} \frac{d s}{s}$.

Now, we introduce inhomogeneity of the mass term into the above heat-kernel expansion. Then, spacetime derivatives for $\varphi(x,z_{f})$ will appear in both expansion coefficients. As far as a disordered phase is concerned, we speculate that this area law would be preserved. It is an interesting question how $\partial_{\mu} \varphi(x,z_{f})$ modify the entanglement entropy of the matter sector at quantum criticality. In this paper, we continue our discussions for the gapped phase.

Finally, we obtain the total entanglement entropy at $z = z_{f}$
\begin{align}
  \mathcal{S}_{EE}(z_{f})
  &= \mathcal{S}_{EE}^{\phi_{\alpha}}(z_{f}) +
    \mathcal{S}_{EE}^{GR}(z_{f})
    \nn
  &
    = \frac{N}{6 (D - 2) (4\pi)^{\frac{D}{2} - 1}}
    \frac{\mathcal{A}[\Sigma(z_{f})]}{\epsilon^{D-2}}
    \nn
  &\qquad  + N \int d^{D-2} x_{\perp} \int_{0}^{2 \pi} d \theta
    \int_{0}^{\infty} d r \int_{0}^{z_{f}} d z \sqrt{g_{n}(r,x_{\perp},z)} ~
    \nonumber \\
  &\qquad
    \quad
    \times
    \Big(
    T_{\mu\nu,n}^{\varphi}(r,x_{\perp},z) + T_{\mu\nu,n}^{g_{\mu\nu}}(r,x_{\perp},z)
    \Big) ~ \frac{\partial g_{n}^{\mu\nu}(r,x_{\perp},z)}{\partial n} \Big|_{n = 1}
%
%    \nn   &\qquad - N \int d^{D-2} x_{\perp} \int_{0}^{2 \pi} d \theta     \int_{0}^{\infty} d r \int_{0}^{z_{f}} d z \sqrt{g(r,x_{\perp},z)} \Big\{     \frac{1}{2u} [\partial_{z} \varphi(r,x_{\perp},z)]^{2}     \nn   &\qquad \quad     + \frac{\mathcal{C}_{\varphi}}{2} g^{\mu\nu}(r,x_{\perp},z)     [\partial_{\mu} \varphi(r,x_{\perp},z)] [\partial_{\nu} \varphi(r,x_{\perp},z)]     + \mathcal{C}_{\xi} R(r,x_{\perp},z) [\varphi(r,x_{\perp},z)]^{2}     \nn   &\qquad \quad     + \frac{1}{2 \kappa} \Big( R(r,x_{\perp},z) - 2 \Lambda \Big) \Big\}
%
. \label{Entanglement_Entropy_Renormalization}
\end{align}
%
%\bqa && \mathcal{S}_{EE}(z_{f}) = \mathcal{S}_{EE}^{\phi_{\alpha}}(z_{f}) + \mathcal{S}_{EE}^{g_{\mu\nu}}(z_{f}) = \frac{N}{6 (D - 2) (4\pi)^{\frac{D}{2} - 1}} \frac{\mathcal{A}[\Sigma(z_{f})]}{\epsilon^{D-2}} \nn && + \frac{N}{2 \kappa} \int d^{D-2} x_{\perp} \int_{0}^{2 \pi} d \theta \int_{0}^{\infty} d r \int_{0}^{z_{f}} d z \sqrt{g(x,z)} \Big[ - 2 \Big( R_{\mu\nu}(x,z) - \frac{1}{2} g_{\mu\nu}(x,z) R(x,z) + \Lambda g_{\mu\nu}(x,z) \Big) \nn && + \frac{1}{g(x,z)} g_{\mu\mu'}(x,z) g_{\nu\nu'}(x,z) \partial_{\mu''} \partial_{\nu''} \Big\{g(x,z) \Big( g^{\mu'\nu'}(x,z) g^{\mu''\nu''}(x,z) - g^{\mu'\mu''}(x,z) g^{\nu'\nu''}(x,z) \Big) \Big\} \Big] \frac{\partial g_{n}^{\mu\nu}(r,x_{\perp},z)}{\partial n} \Big|_{n = 1} \nn && - \frac{N}{2 \kappa} \int d^{D-2} x_{\perp} \int_{0}^{2 \pi} d \theta \int_{0}^{\infty} d r \int_{0}^{z_{f}} d z \sqrt{g(x,z)} \Big( R(x,z) - 2 \Lambda \Big) . \label{Entanglement_Entropy_Renormalization} \eqa
%
In this formula we focus on the fact that the entanglement entropy of the matter
sector is given by the area of the subsystem at a fixed renormalization group
scale $z = z_{f}$ while that of the classical gravity part is given by the
integral of the extra dimensional space. If the emergent geometry is given by a
cap in the extra dimensional space, the area of the subsystem at a given $z =
z_{f}$ vanishes beyond
% the end of
the cap in the extra dimensional space. Instead, the entanglement entropy from the classical geometry is maximized. Actually, the cap geometry has been realized when the corresponding quantum state is gapped \cite{SungSik_Holography_III,Holographic_Description_Kim,Horizon_critical_phenomenon}. Even if the resulting geometry is given by AdS$_{D+1}$, the entanglement entropy is transferred from quantum matter to classical gravity.

%
%\section{Emergent holographic description is transferring quantum entanglement from quantum matter to classical gravity}
%

\subsection{Invariance of entanglement entropy with respect to renormalization group transformations} \label{Entanglement_Entropy_RG_Invariant}

Since the partition function
% of Eq.
\eqref{Dual_Holographic_Partition_Function_Conical_Singularity} has to be invariant with respect to recursive renormalization group transformations, we obtain the following equation
\bqa && \partial_{z_{f}} \mathcal{S}_{EE}(z_{f}) = \partial_{z_{f}} \mathcal{S}_{EE}^{\phi_{\alpha}}(z_{f}) + \partial_{z_{f}} \mathcal{S}_{EE}^{g_{\mu\nu}}(z_{f}) = 0 \Longrightarrow \mathcal{S}_{EE}(z_{f}) = \mathcal{S}_{EE}^{\phi_{\alpha}}(0) . \eqa
In other words, the entanglement entropy is an invariant for renormalization group transformations. Resorting to the full expression Eq.\ (\ref{Entanglement_Entropy_Renormalization}) of the entanglement entropy, we reach the following identity
\bqa
&& \frac{N}{6 (D - 2) (4\pi)^{\frac{D}{2} - 1}} \frac{\partial_{z_{f}}
  \mathcal{A}[\Sigma(z_{f})]}{\epsilon^{D-2}}
\nn &&\quad + N \int d^{D-2} x_{\perp} \int_{0}^{2 \pi} d \theta
\int_{0}^{\infty} d r \sqrt{g_{n}(r,x_{\perp},z_{f})} ~ \Big(
T_{\mu\nu,n}^{\varphi}(r,x_{\perp},z_{f}) +
T_{\mu\nu,n}^{g_{\mu\nu}}(r,x_{\perp},z_{f}) \Big) ~ \frac{\partial
  g_{n}^{\mu\nu}(r,x_{\perp},z_{f})}{\partial n} \Big|_{n = 1}
%
%\nn &&\quad - N \int d^{D-2} x_{\perp} \int_{0}^{2 \pi} d \theta \int_{0}^{\infty} d r \sqrt{g(r,x_{\perp},z_{f})} \Big\{ \frac{1}{2u} [\partial_{z} \varphi(r,x_{\perp},z_{f})]^{2} \nn &&\qquad + \frac{\mathcal{C}_{\varphi}}{2} g^{\mu\nu}(r,x_{\perp},z_{f}) [\partial_{\mu} \varphi(r,x_{\perp},z_{f})] [\partial_{\nu} \varphi(r,x_{\perp},z_{f})] + \mathcal{C}_{\xi} R(r,x_{\perp},z_{f}) [\varphi(r,x_{\perp},z_{f})]^{2} \nn &&\qquad + \frac{1}{2 \kappa} \Big( R(r,x_{\perp},z_{f}) - 2 \Lambda \Big) \Big\}
%
\nn && = 0 .
\eqa
%
%\bqa && \frac{N}{6 (D - 2) (4\pi)^{\frac{D}{2} - 1}} \frac{\partial_{z_{f}} \mathcal{A}[\Sigma(z_{f})]}{\epsilon^{D-2}} \nn && + \frac{N}{2 \kappa} \int d^{D-2} x_{\perp} \int_{0}^{2 \pi} d \theta \int_{0}^{\infty} d r \sqrt{g(x,z_{f})} \Big[ - 2 \Big( R_{\mu\nu}(x,z_{f}) - \frac{1}{2} g_{\mu\nu}(x,z_{f}) R(x,z_{f}) + \Lambda g_{\mu\nu}(x,z_{f}) \Big) \nn && + \frac{1}{g(x,z_{f})} g_{\mu\mu'}(x,z_{f}) g_{\nu\nu'}(x,z_{f}) \partial_{\mu''} \partial_{\nu''} \Big\{g(x,z_{f}) \Big( g^{\mu'\nu'}(x,z_{f}) g^{\mu''\nu''}(x,z_{f}) - g^{\mu'\mu''}(x,z_{f}) g^{\nu'\nu''}(x,z_{f}) \Big) \Big\} \Big] \nn && \times \frac{\partial g_{n}^{\mu\nu}(r,x_{\perp},z_{f})}{\partial n} \Big|_{n = 1} - \frac{N}{2 \kappa} \int d^{D-2} x_{\perp} \int_{0}^{2 \pi} d \theta \int_{0}^{\infty} d r \sqrt{g(x,z_{f})} \Big( R(x,z_{f}) - 2 \Lambda \Big) = 0 . \eqa
%
We suspect that
the bulk gravity contribution may be related with the holographic entanglement
entropy of the Ryu-Takayanagi formula
\cite{Entanglement_Entropy_Ryu_Takayanagi_I,Entanglement_Entropy_Ryu_Takayanagi_II}.
In particular, one may express this identity in terms of fully geometric information such as metric and curvature. In appendix \ref{Entanglement_Entropy_RG_Invariant_General}, we discuss this identity further and verify the first iteration version of this identity.

\section{Discussion} \label{Discussion}

\subsection{Remarks on the appearance of higher-spin fields}

One may suggest an effective higher spin gauge theory as a dual holographic description of the $O(N)$ vector model, which has an infinite tower of higher-spin fields without hierarchy in the spectrum \cite{Higher_Spin_Gauge_Theory_I,Higher_Spin_Gauge_Theory_II,Higher_Spin_Gauge_Theory_III, Higher_Spin_Gauge_Theory_IV}. Such higher-spin fields can result from the procedure to reformulate the non-local quadratic term generated by the renormalization group transformation in a local way, where emergent bi-local fields such as
\bqa && - \int d^{D} x \sqrt{g_{x}^{(0)}} \int d^{D} x' \sqrt{g_{x'}^{(0)}} g^{\mu\nu(0)}_{x}
g^{\mu'\nu'(0)}_{x'} (\partial_{\mu} \phi_{\alpha x}) (\partial_{\nu}
\partial_{\mu'} G_{xx'}^{(0)}) (\partial_{\nu'} \phi_{\alpha x'}) \nonumber \eqa
in Eq. (\ref{Non_Local_Quadratic_Term_Higher_Spin_Field}) are rewritten in terms
of higher-spin fields.
It has been discussed that the origin
% for the appearance
of higher-spin fields is
% in
the nonlocal diffeomorphism invariance \cite{Holography_Higher_Spin_RG_I,Holography_Higher_Spin_RG_II,Holography_Higher_Spin_RG_III}. Here, we show that the nonlocal diffeomorphism invariance is explicitly broken by the existence of effective interactions. Then, we argue that our effective dual holographic mean-field theory in the large $N$ limit gives a
%
%consistent and
%
physically meaningful description for strongly coupled field theories even if we do not take into account higher-spin fields.

First, to discuss the absence of the non-local reparameterization symmetry with
physically simple notations,
we consider a discrete version of the present effective field theory, given by
\bqa && \mathcal{S}_{eff} = - \sum_{ij} t_{ij} \phi_{i} \phi_{j} + m^{2} \sum_{i} \phi_{i}^{2} + \frac{u}{2} \sum_{i} (\phi_{i}^{2})^{2} , \eqa
where we put the continuum field theory on a D-dimensional lattice
and the indices $i,j$ label lattice sites.
The quadratic part enjoys
the invariance under the non-local reparameterization \cite{Holography_Higher_Spin_RG_I,Holography_Higher_Spin_RG_II,Holography_Higher_Spin_RG_III,SungSik_Holography_III}
\bqa && \phi_{i} \longrightarrow \sum_{j} V_{ij} \phi_{j} , ~~~~~ t_{ij} \longrightarrow \sum_{i'j'} V_{ii'} t_{i'j'} V_{jj'} , \eqa
where $V_{ij}$ is an orthogonal matrix
acting on the space of all sites.
%and satisfies $\sum_{i} V_{ki} V_{ij} = \delta_{kj}$.
This
nonlocal diffeomorphism invariance at the quadratic level,
however,
is explicitly broken by
the last self-interaction term:
In this paper, we considered the Hubbard-Stratonovich transformation as follows
\bqa && \frac{u}{2} \sum_{i} (\phi_{i}^{2})^{2} \longrightarrow \frac{1}{2 u} \sum_{i} \varphi_{i}^{2} - i \sum_{i} \varphi_{i} \phi_{i}^{2} , \eqa
where $\varphi_{i}$ is a collective dual order-parameter field. Applying the nonlocal coordinate transformation to $\sum_{i} \varphi_{i} \phi_{i}^{2}$, we obtain
\bqa && \sum_{i} \varphi_{i} \phi_{i}^{2} \longrightarrow \sum_{k} \sum_{j} \phi_{k} \Big( \sum_{i} \varphi_{i} V_{ki} V_{ij} \Big) \phi_{j} . \eqa
It is clear that if the bulk solution of $\varphi_{i}$ does not satisfy $\sum_{i} V_{ki} \varphi_{i} V_{ij} \not= \varphi_{j} \delta_{kj}$, this nonlocal diffeomorphism invariance is broken down explicitly.

However, we would like to point out that the vacuum solution
of the dual order-parameter field is possibly translational invariant,
given by a constant value in the D-dimensional Euclidean space.
Then, we suspect the existence of such higher-spin fields as a vacuum solution.
On the other hand, excitations of higher-spin fields break the nonlocal
diffeomorphism invariance, which would generate gaps for these higher-spin
fields. Still, they may not be ignored in principle because the gap would be at most $O(1)$ in
the unit of the curvature of the bulk.
Frankly speaking, we do not understand what happens in this situation.
In order to see higher-spin fields appearing in interacting theories,
for example,
we may have to properly modify the non-local diffeomorphisms.

Next, we discuss physics of our effective dual holographic mean-field
theory in the large $N$ limit.
We claim that this dual holographic effective field theory still deserves to be investigated,
% more carefully,
considering that
(i)
the partition function derived from this recursive renormalization group transformation is the
same as that resulting from a recursive Kadanoff block-spin transformation,
verified in one spatial dimension,
(ii)
the resulting classical field theory in the large $N$ limit takes into account
quantum corrections in the all-loop order, thus serving as a novel mean-field
theory framework with non-perturbative quantum corrections,
and
(iii)
a comparison
of the current holographic approach
with the Bethe ansatz solution in the Kondo problem
implies that the resulting (dual holographic) effective
field theory takes into account quantum corrections in a non-perturbative way.

\begin{itemize}
\item 
Recently, we studied an effective scalar field theory on a one-dimensional
lattice, by using the recursive Kadanoff block-spin transformation
\cite{Kadanoff_GR_Holography_Kim}.
As a result, we obtained an effective dual holographic partition function, where renormalization group $\beta$-function of all the coupling functions and the bulk effective action for the dual order-parameter field appears in the background of renormalized coupling functions. The emergent extra dimension corresponds to the iteration number of Kadanoff block-spin transformations.

Here, the conventional truncation scheme of the renormalization group transformation was utilized in the recursive Kadanoff block-spin transformation, where only local terms are kept to preserve the original form of the effective Hamiltonian. This appears to be essentially the same approximation we used in our gravitational/geometrical formulation presented in this paper. In fact, by comparing this effective field theory derived from the recursive Kadanoff block-spin transformation with the present geometric formulation (when energy-momentum tensor-type interactions are turned off and thus, the metric tensor is not dynamical, described by the delta-function constraint), we can establish that these two partition functions are equivalent up to a vacuum energy contribution. In particular, we could represent the renormalization group flow of the metric tensor (the present real-space renormalization group scheme) in terms of the running mass parameter (the Kadanoff block-spin renormalization group scheme).

The equivalence between these two formulations indicates that the truncation in the geometrical renormalization group formulation presented in this paper is a physically considerable one or at least as conventional, and often successful, as the truncation in the Kadanoff block spin transformations. With these truncations, higher-spin fields appear neither in the Kadanoff block-spin renormalization group scheme nor in the geometric formulation presented in this paper.

  \item
    Furthermore, we also proposed how to extract
    % out
    renormalization coefficients
  such as field renormalization, mass renormalization, and interaction
  renormalization constants from the present dual holographic effective field
  theory in a general ground. Reformulating the renormalized effective field theory with such renormalized coefficients in an effective holographic way, we
  could determine all these renormalization constants from the present effective
  geometry in a non-perturbative way
  \cite{RG_GR_Holography_Kim}.
  It turns out that this proposal shares essentially the same spirit as the holographic renormalization group formulation \cite{Holographic_Duality_IV,Holographic_Duality_V,Holographic_Duality_VI} for the renormalized effective on-shell action.
  % pointed out by the referee who reviewed this manuscript.

  Finally, let us recapitulate the essential ingredients of this effective mean-field theory in the large $N$ limit. The crucial step in this recursive renormalization group transformation is the
  renormalization group transformation given by the path integral of the heavy (large mass) scalar
  (original matter) fields. As a result of this renormalization group transformation, the metric tensor and the dual order-parameter field become truly renormalized. As
  discussed above, the renormalization group flow of the metric tensor correspond to the renormalization group
  $\beta-$functions of the coupling functions.
  An interesting point is that the renormalization group flow of the metric tensor is given by the Green's function of the high-energy scalar
  fields, which is the only dynamical information in this renormalization group transformation.
  If one looks into this Green's function more carefully,
  it depends not only on the dual order parameter field but also on the metric tensor.
  In other words, self-energy
  corrections given by the renormalization group flow of the dual order parameter field and vertex
  corrections given by that of the metric tensor are self-consistently and
  non-linearly intertwined, which takes into account renormalization effects
  non-perturbatively.
  Indeed, the IR boundary condition of the dual order
  parameter field corresponds to a mean-field equation of the order parameter
  field but with full renormalization effects given by the self-consistent renormalization group flow
  of the coupling functions
  \cite{Kadanoff_GR_Holography_Kim,RG_GR_Holography_Kim}.

  \item
 % Before closing our response to this criticism, we would also like to point out our recent study on the Kondo effect.
    In Ref. \cite{Holographic_Liquid_Kim},
    we applied momentum-space renormalization group transformations to the Kondo problem.
    Here, we separate slow and fast modes for each field in frequency space,
    and perform renormalization group transformations in a recursive way.
  As a result, we were able to describe the crossover regime from the high-temperature
  decoupled local-moment fluctuating regime to the low-temperature Kondo-singlet
  local-Fermi-liquid state, where log-divergences are fully resummed through this
  recursive renormalization group framework.
 Unfortunately, we could not introduce the gravitational field to this
  momentum-space (or frequency-space) renormalization group transformation scheme,
  where the resulting effective action should be non-local in the reciprocal space,
  which prohibits us from applying the conventional renormalization group transformation scheme.
  Nevertheless, it turned out that both the specific heat and the spin
  susceptibility for the impurity dynamics
  show a reasonable match to those of the Bethe ansatz solution.
  This remarkable result implies that the resulting (dual holographic) effective
  field theory takes into account quantum corrections in a non-perturbative way,
  i.e., in the all-loop order, although this does not mean that this solution is exact.

  %
  % Of course, we admit that all these discussions did not completely justify the absence of the higher-spin fields.
  %To confirm that the present dual holographic effective field theory serves as a physically meaningful description,
  %we have to check out how much the entropy is carried by such higher-spin fields.
  %In other words, we have to compare the thermodynamic entropy between the field-theory calculation (maybe based on simulations) and the present mean-field theory one (in the large $N$ limit) at least. Or, it would be good to consider the entanglement entropy, where a general formulation has been shown in the present study. This is certainly an important future direction. This is what we can answer the referee's criticism.
  %

\end{itemize}

\subsection{Remarks on the linear approximation for the dynamics of the metric tensor in the recursive renormalization group transformation}

We point out that our linear approximation in the recursive renormalization group transformation for the gravity sector
gives rise to a linearized Einstein-scalar theory along the $z-$directional emergent space.
However, we argue below that
%Resorting to the Ricci flow in the holographic renormalization,
%we conclude that
%Here, we argue that
the effective bulk gravity theory contains more than that
of a linearized Einstein-scalar theory.
In particular, we claim that this effective action
% to describe the RG flow of the metric tensor
allows a higher derivative curvature term.
%The linear approximation for the metric tensor in the recursive
%RG transformation gives rise to a higher-order curvature term such as
%$R^{\mu\nu}(x,z) R_{\mu\nu}(x,z)$ in the Einstein-Hilbert action.
To overcome the
limitation of this linear approximation, we suspect that the full diffeomorphism invariance
including the extra-dimensional space has to be introduced into the bulk effective action.
Although we do not have any concrete formulation yet, we discuss how the full nonlinearity of gravity
can be taken into account consistently with the IR boundary condition for matching
between microscopic and macroscopic degrees of freedom.

Previously, we discussed that the renormalization group flow of the metric tensor gives rise to all-loop order renormalization effects of the coupling functions, non-linearly intertwined with the renormalization group flow of the dual order-parameter field, even if the metric tensor is not dynamical. We recall the renormalization group flow of the metric tensor, given by $\partial_{z} g^{\mu\nu}(x,z) = \beta^{\mu\nu}_{g}[g^{\mu\nu}(x,z),\varphi(x,z)]$, where the $\beta-$function is $\beta^{\mu\nu}_{g}[g^{\mu\nu}(x,z),\varphi(x,z)] = g^{\mu\nu'}(x,z)\big(\partial_{\nu'} \partial_{\mu'} G_{xx'}[g^{\mu\nu}(x,z),\varphi(x,z)]\big)_{x' \rightarrow x} g^{\mu'\nu}(x,z)$.

First, to see how higher derivative terms can arise,
we consider the case when the metric tensor is dynamical,
where the effective action of the metric tensor is given by
\begin{align}
 \mathcal{S}_{grav.} &= N \int_{0}^{z_{f}} d z \int d^{D} x \sqrt{g(x,z)} \Big\{ - \frac{1}{2 \lambda} \Big(\partial_{z} g^{\mu\nu}(x,z) - \beta^{\mu\nu}_{g}[g^{\mu\nu}(x,z),\varphi(x,z)] \Big)^{2}
  \nonumber \\
  &\qquad
     + \frac{1}{2 \kappa} \Big( R(x,z) - 2 \Lambda \Big) \Big\} .
\end{align}
To develop an
intuitive picture on the role of the first term,
%in the dynamics of the metric tensor,
it is tempting to regard $\beta^{\mu\nu}_{g}[g^{\mu\nu}(x,z),\varphi(x,z)]$
as the Ricci tensor, $R_{\mu\nu}(x,z)$;
we consider
\begin{align}
     \mathcal{S}_{grav.} &= N \int_{0}^{z_{f}} d z \int d^{D} x \sqrt{g(x,z)} \Big\{ - \frac{1}{2 \lambda} \Big(\partial_{z} g^{\mu\nu}(x,z) + 2 R^{\mu\nu}(x,z) \Big) \Big(\partial_{z} g_{\mu\nu}(x,z) + 2 R_{\mu\nu}(x,z) \Big)
     \nonumber \\
  &\qquad + \frac{1}{2 \kappa} \Big( R(x,z) - 2 \Lambda \Big) \Big\} .
\end{align}
This is natural in the sense that the resulting
evolution equation for the metric tensor in the $\lambda \rightarrow 0$ limit is given by the Ricci flow equation
\cite{Ricci_Flow_0,Ricci_Flow_I,Ricci_Flow_II,Ricci_Flow_III,Ricci_Flow_IV,Ricci_Flow_V}
\begin{align}
 \partial_{z} g_{\mu\nu}(x,z) = - 2 R_{\mu\nu}(x,z) .
\end{align}
The Ricci flow equation is to describe the deformation of a Riemannian metric
$g_{\mu\nu}(x,z)$ with an extra-dimensional space coordinate $z$, here, which
plays the same role as time. $\mu$ and $\nu$ cover the $D-$dimensional spacetime
coordinate $0, ..., D-1$. This evolution equation may be regarded as an analog
of the diffusion equation for geometries, given by a parabolic partial
differential equation. The deformation is governed by Ricci curvature, which
leads to homogeneity of geometry.
In principle, one may consider that this Ricci flow equation
arises from the gradient expansion of the Green's function with respect to the
mass parameter, where additional terms given by curvature gradients and dual
order-parameter fields are all neglected. See the Green's function formula of
Eq. (B.4) in appendix B.1.

%
%As a result, the effective bulk action of the metric tensor is given by
%\bqa && \mathcal{S}_{grav.} = N \int_{0}^{z_{f}} d z \int d^{D} x \sqrt{g(x,z)} \Big\{ - \frac{1}{2 \lambda} \Big(\partial_{z} g^{\mu\nu}(x,z) + 2 R^{\mu\nu}(x,z) \Big)^{2} \nn && + \frac{1}{2 \kappa} \Big( R(x,z) - 2 \Lambda \Big) \Big\} . \eqa
%Here, we have to point out that the holographic renormalization group flow gives rise to the Ricci flow naturally \cite{Holographic_RG_Flow_Ricci_Flow_I,Holographic_RG_Flow_Ricci_Flow_II}. Actually, the present holographic dual effective field theory turns out to reproduce this result at the fixed point of the $z_{f} \rightarrow \infty$ limit.
%

Here,
we would also like to
point out that the holographic renormalization group flow gives rise to the Ricci
flow naturally \cite{Holographic_RG_Flow_Ricci_Flow_I,Holographic_RG_Flow_Ricci_Flow_II}.
Actually, our holographic dual effective field theory turns out to reproduce this result at the fixed point of the $z_{f} \rightarrow \infty$ limit.

To verify the above statement,
we consider the Hamilton-Jacobi formulation for the holographic dual effective field theory, given by
\bqa && \frac{1}{\sqrt{g(x,z_{f})}} \frac{\partial}{\partial z_{f}} \mathcal{I}[\varphi(x,z_{f}), g_{\mu\nu}(x,z_{f})] = \frac{N \lambda}{2} \Big\{ \frac{1}{\sqrt{g(x,z_{f})}} \frac{\partial \mathcal{I}[\varphi(x,z_{f}), g_{\mu\nu}(x,z_{f})]}{\partial g^{\mu\nu}(x,z_{f})} \Big\}^{2} \nn && + N \beta_{\mu\nu}^{g}[\varphi(x,z_{f}), g_{\mu\nu}(x,z_{f})] \Big\{ \frac{1}{\sqrt{g(x,z_{f})}} \frac{\partial \mathcal{I}[\varphi(x,z_{f}), g_{\mu\nu}(x,z_{f})]}{\partial g^{\mu\nu}(x,z_{f})} \Big\} + \frac{N}{2 \kappa} \Big( R(x,z_{f}) - 2 \Lambda \Big) , \nn \eqa
where $\mathcal{I}[\varphi(x,z_{f}), g_{\mu\nu}(x,z_{f})]$ is an IR effective on-shell action. We recall our gauge fixing for the metric tensor, given by $g_{DD}(x,z) = 1$ and $g_{\mu D}(x,z) = 0$ with $\mu = 0, \ldots, D-1$. Here, we focused on the gravitational part only. We point out that the $\beta-$function has to vanish in the $z_{f} \rightarrow \infty$ limit corresponding to a fixed point. When this $\beta-$function of the metric tensor vanishes at the fixed point, this Hamilton-Jacobi equation is essentially reduced into that of the holographic duality conjecture \cite{Holographic_Duality_IV,Holographic_Duality_V, Holographic_Duality_VI}, which reproduces the Ricci flow \cite{Holographic_RG_Flow_Ricci_Flow_I,Holographic_RG_Flow_Ricci_Flow_II}. On the other hand, when the $\beta_{g}^{\mu\nu}[\varphi(x,z_{f}), g_{\mu\nu}(x,z_{f})]$ function does not vanish, it depends on both $\varphi(x,z_{f})$ and $g_{\mu\nu}(x,z_{f})$ through the Green's function in an intertwined and nonlinear way, and this nonlinear intertwined renormalization structure gives rise to difficulty in solving this Hamilton-Jacobi equation. The present prescription generalizes
the holographic renormalization group flow of the AdS geometry towards that without conformal
symmetry. More detailed discussions can be found in section IV. Discussion: Ricci Flow of Ref. \cite{RG_GR_Holography_Kim}.

Resorting to the Ricci flow in the holographic renormalization,
we conclude that the linear approximation for the metric tensor in the recursive
renormalization group transformation gives rise to a higher-order curvature term such as
$R^{\mu\nu}(x,z) R_{\mu\nu}(x,z)$ in the Einstein-Hilbert action.
We currently do not know how to
justify the linear approximation for the evolution of the
metric tensor in the recursive renormalization group procedure,
except that we use the same approximation for the heavy dual scalar field.
%-- a systematic study of how good/bad this approximation is left for the future.
Nevertheless, even within the linear approximation,
the appearance of the higher derivative curvature term is an interesting
feature, rather unexpected.

%
%Resorting to the Ricci flow in the holographic renormalization, we conclude that the linear approximation for the metric tensor in the recursive renormalization group transformation gives rise to a higher-order curvature term such as $R^{\mu\nu}(x,z) R_{\mu\nu}(x,z)$ in the Einstein-Hilbert action. Although we cannot justify the linear approximation for the evolution of the metric tensor in the recursive renormalization group procedure, the appearance of the higher derivative curvature term is an interesting feature, rather unexpected. To show internal consistency for the linear approximation in the metric-tensor evolution at least, we have to investigate the Ward identity in this emergent dual holographic description. It turns out that the Ward identity of an effective on-shell action \cite{Holographic_Ward_Identity} is given by the coefficient of the shift vector in the ADM Hamiltonian formulation, which is nothing but the generator of diffeomorphism \cite{ADM_Hamiltonian_Formulation}. This requires us to extend the present gauge-fixed version to a fully covariant formulation, where the shift vector has to be introduced explicitly. This is certainly a future direction of our research, which cannot be covered in the present study.
%

Second, we discuss how to introduce full nonlinearity into the above dual holographic effective field theory.
Although we do not have any concrete idea for the first-principle derivation in this issue, we suspect that
full diffeomorphism invariance has to be satisfied, which requires to extend the present formulation,
including both spacetime dependent lapse function and shift vector fields \cite{ADM_Hamiltonian_Formulation}. In this respect we suggest
the following dual holographic effective field theory
\bqa && Z = \int D \phi_{\alpha}(x) D \mathcal{G}^{MN}(x,z) D \varphi(x,z) \exp\Big\{ - \mathcal{S}_{Bulk} - \mathcal{S}_{IR} - \mathcal{S}_{UV} \Big\} . \eqa
Here, one of the two key points is that the bulk effective action enjoys the full $(D+1)-$dimensional diffeomorphism invariance, given by
\bqa && \mathcal{S}_{Bulk} = N \int_{0}^{z_{f}} d z \int d^{D} x \sqrt{\mathcal{G}(x,z)} \Bigg( \frac{1}{2} \mathcal{G}^{MN} \Big(\partial_{M} \varphi(x,z) \Big) \Big( \partial_{N} \varphi(x,z) \Big) + \mathcal{C}_{\xi} \mathcal{R}(x,z) [\varphi(x,z)]^{2} \nn && + \frac{1}{2 \kappa} \Big\{ \mathcal{R}(x,z) - 2 \Lambda \Big\} + ~ \mbox{higher-derivative curvature terms} \Bigg) , \eqa
where $M$ and $N$ are $(D+1)-$dimensional coordinates including the emergent extra-dimensional space. This bulk effective action may be regarded as a conventional holographic bulk action although it is extended to include higher-derivative curvature terms as discussed above. The IR boundary effective action is
\bqa && \mathcal{S}_{IR} = \int d^{D} x \sqrt{g(x,z_{f})} \Bigg( g^{\mu\nu}(x,z_{f}) \Big(\partial_{\mu} \phi_{\alpha}(x) \Big) \Big(\partial_{\nu} \phi_{\alpha}(x) \Big) + \Big( m^{2} - i \varphi(x,z_{f}) \Big) [\phi_{\alpha}(x)]^{2} \nn && + \xi R(x,z_{f}) [\phi_{\alpha}(x)]^{2} \Bigg) + \mbox{GHY boundary term } , \eqa
where $g^{\mu\nu}(x,z_{f})$ represent fully renormalized coupling functions as discussed before, which have to be identified with an induced metric tensor from the bulk metric. Obviously, $\mu$ and $\nu$ span the boundary $D-$dimensional spacetime coordinate at $z = z_{f}$. In addition, the Gibbons-Hawking-York boundary term has to be taken into account from the bulk effective action \cite{Gibbons_Hawking_York_I,Gibbons_Hawking_York_II}. Finally, the UV boundary action is given by
\bqa && \mathcal{S}_{UV} = N \int d^{D} x \sqrt{g(x,0)} \Bigg( \frac{1}{2 u} \Big(\varphi(x,0) - \varphi^{ext}(x)\Big)^{2} + \mathcal{C}_{\xi} R(x,0) [\varphi(x,0)]^{2} \Bigg) \nn && + ~ \mbox{GHY boundary term } , \eqa
essentially the same as the presently derived version of the UV boundary condition, where the UV boundary metric tensor is nothing but Euclidean. Here, we introduced an external source term $\varphi^{ext}(x)$ to measure density-fluctuation correlations.

Compared to the conventional approach in the dual holographic theory, the only novel aspect is to introduce the IR boundary effective action explicitly into the framework. Thus, the last of the two key points is on how to incorporate the IR boundary condition, which has to be consistent with the bulk effective action. An idea is to consider the ADM Hamiltonian gravity formulation \cite{ADM_Hamiltonian_Formulation} for the bulk effective action. Then, the bulk effective action is rewritten as follows formally
\bqa && \mathcal{S}_{Bulk}^{ADM} = N \int_{0}^{z_{f}} d z \int d^{D} x \Bigg( \pi_{\mu\nu}(x,z) \partial_{z} g^{\mu\nu}(x,z) - N(x,z) H(x,z) - N^{\mu}(x,z) H_{\mu}(x,z) \Bigg) , \nn \eqa
where $\pi_{\mu\nu}(x,z)$ is the canonical conjugate variable to $g^{\mu\nu}(x,z)$, $N(x,z)$ is the lapse function, and $N^{\mu}(x,z)$ is the shift vector field. As a result, we have two constraint equations, given by $H(x,z) = 0$, which can be identified with the Hamilton-Jacobi equation in the holographic renormalization framework, and $H^{\mu}(x,z) = 0$, which gives rise to the Ward identity involved with the $D-$dimensional diffeomorphism invariance \cite{Holographic_Ward_Identity}. Solving the Hamilton-Jacobi equation, we suspect that the resulting on-shell effective action has to be consistent with the IR boundary effective action, where their consistency is guaranteed by the corresponding Ward identity of $H^{\mu}(x,z) = 0$. This IR-boundary matching procedure makes the connection between UV microscopic degrees of freedom and IR emergent macroscopic ones be complete. However, we would like to point out that actual implementations have to be performed carefully.

One may concern that the IR boundary effective action may not be uniquely determined. Suppose that the IR boundary action is given by \cite{Holographic_Description_Einstein_Maxwell}
\bqa && \mathcal{S}_{IR} = \int d^{D} x \sqrt{g(x,z_{f})} \Bigg( \bar{\psi}_{\alpha}(x) \gamma^{a} e_{a}^{\mu}(x,z_{f}) \Big( \partial_{\mu} - \frac{i}{4} \omega_{\mu}^{a'b'}(x,z_{f}) \sigma_{a'b'} \Big) \psi_{\alpha}(x) \nn && + [m - i \varphi(x,z_{f})] \bar{\psi}_{\alpha}(x) \psi_{\alpha}(x) \Bigg) + ~ \mbox{GHY boundary term } , \eqa
where ``chiral symmetry breaking" in interacting Dirac-fermion systems \cite{X_Breaking} has been considered.
Here, $\psi_{\alpha}(x)$ is a Dirac spinor at $x$ in $D$ spacetime dimensions.
$\alpha$ runs from $1$ to $N$, denoting the flavor degeneracy of Dirac fermions.
$\gamma^{a}$ is a Dirac $\gamma$ matrix, defined in a local rest frame at $x$
and satisfying the Clifford algebra $\{\gamma^{a}, \gamma^{b}\} = 2 \delta^{ab}$
with the Euclidean signature. $e_{a}^{\mu}(x,z_{f})$ defines the local rest
frame given by the tangent manifold at $x$, called vierbein. The corresponding
background metric is given by the vierbein as follows $g_{\mu\nu}(x,z_{f}) =
e_{\mu}^{a}(x,z_{f}) e_{\nu}^{b}(x,z_{f}) \delta_{ab}$.
$\omega_{\mu}^{ab}(x,z_{f}) = e_{\nu}^{a}(x,z_{f}) \partial_{\mu} e^{\nu
  b}(x,z_{f}) + e_{\nu}^{a}(x,z_{f}) \Gamma^{\nu}_{\sigma\mu}(x,z_{f}) e^{\sigma
  b}(x,z_{f})$ is a spin connection and $\sigma_{ab} = \frac{i}{2} [\gamma^{a},
\gamma^{b}]$ is a commutator of Dirac gamma
matrices in the local rest frame. Here, $\Gamma_{\sigma\mu}^{\nu}(x,z_{f}) = \frac{1}{2} g^{\nu\delta}(x,z_{f}) \Big(\partial_{\sigma} g_{\delta\mu}(x,z_{f}) + \partial_{\mu} g_{\sigma\delta}(x,z_{f}) - \partial_{\delta} g_{\sigma\mu}(x,z_{f}) \Big)$ is the Christoffel symbol. $m$ represents a mass of Dirac fermions. $\varphi(x,z_{f})$ is an order-parameter field dual to $\frac{1}{N} \sum_{\alpha = 1}^{N} \bar{\psi}_{\alpha}(x) \psi_{\alpha}(x)$ at $z = z_{f}$.
Resorting to the ADM Hamiltonian gravity formulation, we assign the IR boundary condition to the same bulk effective action but in terms of microscopic fermion degrees of freedom instead of bosons. Solving two coupled equations for dual scalar fields and metric tensor fields modified by higher-derivative curvature terms, we would find effective geometry as a function of the mass parameter or more precisely, the ratio between the mass parameter and the interaction strength. Although we resort to the same bulk effective action, we suspect that both aspects of the difference in the IR boundary condition and the presence of higher-derivative curvature terms may lead to different geometries or different conformal dimensions at least, when the mass parameter is tuned to cause a quantum phase transition. This would be certainly a future direction of our research.

\section{Conclusion}

In the present study we tried to clarify the underlying physics of the dual
holographic formulation for quantum field theories as a non-perturbative description.
%As a result, we could figure out
We uncovered
how the emergent geometry takes into account
the information of quantum entanglement.
Let us recapitulate the essential features of
the dual holographic formulation.
First, the infinitesimal distance in the emergent extra dimensional space is
identified with the renormalization group scale
in the recursive renormalization group transformation.
Second, the IR boundary condition describes the renormalization group
$\beta$-functions
of all the coupling parameters in the corresponding quantum field theory.
In particular, the IR boundary conditions given by metric tensor fields are
identified with the renormalization group equations
of all interaction vertices and that given by the scalar field is nothing but
the Callan-Symanzik equation for an order parameter
field in the dual description.
Finally, the bulk Einstein-Klein-Gordon type equation describes the entanglement transfer from quantum matter to classical geometry in the large $N$ limit, where $N$ is the number of flavors of scalar fields.

The future direction in this line of research would be to solve the resulting
coupled equations of motion
in the emergent curved spacetime with an extra dimension and to show the
appearance of hydrodynamic behaviors at IR,
for example, calculating
the ratio
$\eta / s$, where $\eta$ is shear viscosity and $s$ is
thermodynamic entropy
\cite{Holographic_Liquid_Son_I,Holographic_Liquid_Son_II,Holographic_Liquid_Son_III,Holographic_Liquid_Son_IV}.
Recently, one of the authors has investigated the Kondo effect based on this
emergent geometric description
and
%could
was able to
describe the crossover behavior around the Kondo temperature from the
UV local-moment fixed point to the IR local Fermi-liquid one in a
non-perturbative way \cite{Holographic_Liquid_Kim}.
The so-called holographic liquid state may serve as a novel quantum liquid beyond the perturbation theoretical framework.

\acknowledgments

K.-S. Kim was supported by the Ministry of Education, Science, and Technology (No. 2011-0030046 and NRF-2021R1A2C1006453) of the National Research Foundation of Korea (NRF) and by TJ Park Science Fellowship of the POSCO TJ Park Foundation.
This work was supported by a grant from
the Simons Foundation (566116, SR).

\appendix

\section{Emergent geometric description for a scalar lattice field theory in one dimension} \label{RG_GR_Correspondence_Appendix}

%
%\subsection{UV lattice model in one dimension}
%

We start from the following one-dimensional lattice field theory
\begin{align}
  Z = \int D \Phi_{i} \exp\Big[ - \int_{0}^{\beta} d \tau \sum_{i = 1}^{M} \Big\{ (\partial_{\tau} \Phi_{i})^{2} - t (\Phi_{i} \Phi_{i+1} + \Phi_{i+1} \Phi_{i}) + m^{2} \Phi_{i}^{2} + \frac{u}{2} \Phi_{i}^{4} \Big\} \Big] .
\end{align}
Here, $\Phi_{i}$ is a real scalar field at site $i$. $t$ is a hopping integral and $m$ is a mass parameter. $u$ represents the strength of self-interactions between these scalar fields.

%
%\subsection{Preparation for the first renormalization group transformation}
%

To implement recursive Kadanoff block-spin transformations \cite{Kadanoff_RG} in this lattice field theory, we uplift all coupling parameters into dynamical field variables as follows
\begin{align}
  Z
  &= \int D \Phi_{i} D \varphi_{i}^{(0)} D \rho_{i}^{(0)} D s_{i
  i+1}^{(0)} D t_{i i+1}^{(0)} D \chi_{i}^{(0)} D m_{i}^{2(0)} D v_{i}^{(0)} D
u_{i}^{(0)}
    \nn
  & \quad
\times \exp\Big[ - \int_{0}^{\beta} d \tau \sum_{i = 1}^{M} \Big\{
(\partial_{\tau} \Phi_{i})^{2} - t_{i i+1}^{(0)} (\Phi_{i} \Phi_{i+1} +
\Phi_{i+1} \Phi_{i}) + m_{i}^{2(0)} \Phi_{i}^{2} + \frac{u_{i}^{(0)}}{2}
\rho_{i}^{(0) 2}
    \nn
  & \quad + i \varphi_{i}^{(0)} (\rho_{i}^{(0)} - \Phi_{i}^{2}) + i s_{i i+1}^{(0)} (t_{i i+1}^{(0)} - t) + i \chi_{i}^{(0)} (m_{i}^{2(0)} - m^{2}) + i v_{i}^{(0)} (u_{i}^{(0)} - u) \Big\} \Big] .
\end{align}
Here, $s_{i i+1}^{(0)}$, $\chi_{i}^{(0)}$, and $v_{i}^{(0)}$ are Lagrange multiplier fields to impose each constraint for hopping, mass, and self-interaction parameters, respectively. $\varphi_{i}^{(0)}$ is a Lagrange multiplier field to realize the Hubbard-Stratonovich transformation for the self-interaction channel.

%
%\subsection{Kadanoff block-spin transformation and rescaling}
%

Separating all dynamical fields at each site into those at even and odd sites, integrating over all field variables in even sites, and rescaling to return the lattice to the original one, we obtain an effective lattice field theory with renormalized hopping, mass, self-interaction, and dual scalar fields as follows
\begin{align}
& Z = \int D \Phi_{i} D \varphi_{i}^{(0)} D \rho_{i}^{(0)} D s_{i i+1}^{(0)} D
t_{i i+1}^{(0)} D \chi_{i}^{(0)} D m_{i}^{2(0)} D v_{i}^{(0)} D u_{i}^{(0)}
                \nn
  &\quad \times \exp\Bigg[ - \frac{1}{4} \sum_{i = 1}^{M} \mbox{tr}_{\tau\tau'} \ln
    \Big( - \partial_{\tau}^{2} + m_{i}^{2(0)} \Big)
    \nonumber \\
  &\quad
    - \int_{0}^{\beta} d \tau
\sum_{i = 1}^{M} \Bigg\{ (\partial_{\tau} \Phi_{i})^{2} + \Big( m_{i}^{2(0)} -
\frac{2m_{i}^{2(0)} t_{i i+1}^{(0) 2}}{ 2 m_{i}^{4(0)} + u_{i}^{(0)}} \Big)
\Phi_{i}^{2} - i \varphi_{i}^{(0)} \Phi_{i}^{2}
    \nn
 &\qquad + \frac{u_{i}^{(0)}}{2} \rho_{i}^{(0) 2} + i \varphi_{i}^{(0)}
\rho_{i}^{(0)} + i s_{i i+1}^{(0)} (t_{i i+1}^{(0)} - t) + i \chi_{i}^{(0)}
(m_{i}^{2(0)} - m^{2}) + i v_{i}^{(0)} (u_{i}^{(0)} - u) \Bigg\}
   \nn
  &\quad - \int_{0}^{\beta} d \tau \sum_{i = 1}^{M} \Bigg\{ - \frac{t_{i
    i+1}^{(0) 2}}{2 m_{i}^{4(0)} + u_{i}^{(0)}} \Big\{ m_{i}^{2(0)} -
\frac{u_{i}^{(0)} t_{i i+1}^{(0)2}}{m_{i}^{4 (0)}} \Big( \Phi_{i}^{2} +
\Phi_{i+1}^{2} \Big) \Big\} \Big( \Phi_{i} \Phi_{i+1} + \Phi_{i+1} \Phi_{i}
\Big)
    \nn
  &\qquad  + \frac{u_{i}^{(0)}}{2 m_{i}^{4(0)} + u_{i}^{(0)}} \Big\{ \frac{t_{i i+1}^{(0)4}}{m_{i}^{4(0)}} \Phi_{i}^{4} + 3 \frac{t_{i i+1}^{(0)4}}{m_{i}^{4 (0)}} \Phi_{i}^{2} \Phi_{i+1}^{2} \Big\} + \frac{1}{16 m_{i}^{4(0)}} \Big( \frac{1}{u_{i}^{(0)}} + \frac{1}{2 m_{i}^{4(0)}} \Big)^{-1} \Bigg\} \Bigg] .
\end{align}

%
%\subsection{Preparation for the second renormalization group transformation}
%

To prepare for the second renormalization group transformation, we rewrite this partition function in terms of updated coupling fields in the following way
\begin{align}
  & Z = \int D \Phi_{i}
    D \varphi_{i}^{(0)} D \rho_{i}^{(0)} D s_{i i+1}^{(0)} D
    t_{i i+1}^{(0)} D \chi_{i}^{(0)} D m_{i}^{2(0)} D v_{i}^{(0)} D u_{i}^{(0)}
    \nonumber \\
  &\qquad
    \times
    D \varphi_{i}^{(1)} D \rho_{i}^{(1)} D s_{i i+1}^{(1)} D t_{i i+1}^{(1)} D
\chi_{i}^{(1)} D m_{i}^{2(1)} D v_{i}^{(1)} D u_{i}^{(1)}
                \nn
  &\quad
\times \exp\Bigg[ - \frac{1}{4} \sum_{i = 1}^{M} \mbox{tr}_{\tau\tau'} \ln \Big(
    - \partial_{\tau}^{2} + m_{i}^{2(0)} \Big)
    \nonumber \\
  &\quad
    - \int_{0}^{\beta} d \tau \sum_{i =
  1}^{M} \Bigg\{ (\partial_{\tau} \Phi_{i})^{2} + m_{i}^{2(1)} \Phi_{i}^{2} - i
\varphi_{i}^{(0)} \Phi_{i}^{2} - t_{i i+1}^{(1)} \Big( \Phi_{i} \Phi_{i+1} +
\Phi_{i+1} \Phi_{i} \Big)
    \nn
  & \quad + \frac{u_{i}^{(0)}}{2} \rho_{i}^{(0) 2} + i \varphi_{i}^{(0)}
\rho_{i}^{(0)} + i s_{i i+1}^{(0)} (t_{i i+1}^{(0)} - t) + i \chi_{i}^{(0)}
(m_{i}^{2(0)} - m^{2}) + i v_{i}^{(0)} (u_{i}^{(0)} - u)
    \nn
  & \quad + \frac{u_{i}^{(1)}}{2} \Big( \rho_{i}^{(1) 2} + 3 \rho_{i}^{(1)}
    \rho_{i+1}^{(1)} \Big)
    \nonumber \\
  &\quad
    + i \varphi_{i}^{(1)} ( \rho_{i}^{(1)} - \Phi_{i}^{2} ) +
i s_{i i+1}^{(1)} \Big[ t_{i i+1}^{(1)} - \frac{t_{i i+1}^{(0) 2}}{2
  m_{i}^{4(0)} + u_{i}^{(0)}} \Big\{ m_{i}^{2(0)} - \frac{u_{i}^{(0)} t_{i
    i+1}^{(0)2}}{m_{i}^{4 (0)}} \Big( \rho_{i}^{(1)} + \rho_{i+1}^{(1)} \Big)
\Big\} \Big]
    \nn
  & \quad + i \chi_{i}^{(1)} \Big\{ m_{i}^{2(1)} - \Big( m_{i}^{2(0)} - \frac{2m_{i}^{2(0)} t_{i i+1}^{(0) 2}}{ 2 m_{i}^{4(0)} + u_{i}^{(0)}} \Big) \Big\}
    \nonumber\\
  &\quad
    + i v_{i}^{(1)} \Big( u_{i}^{(1)} - \frac{2 u_{i}^{(0)} t_{i i+1}^{(0)4}}{m_{i}^{4(0)}(2 m_{i}^{4(0)} + u_{i}^{(0)})} \Big) + \frac{1}{16 m_{i}^{4(0)}} \Big( \frac{1}{u_{i}^{(0)}} + \frac{1}{2 m_{i}^{4(0)}} \Big)^{-1} \Bigg\} \Bigg] .
\end{align}
Here, $s_{i i+1}^{(1)}$, $\chi_{i}^{(1)}$, and $v_{i}^{(1)}$ are Lagrange multiplier fields to impose the renormalization group flow equations for hopping, mass, and self-interaction fields, respectively. $\varphi_{i}^{(1)}$ is a Lagrange multiplier field to realize the Hubbard-Stratonovich transformation for newly generated effective interactions.

%
%\subsection{Shift of $\varphi_{i}^{(1)}$ and gradient-expansion approximation}
%

%Taking a shift in
Shifting
$\varphi_{i}^{(1)}$ as
\bqa && \varphi_{i}^{(1)} \Longrightarrow \varphi_{i}^{(1)} - \varphi_{i}^{(0)} \eqa
and keeping only local interactions in the sense of the gradient expansion, we obtain
\begin{align}
  &
    Z = \int D \Phi_{i} D \varphi_{i}^{(0)} D \rho_{i}^{(0)} D s_{i i+1}^{(0)} D t_{i i+1}^{(0)} D \chi_{i}^{(0)} D m_{i}^{2(0)} D v_{i}^{(0)} D u_{i}^{(0)}
    \nonumber \\
  &\quad\times
    D \varphi_{i}^{(1)} D \rho_{i}^{(1)} D s_{i i+1}^{(1)} D t_{i i+1}^{(1)} D \chi_{i}^{(1)} D m_{i}^{2(1)} D v_{i}^{(1)} D u_{i}^{(1)}
    \nonumber\\
  & \quad \times \exp\Bigg[ - \frac{1}{4} \sum_{i = 1}^{M} \mbox{tr}_{\tau\tau'} \ln \Big( - \partial_{\tau}^{2} + m_{i}^{2(0)} \Big)
    \nonumber \\
  &\quad
    - \int_{0}^{\beta} d \tau \sum_{i = 1}^{M} \Bigg\{ (\partial_{\tau} \Phi_{i})^{2} + m_{i}^{2(1)} \Phi_{i}^{2} - i \varphi_{i}^{(1)} \Phi_{i}^{2} - t_{i i+1}^{(1)} \Big( \Phi_{i} \Phi_{i+1} + \Phi_{i+1} \Phi_{i} \Big)
    \nonumber \\
  &\quad
    + \frac{u_{i}^{(0)}}{2} \rho_{i}^{(0) 2} + i \varphi_{i}^{(0)} \rho_{i}^{(0)} + i s_{i i+1}^{(0)} (t_{i i+1}^{(0)} - t) + i \chi_{i}^{(0)} (m_{i}^{2(0)} - m^{2}) + i v_{i}^{(0)} (u_{i}^{(0)} - u)
    \nonumber \\
  &\quad
    + \frac{u_{i}^{(1)}}{2} \rho_{i}^{(1) 2} + i \rho_{i}^{(1)} (\varphi_{i}^{(1)} - \varphi_{i}^{(0)})
    \nonumber \\
  &\quad
    + i s_{i i+1}^{(1)} \Big\{ t_{i i+1}^{(1)} - \frac{t_{i i+1}^{(0) 2}}{2 m_{i}^{4(0)} + u_{i}^{(0)}} \Big( m_{i}^{2(0)} - \frac{2 u_{i}^{(0)} t_{i i+1}^{(0)2}}{m_{i}^{4 (0)}} \rho_{i}^{(1)} \Big) \Big\}
    \nonumber \\
  &\quad
    + i \chi_{i}^{(1)} \Big\{ m_{i}^{2(1)} - \Big( m_{i}^{2(0)} - \frac{2m_{i}^{2(0)} t_{i i+1}^{(0) 2}}{ 2 m_{i}^{4(0)} + u_{i}^{(0)}} \Big) \Big\}
    \nonumber \\
  &\quad
    + i v_{i}^{(1)} \Big( u_{i}^{(1)} - \frac{u_{i}^{(0)} t_{i i+1}^{(0)4}}{2 m_{i}^{4(0)}(2 m_{i}^{4(0)} + u_{i}^{(0)})} \Big) + \frac{1}{16 m_{i}^{4(0)}} \Big( \frac{1}{u_{i}^{(0)}} + \frac{1}{2 m_{i}^{4(0)}} \Big)^{-1} \Bigg\} \Bigg] .
\end{align}
This completes the first renormalization group transformation.

%
%\subsection{Recursive renormalization group transformations}
%

It is straightforward to extend this expression in a recursive way, given by
\begin{align}
  & Z = \int D \Phi_{i} \Pi_{l = 0}^{f} D \varphi_{i}^{(l)} D \rho_{i}^{(l)} D s_{i i+1}^{(l)} D t_{i i+1}^{(l)} D \chi_{i}^{(l)} D m_{i}^{2(l)} D v_{i}^{(l)} D u_{i}^{(l)}
    \nn
  &
    \exp\Bigg[ - \frac{1}{4} \sum_{l = 1}^{f} \sum_{i = 1}^{M} \mbox{tr}_{\tau\tau'} \ln \Big( - \partial_{\tau}^{2} + m_{i}^{2(l-1)} \Big)
    \nonumber \\
  &
    - \int_{0}^{\beta} d \tau \sum_{i = 1}^{M} \Bigg\{ (\partial_{\tau} \Phi_{i})^{2} + m_{i}^{2(f)} \Phi_{i}^{2} - i \varphi_{i}^{(f)} \Phi_{i}^{2} - t_{i i+1}^{(f)} \Big( \Phi_{i} \Phi_{i+1} + \Phi_{i+1} \Phi_{i} \Big)
    \nn
  & + \frac{u_{i}^{(0)}}{2} \rho_{i}^{(0) 2} + i \varphi_{i}^{(0)} \rho_{i}^{(0)} + i s_{i i+1}^{(0)} (t_{i i+1}^{(0)} - t) + i \chi_{i}^{(0)} (m_{i}^{2(0)} - m^{2}) + i v_{i}^{(0)} (u_{i}^{(0)} - u)
    \nonumber \\
  & + \sum_{l = 1}^{f} \Big[ \frac{u_{i}^{(l)}}{2} \rho_{i}^{(l) 2} + i \rho_{i}^{(l)} (\varphi_{i}^{(l)} - \varphi_{i}^{(l-1)})
    \nonumber \\
  &
    + i s_{i i+1}^{(l)} \Big\{ t_{i i+1}^{(l)} - \frac{t_{i i+1}^{(l-1) 2}}{2 m_{i}^{4(l-1)} + u_{i}^{(l-1)}} \Big( m_{i}^{2(l-1)} - \frac{2 u_{i}^{(l-1)} t_{i i+1}^{(l-1)2}}{m_{i}^{4 (l-1)}} \rho_{i}^{(l)} \Big) \Big\}
    \nonumber \\
  & + i \chi_{i}^{(l)} \Big\{ m_{i}^{2(l)} - \Big( m_{i}^{2(l-1)} - \frac{2m_{i}^{2(l-1)} t_{i i+1}^{(l-1) 2}}{ 2 m_{i}^{4(l-1)} + u_{i}^{(l-1)}} \Big) \Big\}
    \nonumber \\
  &
    + i v_{i}^{(l)} \Big( u_{i}^{(l)} - \frac{u_{i}^{(l-1)} t_{i i+1}^{(l-1)4}}{2 m_{i}^{4(l-1)}(2 m_{i}^{4(l-1)} + u_{i}^{(l-1)})} \Big)
   + \frac{1}{16 m_{i}^{4(l-1)}} \Big( \frac{1}{u_{i}^{(l-1)}} + \frac{1}{2 m_{i}^{4(l-1)}} \Big)^{-1} \Big] \Bigg\} \Bigg] .
\end{align}
Resorting to
\bqa && a \sum_{l = 1}^{f} \Longrightarrow \int_{0}^{z_{f}} d z \label{a_definition_I} \eqa
and
\bqa && \frac{\varphi_{i}^{(l)} - \varphi_{i}^{(l-1)}}{a} \Longrightarrow \partial_{z} \varphi(i,\tau,z) , \label{a_definition_II} \eqa
where $a$ is a scale for the renormalization group transformation, we rewrite the above partition function in the following way
\begin{align}
  Z &= \int D \Phi(i,\tau) \int D \varphi(i,\tau,z) D \rho(i,\tau,z)
      \nonumber \\
  &\quad
    \times
      \int D
s(i,i+1,\tau,z) D t(i,i+1,\tau,z) D \chi(i,\tau,z) D m^{2}(i,\tau,z)
    D v(i,\tau,z) D u(i,\tau,z)
    \nonumber \\
  &\quad \times
    \exp \Big\{- S_{IR} - S_{UV} - S_{Bulk} \Big\} .
\end{align}
Here, both IR and UV effective action are given by
\begin{align}
& S_{IR} + S_{UV} = \int_{0}^{\beta} d \tau \sum_{i = 1}^{M} \Bigg\{
\Big(\partial_{\tau} \Phi(i,\tau)\Big)^{2} + m^{2}(i,\tau,z_{f})
[\Phi(i,\tau)]^{2} - i \varphi(i,\tau,z_{f}) [\Phi(i,\tau)]^{2}
                \nn
  &\quad  - t(i,i+1,\tau,z_{f}) \Big( \Phi(i,\tau) \Phi(i+1,\tau) +
\Phi(i+1,\tau) \Phi(i,\tau) \Big) + \frac{u(i,\tau,0)}{2} [\rho(i,\tau,0)]^{2} +
i \varphi(i,\tau,0) \rho(i,\tau,0)
    \nn
  &\quad + i s(i,i+1,\tau,0) \Big(t(i,i+1,\tau,0) - t\Big) + i \chi(i,\tau,0) \Big(m^{2}(i,\tau,0) - m^{2}\Big) + i v(i,\tau,0) \Big(u(i,\tau,0) - u\Big) \Bigg\} .
              \end{align}
The bulk effective action is
\bqa && S_{Bulk} = \int_{0}^{z_{f}} d z \int_{0}^{\beta} d \tau \sum_{i = 1}^{M}
\Bigg\{ \frac{u(i,\tau,z)}{2 a} [\rho(i,\tau,z)]^{2} + i \rho(i,\tau,z)
\partial_{z} \varphi(i,\tau,z)
\nn &&\quad
+ i s(i,i+1,\tau,z) \Big\{ \partial_{z}
t(i,i+1,\tau,z)
+ \frac{1}{a} t(i,i+1,\tau,z)
\nn &&
\quad \quad
- \frac{[t(i,i+1,\tau,z)]^{2}}{a\Big(2
  [m^{2}(i,\tau,z)]^{2} + u(i,\tau,z)\Big)} \Big( m^{2}(i,\tau,z) - \frac{2
  u(i,\tau,z) [t(i,i+1,\tau,z)]^{2}}{[m^{2}(i,\tau,z)]^{2}} \rho(i,\tau,z) \Big)
\Big\}
\nn &&
\quad + i \chi(i,\tau,z) \Big\{ \partial_{z} m^{2}(i,\tau,z) + \frac{2
  m^{2}(i,\tau,z) [t(i,i+1,\tau,z)]^{2}}{a \Big(2 [m^{2}(i,\tau,z)]^{2} +
  u(i,\tau,z)\Big) } \Big\}
\nn &&
\quad + i v(i,\tau,z) \Bigg( \partial_{z} u(i,\tau,z) + \frac{1}{a}
u(i,\tau,z) - \frac{u(i,\tau,z) [t(i,i+1,\tau,z)]^{4}}{2 a [m^{2}(i,\tau,z)]^{2}
  \Big(2 [m^{2}(i,\tau,z)]^{2} + u(i,\tau,z)\Big)} \Bigg)
\nn &&
\quad + \frac{u(i,\tau,z)}{8 a \Big(2 [m^{2}(i,\tau,z)]^{2} + u(i,\tau,z)\Big)} + \frac{1}{4 a} \ln \Big( - \partial_{\tau}^{2} + m^{2}(i,\tau,z) \Big) \Bigg\} . \eqa

%
%\subsection{Gaussian integrations for $\rho(i,\tau,z)$, $s(i,i+1,\tau,z)$, $\chi(i,\tau,z)$, and $v(i,\tau,z)$}
%

Performing Gaussian integrals for $\rho(i,\tau,z)$, $s(i,i+1,\tau,z)$, $\chi(i,\tau,z)$, and $v(i,\tau,z)$, one can further simplify these expressions as
\begin{align}
  &
    Z = \int D \Phi(i,\tau) D \varphi(i,\tau,z) D t(i,i+1,\tau,z) D m^{2}(i,\tau,z) D u(i,\tau,z)
    \nn
  &
    \times
    \delta \Big(t(i,i+1,\tau,0) - t\Big) \delta \Big(m^{2}(i,\tau,0) - m^{2}\Big) \delta \Big(u(i,\tau,0) - u\Big)
    \nonumber \\
  &\times
    \delta \Bigg( \partial_{z} m^{2}(i,\tau,z) + \frac{2 m^{2}(i,\tau,z) [t(i,i+1,\tau,z)]^{2}}{a \Big(2 [m^{2}(i,\tau,z)]^{2} + u(i,\tau,z)\Big) } \Bigg)
    \nn
  & \times \delta \Bigg( \partial_{z} u(i,\tau,z) + \frac{1}{a} u(i,\tau,z) - \frac{u(i,\tau,z) [t(i,i+1,\tau,z)]^{4}}{2 a [m^{2}(i,\tau,z)]^{2} \Big(2 [m^{2}(i,\tau,z)]^{2} + u(i,\tau,z)\Big)} \Bigg)
    \nonumber\\
  &\times
    \exp \Big\{- S_{IR} - S_{UV} - S_{Bulk} \Big\} .
\end{align}
Both IR and UV effective actions are given by
\begin{align}
& S_{IR} + S_{UV} = \int_{0}^{\beta} d \tau \sum_{i = 1}^{M} \Bigg\{
\Big(\partial_{\tau} \Phi(i,\tau)\Big)^{2} + m^{2}(i,\tau,z_{f})
[\Phi(i,\tau)]^{2} - i \varphi(i,\tau,z_{f}) [\Phi(i,\tau)]^{2}
                \nn
  & \qquad
       - t(i,i+1,\tau,z_{f}) \Big( \Phi(i,\tau) \Phi(i+1,\tau) + \Phi(i+1,\tau) \Phi(i,\tau) \Big) + \frac{1}{2u(i,\tau,0)} [\varphi(i,\tau,0)]^{2} \Bigg\}
\end{align}
and the bulk effective action is
\begin{align}
  &
    S_{Bulk} = \int_{0}^{z_{f}} d z \int_{0}^{\beta} d \tau \sum_{i = 1}^{M} \Bigg\{ \frac{3 a}{8 u(i,\tau,z)} [\partial_{z} \varphi(i,\tau,z)]^{2}
    \nn
  & + \frac{a [m^{2}(i,\tau,z)]^{4} \Big(2 [m^{2}(i,\tau,z)]^{2} + u(i,\tau,z)\Big)^{2}}{8 u(i,\tau,z) [t(i,i+1,\tau,z)]^{8}}
    \nonumber \\
  &\qquad
    \times
    \Big\{ \partial_{z} t(i,i+1,\tau,z) + \frac{1}{a} t(i,i+1,\tau,z) - \frac{m^{2}(i,\tau,z) [t(i,i+1,\tau,z)]^{2}}{a\Big(2 [m^{2}(i,\tau,z)]^{2} + u(i,\tau,z)\Big)} \Big\}^{2}
    \nn
  & + \frac{a [m^{2}(i,\tau,z)]^{2} \Big(2 [m^{2}(i,\tau,z)]^{2} + u(i,\tau,z)\Big) }{8 u(i,\tau,z) [t(i,i+1,\tau,z)]^{4}}
    \Big(- i \partial_{z} \varphi(i,\tau,z)\Big)
    \nonumber \\
  &\qquad
    \times
    \Big\{ \partial_{z} t(i,i+1,\tau,z) + \frac{1}{a} t(i,i+1,\tau,z)
   - \frac{m^{2}(i,\tau,z) [t(i,i+1,\tau,z)]^{2}}{a\Big(2 [m^{2}(i,\tau,z)]^{2} + u(i,\tau,z)\Big)} \Big\}
    \nonumber \\
  &
    + \frac{u(i,\tau,z)}{8 a \Big(2 [m^{2}(i,\tau,z)]^{2} + u(i,\tau,z)\Big)} + \frac{1}{4 a} \ln \Big( - \partial_{\tau}^{2} + m^{2}(i,\tau,z) \Big) \Bigg\} .
\end{align}

%
%\subsection{Uniform approximation}
%

For the presentation and as a natural approximation, we take a uniform ansatz in the following way
\bqa && Z = \int D \Phi(i,\tau) D \varphi(z) D t(z) D m^{2}(z) D u(z) \delta
\Big(t(0) - t\Big) \delta \Big(m^{2}(0) - m^{2}\Big) \delta \Big(u(0) - u\Big)
\nn && \quad \times \delta \Bigg( \partial_{z} m^{2}(z) + \frac{2 m^{2}(z)
  [t(z)]^{2}}{a \Big(2 [m^{2}(z)]^{2} + u(z)\Big) } \Bigg) \delta \Bigg(
\partial_{z} u(z) + \frac{1}{a} u(z) - \frac{u(z) [t(z)]^{4}}{2 a [m^{2}(z)]^{2}
  \Big(2 [m^{2}(z)]^{2} + u(z)\Big)} \Bigg)
\nn &&\quad \times \exp \Big\{- S_{IR} - S_{UV} - S_{Bulk} \Big\} , \eqa
where the IR and UV effective actions are
\bqa && S_{IR} + S_{UV} = \int_{0}^{\beta} d \tau \sum_{i = 1}^{M} \Bigg\{
\Big(\partial_{\tau} \Phi(i,\tau)\Big)^{2} + m^{2}(z_{f}) [\Phi(i,\tau)]^{2} - i
\varphi(z_{f}) [\Phi(i,\tau)]^{2}
\nn &&\quad - t(z_{f}) \Big( \Phi(i,\tau) \Phi(i+1,\tau) + \Phi(i+1,\tau) \Phi(i,\tau) \Big) + \frac{1}{2u(0)} [\varphi(0)]^{2} \Bigg\} \eqa
and the bulk effective action is
\begin{align}
& S_{Bulk} = \beta M \int_{0}^{z_{f}} d z \Bigg\{ \frac{3 a}{8 u(z)}
                [\partial_{z} \varphi(z)]^{2}
                \nonumber \\
  & \quad
    + \frac{a [m^{2}(z)]^{4} \Big(2 [m^{2}(z)]^{2} +
  u(z)\Big)^{2}}{8 u(z) [t(z)]^{8}} \Big\{ \partial_{z} t(z) + \frac{1}{a} t(z)
- \frac{m^{2}(z) [t(z)]^{2}}{a\Big(2 [m^{2}(z)]^{2} + u(z)\Big)} \Big\}^{2}
                \nn
  &\quad  + \frac{a [m^{2}(z)]^{2} \Big(2 [m^{2}(z)]^{2} + u(z)\Big) }{8 u(z)
  [t(z)]^{4}} \Big(- i \partial_{z} \varphi(z)\Big) \Big\{ \partial_{z} t(z) +
\frac{1}{a} t(z) - \frac{m^{2}(z) [t(z)]^{2}}{a\Big(2 [m^{2}(z)]^{2} +
  u(z)\Big)} \Big\}
    \nn
  &\quad + \frac{u(z)}{8 a \Big(2 [m^{2}(z)]^{2} + u(z)\Big)} + \frac{1}{4 a} \frac{1}{\beta} \sum_{i \omega_{n}} \ln \Big( \omega_{n}^{2} + m^{2}(z) \Big) \Bigg\} .
\end{align}

%
%\subsection{Free energy density}
%

Finally, we perform the mean-field analysis, where quantum fluctuations of $\varphi(z)$ and $t(z)$ are neglected to be their saddle-point values. Considering
\bqa && \varphi(z) \Longrightarrow - i \varphi(z) \eqa
for real saddle-point values in $\varphi(z)$, we obtain the mean-field free-energy functional for $\varphi(z)$ and $t(z)$
\begin{align}
  & \frac{1}{M} F[\varphi(z), t(z);z_{f}] =
    \nonumber\\
  &\quad
    \frac{1}{2 \beta} \sum_{i\omega_{n}} \frac{1}{M} \sum_{q} \ln \Big\{ \omega_{n}^{2} - 2 t(z_{f}) \cos q + m^{2}(z_{f}) - \varphi(z_{f}) \Big\} - \frac{1}{2u(0)} [\varphi(0)]^{2}
    \nn
  & \quad + \int_{0}^{z_{f}} d z \Bigg\{- \frac{3 a}{8 u(z)} [\partial_{z} \varphi(z)]^{2}
    \nonumber \\
  &\qquad
    + \frac{a [m^{2}(z)]^{4} \Big(2 [m^{2}(z)]^{2} + u(z)\Big)^{2}}{8 u(z) [t(z)]^{8}} \Big\{ \partial_{z} t(z) + \frac{1}{a} t(z) - \frac{m^{2}(z) [t(z)]^{2}}{a\Big(2 [m^{2}(z)]^{2} + u(z)\Big)} \Big\}^{2}
    \nn
  & \qquad
    - \frac{a [m^{2}(z)]^{2} \Big(2 [m^{2}(z)]^{2} + u(z)\Big) }{8 u(z) [t(z)]^{4}} \Big( \partial_{z} \varphi(z)\Big) \Big\{ \partial_{z} t(z) + \frac{1}{a} t(z) - \frac{m^{2}(z) [t(z)]^{2}}{a\Big(2 [m^{2}(z)]^{2} + u(z)\Big)} \Big\}
    \nn
  &\qquad + \frac{u(z)}{8 a \Big(2 [m^{2}(z)]^{2} + u(z)\Big)} + \frac{1}{4 a} \frac{1}{\beta} \sum_{i \omega_{n}} \ln \Big( \omega_{n}^{2} + m^{2}(z) \Big) \Bigg\} .
\end{align}
The renormalization group equation for the mass parameter is
\bqa && \partial_{z} m^{2}(z) = - \frac{2 m^{2}(z) [t(z)]^{2}}{a \Big(2 [m^{2}(z)]^{2} + u(z)\Big) } \eqa
and that for the interaction vertex is
\bqa && \partial_{z} u(z) = - \frac{1}{a} u(z) + \frac{u(z) [t(z)]^{4}}{2 a [m^{2}(z)]^{2} \Big(2 [m^{2}(z)]^{2} + u(z)\Big)} . \eqa
Minimizing this free-energy functional with respect to both $\varphi(z)$ and $t(z)$, we obtain their equations of motion, given by the second-order derivatives in the coordinate of the extra dimensional space. On the other hand, their IR boundary conditions are given by the linear derivatives in $z$, corresponding to the renormalization group equation for the hopping integral and the Callan-Symanzik equation for the order parameter field, respectively.

\section{Further developments for the invariance of entanglement entropy with respect to renormalization group transformations} \label{Entanglement_Entropy_RG_Invariant_General}

%
%\subsection{Formulation of $\partial_{z_{f}} \mathcal{A}[\Sigma(z_{f})]$}
%

\subsection{An identity for entanglement entropy at $z = z_{f}$}

Resorting to the evolution equation for the metric tensor along the extra-dimensional space, we rewrite the area law of the matter sector at $z = z_{f}$ in terms of the Green's function as follows
\begin{align}
  \partial_{z_{f}} \mathcal{A}[\Sigma(z_{f})]
  &= \int_{\Sigma(z_{f})}
    d^{D-2} x_{\perp} \partial_{z_{f}} \sqrt{\gamma(x,z_{f})}
    \nonumber \\
  & = \frac{1}{2}
\int_{\Sigma(z_{f})} d^{D-2} x_{\perp} \sqrt{\gamma(x,z_{f})}
\gamma^{ij}(x,z_{f}) \partial_{z_{f}} \gamma_{ij}(x,z_{f})
    \nn
  & = - \frac{1}{4 d z} \int_{\Sigma(z_{f})} d^{D-2} x_{\perp} \sqrt{\gamma(x,z_{f})} \gamma^{ij}(x,z_{f}) \big(\partial_{i} \partial_{j} G[x,x;g_{\mu\nu}(x,z_{f})]\big) . \label{Area_Law_Green_Function}
\end{align}
%
%\subsection{Green's function in the heat kernel expansion}
%
We recall that the Green's function is given by
\begin{align}
  & \Big\{- \frac{1}{\sqrt{g(x,z_{f})}} \partial_{\mu} \Big( \sqrt{g(x,z_{f})} g^{\mu\nu}(x,z_{f}) \partial_{\nu} \Big) + \frac{m^{2}}{e^{2 d z} - 1} \Big\} G[x,x';g_{\mu\nu}(x,z_{f})]
    \nonumber \\
  &
    = \frac{1}{\sqrt{g(x,z_{f})}} \delta^{(D)}(x-x') .
\end{align}

One can reformulate this Green's function in terms of the geometric information such as metric and curvature, considering the heat kernel expansion. Introducing the heat kernel $K(s,x,x';z) \equiv \langle x z | e^{- s \mathcal{D}} | x' z \rangle$ with the differential operator $\mathcal{D} \equiv - \frac{1}{\sqrt{g(x,z)}} \partial_{\mu} \Big( \sqrt{g(x,z)} g^{\mu\nu}(x,z) \partial_{\nu} \Big) + \frac{m^{2}}{e^{2 d z} - 1}$, which satisfies the diffusion equation $(\partial_{s} + \mathcal{D}) K(s,x,x';z) = 0$ with an initial condition $K(s = 0, x, x';z) = \frac{1}{\sqrt{g(x,z)}} \delta^{(D)}(x-x')$, we obtain
\bqa && G(x,x';z) = \int_{\epsilon^{2}}^{\infty} d s ~ K(s,x,x';z) . \eqa
%
%\bqa && \mbox{tr}_{xx'} K(s;z) = \sum_{n = 0}^{\infty} a_{n}(z) s^{\frac{n-D}{2}} \eqa
%
Here, $\epsilon$ is a UV cutoff to cure the UV divergence of the entanglement entropy, introduced in subsection \ref{Heat_kernel_method_for_entanglement_entropy}.
As discussed in the heat-kernel expansion Eq. (\ref{Heat_Kernel_Expansion}), we express the Green's function in terms of the metric and curvature at $z = z_{f}$ as
\bqa
&& G(x,x;z_{f}) = \int_{\epsilon^{2}}^{\infty} d s ~ K(s,x,x;z_{f}) = \int_{\epsilon^{2}}^{\infty} d s
~ \sum_{n = 0}^{\infty} a_{n}(x,z_{f}) s^{\frac{n-D}{2}}
\nn &&\quad = \int_{\epsilon^{2}}^{\infty} d s ~ \Big\{ a_{0}(x,z_{f}) s^{-\frac{D}{2}} +
a_{2}(x,z_{f}) s^{\frac{2 - D}{2}} + a_{4}(x,z_{f}) s^{\frac{4 - D}{2}} +
\ldots
\Big\}
\nn &&\quad = \int_{\epsilon^{2}}^{\infty} d s ~ \frac{1}{(4 \pi)^{\frac{D}{2}}} \Big[
s^{-\frac{D}{2}}
+ \Big( \frac{1}{6} R(x,z_{f}) - \frac{m^{2}}{e^{2 d z} - 1}
\Big) s^{\frac{2 - D}{2}}
\nn &&\qquad
+ \Big\{ \frac{1}{180}
R^{2}_{\mu\nu\alpha\beta}(x,z_{f}) - \frac{1}{180} R_{\mu\nu}^{2}(x,z_{f})
\nn &&\qquad + \frac{1}{6} \frac{1}{\sqrt{g(x,z_{f})}} \partial_{\mu} \Big(
\sqrt{g(x,z_{f})} g^{\mu\nu}(x,z_{f}) \partial_{\nu} \Big) \Big( \frac{1}{5}
R(x,z_{f}) - \frac{m^{2}}{e^{2 d z} - 1} \Big)
\nn &&\qquad
+ \frac{1}{2} \Big( \frac{1}{6}
R(x,z_{f}) - \frac{m^{2}}{e^{2 d z} - 1} \Big)^{2} \Big\} s^{\frac{4 - D}{2}} +
\ldots \Big] . \eqa

%
%\subsection{Evaluation of $\partial_{z_{f}} \mathcal{A}[\Sigma(z_{f})]$}
%
%\bqa && \partial_{z_{f}} \mathcal{A}[\Sigma(z_{f})] = - \frac{1}{4 d z} \int_{\Sigma(z_{f})} d^{D-2} x_{\perp} \sqrt{\gamma(x,z_{f})} \gamma^{ij}(x,z_{f}) [\partial_{i} \partial_{j} G(x,x;z_{f})] \nn && = - \frac{1}{4 d z} \int_{\Sigma(z_{f})} d^{D-2} x_{\perp} \sqrt{\gamma(x,z_{f})} \gamma^{ij}(x,z_{f}) \partial_{i} \partial_{j} \Bigg\{ \int_{0}^{\infty} d s ~ \frac{1}{(4 \pi)^{\frac{D}{2}}} \Big[ s^{-\frac{D}{2}} + \Big( \frac{1}{6} R(x,z_{f}) - \frac{m^{2}}{e^{2 d z} - 1} \Big) s^{\frac{2 - D}{2}} \nn && + \Big\{ \frac{1}{180} R^{2}_{\mu\nu\alpha\beta}(x,z_{f}) - \frac{1}{180} R_{\mu\nu}^{2}(x,z_{f}) + \frac{1}{6} \frac{1}{\sqrt{g(x,z_{f})}} \partial_{\mu} \Big( \sqrt{g(x,z_{f})} g^{\mu\nu}(x,z_{f}) \partial_{\nu} \Big) \Big( \frac{1}{5} R(x,z_{f}) - \frac{m^{2}}{e^{2 d z} - 1} \Big) \nn && + \frac{1}{2} \Big( \frac{1}{6} R(x,z_{f}) - \frac{m^{2}}{e^{2 d z} - 1} \Big)^{2} \Big\} s^{\frac{4 - D}{2}} + ... ... \Big] \Bigg\} \eqa
%

Inserting this expression into Eq.\ (\ref{Area_Law_Green_Function}), we obtain
\begin{align}
  &
    \partial_{z_{f}} \mathcal{A}[\Sigma(z_{f})] =
    \nonumber \\
  &
    - \frac{1}{4 d z} \Bigg( \frac{1}{(4 \pi)^{\frac{D}{2}}} \int_{\epsilon^{2}}^{\infty} d s ~ s^{\frac{2 - D}{2}} \Bigg) \int_{\Sigma(z_{f})} d^{D-2} x_{\perp} \sqrt{\gamma(x,z_{f})} \gamma^{ij}(x,z_{f}) \partial_{i} \partial_{j} \Big( \frac{1}{6} R(x,z_{f}) - \frac{m^{2}}{e^{2 d z} - 1} \Big)
    \nn
  & - \frac{1}{4 d z} \Bigg( \frac{1}{(4 \pi)^{\frac{D}{2}}} \int_{\epsilon^{2}}^{\infty} d s ~ s^{\frac{4 - D}{2}} \Bigg) \int_{\Sigma(z_{f})} d^{D-2} x_{\perp} \sqrt{\gamma(x,z_{f})} \gamma^{ij}(x,z_{f})
    \nonumber \\
  &\quad
    \times
    \partial_{i} \partial_{j} \Big\{ \frac{1}{180} R^{2}_{\mu\nu\alpha\beta}(x,z_{f}) - \frac{1}{180} R_{\mu\nu}^{2}(x,z_{f})
    \nn
  & \qquad + \frac{1}{6} \frac{1}{\sqrt{g(x,z_{f})}} \partial_{\mu} \Big( \sqrt{g(x,z_{f})} g^{\mu\nu}(x,z_{f}) \partial_{\nu} \Big) \Big( \frac{1}{5} R(x,z_{f}) - \frac{m^{2}}{e^{2 d z} - 1} \Big)
    \nonumber \\
  &\qquad
    + \frac{1}{2} \Big( \frac{1}{6} R(x,z_{f}) - \frac{m^{2}}{e^{2 d z} - 1} \Big)^{2} \Big\} .
\end{align}
%
%\subsection{Invariance of entanglement entropy with respect to renormalization group transformations}
%
Combining this expression with that of the gravity sector, we reach the following identity for the entanglement entropy
\begin{align}
& \frac{N}{6 (D - 2) (4\pi)^{\frac{D}{2} - 1}} \frac{1}{\epsilon^{D-2}}
\frac{1}{4 d z}
\frac{1}{(4 \pi)^{\frac{D}{2}}}
\int_{\Sigma(z_{f})} d^{D-2} x_{\perp} \sqrt{\gamma(x,z_{f})} \gamma^{ij}(x,z_{f})
\Big[
                \nn
  &\quad
-
\Big(  \int_{\epsilon^{2}}^{\infty} d s ~ s^{\frac{2 - D}{2}} \Big)
\partial_{i} \partial_{j} \Big(
\frac{1}{6} R(x,z_{f})
- \frac{m^{2}}{e^{2 d z} - 1} \Big)
    \nonumber \\ &\quad
-
\Big(
\int_{\epsilon^{2}}^{\infty} d s ~ s^{\frac{4 - D}{2}}
\Big)
%\int_{\Sigma(z_{f})} d^{D-2} x_{\perp} \sqrt{\gamma(x,z_{f})} \gamma^{ij}(x,z_{f})
\partial_{i} \partial_{j} \Big\{ \frac{1}{180}
R^{2}_{\mu\nu\alpha\beta}(x,z_{f})
- \frac{1}{180} R_{\mu\nu}^{2}(x,z_{f})
    \nn
  &\qquad
+ \frac{1}{6}
\frac{1}{\sqrt{g(x,z_{f})}} \partial_{\mu} \Big( \sqrt{g(x,z_{f})}
g^{\mu\nu}(x,z_{f}) \partial_{\nu} \Big) \Big( \frac{1}{5} R(x,z_{f}) -
\frac{m^{2}}{e^{2 d z} - 1} \Big) + \frac{1}{2} \Big( \frac{1}{6} R(x,z_{f}) -
\frac{m^{2}}{e^{2 d z} - 1} \Big)^{2} \Big\} \Big]
    \nn
  & + N \int d^{D-2} x_{\perp} \int_{0}^{2 \pi} d \theta
\int_{0}^{\infty} d r \sqrt{g_{n}(r,x_{\perp},z_{f})} ~
%
%\Big[
%
\Big(
T_{\mu\nu,n}^{\varphi}(r,x_{\perp},z_{f}) +
T_{\mu\nu,n}^{g_{\mu\nu}}(r,x_{\perp},z_{f}) \Big) ~ \frac{\partial
  g_{n}^{\mu\nu}(r,x_{\perp},z_{f})}{\partial n} \Big|_{n = 1}
%    \nn  &\quad - \Big\{ \frac{1}{2u} [\partial_{z} \varphi(r,x_{\perp},z_{f})]^{2} + \frac{\mathcal{C}_{\varphi}}{2} g^{\mu\nu}(r,x_{\perp},z_{f}) [\partial_{\mu} \varphi(r,x_{\perp},z_{f})] [\partial_{\nu} \varphi(r,x_{\perp},z_{f})] + \mathcal{C}_{\xi} R(r,x_{\perp},z_{f}) [\varphi(r,x_{\perp},z_{f})]^{2}     \nn  &\qquad + \frac{1}{2 \kappa} \Big( R(r,x_{\perp},z_{f}) - 2 \Lambda \Big) \Big\} \Big]
= 0 .
\end{align}
We point out that both integrals of
$\int^{\infty}_{\epsilon^{2}} ds\, s^{(2-D)/2}$
and
$\int^{\infty}_{\epsilon^{2}} ds\, s^{(4-D)/2}$
diverge for large $s$.
Existence of these IR divergences may be an artifact of the small $s-$expansion.
One may introduce the correlation length into the upper cutoff of these integrals
as $\int^{\xi^{2}}_{\epsilon^{2}} ds\, s^{(2-D)/2}$
and
$\int^{\xi^{2}}_{\epsilon^{2}} ds\, s^{(4-D)/2}$.
We recall that this IR cutoff reproduces the entanglement entropy in $D = 2$, as shown in Eq. (\ref{EE_2D}).
Now, the identity is expressed in a fully geometric way.

\subsection{Entanglement entropy after the first iteration}

%
%\subsubsection{Effective field theory after the first iteration}
%

%Frankly speaking,
While it is written in terms of all the geometrical information,
it is not clear how to interpret the above lengthy expression.
%although it is written in terms of all the geometrical information.
In addition, it is not easy to check out this identity in a brute force way. In this respect we consider the entanglement entropy after the first iteration of the renormalization group transformation and verify the invariance explicitly for the case without all types of interactions.

%
%\bqa && Z^{(1)} = \int D \phi_{\alpha} D g_{\mu\nu}^{(0)} D T_{\mu\nu}^{(0)} D g_{\mu\nu}^{(1)} D T_{\mu\nu}^{(1)} \exp\Big[ - \int d^{D} x \sqrt{g^{(1)}} \Big\{ g^{(1)\mu\nu} (\partial_{\mu} \phi_{\alpha}) (\partial_{\nu} \phi_{\alpha}) + m^{2} \phi_{\alpha}^{2} \Big\} \nn && - N \int d^{D} x \sqrt{g^{(0)}} \Big\{ 2 \alpha^{(0)} d z \Big( - \mathcal{C}_{\Lambda} + \mathcal{C}_{R} R^{(0)} \Big) - T_{\mu\nu}^{(0)} (g^{(0)\mu\nu} - g_{B}^{\mu\nu}) \Big\} \nn && - N \int d^{D} x \sqrt{g^{(1)}} \Big\{- T_{\mu\nu}^{(1)} \Big(g^{(1)\mu\nu} - g^{(0)\mu\nu} - g^{(0)\mu\nu'} (\partial_{\nu'} \partial_{\mu'} G_{xx'}^{(0)})_{x' \rightarrow x} g^{(0)\mu'\nu}\Big) \Big\} \Big] \eqa
%
%\bqa && \alpha^{(0)} = 1 \eqa
%
%\bqa && Z^{(1)} = \int D \phi_{\alpha} D g_{\mu\nu}^{(0)} D g_{\mu\nu}^{(1)} \delta(g^{(0)\mu\nu} - g_{B}^{\mu\nu}) \delta\Big(g^{(1)\mu\nu} - g^{(0)\mu\nu} - g^{(0)\mu\nu'} (\partial_{\nu'} \partial_{\mu'} G_{xx'}^{(0)})_{x' \rightarrow x} g^{(0)\mu'\nu}\Big) \nn && \exp\Big[ - \int d^{D} x \sqrt{g^{(1)}} \Big\{ g^{(1)\mu\nu} (\partial_{\mu} \phi_{\alpha}) (\partial_{\nu} \phi_{\alpha}) + m^{2} \phi_{\alpha}^{2} \Big\} - N \int d^{D} x \sqrt{g^{(0)}} \Big\{ 2 d z \Big( - \mathcal{C}_{\Lambda} + \mathcal{C}_{R} R^{(0)} \Big) \Big\} \Big] \eqa
%

We recall the partition function after the renormalization group transformation
\begin{align}
& Z^{(1)} = \int D \phi_{\alpha} D g_{\mu\nu}^{(1)}
\delta\Big(g^{(1)\mu\nu} - g_{B}^{\mu\nu} - g_{B}^{\mu\nu'} (\partial_{\nu'}
\partial_{\mu'} G_{xx'})_{x' \rightarrow x} g_{B}^{\mu'\nu}\Big)
                \nn
  &\quad\times \exp\Big[ - \int d^{D} x \sqrt{g^{(1)}} \Big\{ g^{(1)\mu\nu} (\partial_{\mu} \phi_{\alpha}) (\partial_{\nu} \phi_{\alpha}) + m^{2} \phi_{\alpha}^{2} \Big\} - \frac{N}{2 \kappa} \int d^{D} x \sqrt{g_{B}} \Big( R_{B} - 2 \Lambda \Big) \Big] ,
\end{align}
where the coefficients of the gravity action are $2 d z \mathcal{C}_{R} = \frac{1}{2 \kappa}$ and $\frac{\mathcal{C}_{\Lambda}}{\mathcal{C}_{R}} = 2 \Lambda$. Then, the entanglement entropy is
\bqa
&& \mathcal{S}_{EE}^{(1)} = \frac{N}{6 (D - 2) (4\pi)^{\frac{D}{2} - 1}}
\frac{1}{\epsilon^{D-2}} \int_{\Sigma^{(1)}} d^{D-2} x_{\perp}
\sqrt{g_{\perp}^{(1)}(x)}
\nn && \quad + \frac{N}{2 \kappa} \int d^{D-2} x_{\perp} \int_{0}^{2 \pi} d
\theta \int_{0}^{\infty} d r \sqrt{g_{B,n}(x)} \Big[ 2 \Big( R^{B}_{\mu\nu,n}(x) -
\frac{1}{2} g^{B}_{\mu\nu,n}(x) R_{B,n}(x) + \Lambda g^{B}_{\mu\nu,n}(x) \Big)
\nn && \quad + \frac{1}{g_{B,n}(x)} g^{B}_{\mu\mu',n}(x) g^{B}_{\nu\nu',n}(x)
\partial_{\mu''} \partial_{\nu''} \Big\{g_{B,n}(x) \Big( g_{B,n}^{\mu'\nu'}(x)
g_{B,n}^{\mu''\nu''}(x) - g_{B,n}^{\mu'\mu''}(x) g_{B,n}^{\nu'\nu''}(x) \Big) \Big\}
\Big] \frac{\partial g_{B, n}^{\mu\nu}(r,x_{\perp})}{\partial n} \Big|_{n = 1}
%
%\nn && \quad - \frac{N}{2 \kappa} \int d^{D-2} x_{\perp} \int_{0}^{2 \pi} d \theta \int_{0}^{\infty} d r \sqrt{g_{B}(x)} \Big( R_{B}(x) - 2 \Lambda \Big)
%
. \nn \eqa
This entanglement entropy has to be identified with
%
%\subsubsection{Entanglement entropy of the free scalar field theory}
%
\bqa && \mathcal{S}_{EE} = \frac{N}{6 (D - 2) (4\pi)^{\frac{D}{2} - 1}} \frac{1}{\epsilon^{D-2}} \int_{\Sigma_{B}} d^{D-2} x_{\perp} \sqrt{g^{B}_{\perp}(x)} . \eqa
%
%\subsubsection{Entanglement entropy after the first iteration}
%
%\bqa && \mathcal{S}_{EE}^{(1)} = \frac{N}{6 (D - 2) (4\pi)^{\frac{D}{2} - 1}} \frac{1}{\epsilon^{D-2}} \int_{\Sigma^{(1)}} d^{D-2} x_{\perp} \sqrt{g_{\perp}^{(1)}(x)} \nn && + \frac{N}{2 \kappa} \int d^{D-2} x_{\perp} \int_{0}^{2 \pi} d \theta \int_{0}^{\infty} d r \sqrt{g_{B}(x)} \Big[ - 2 \Big( R^{B}_{\mu\nu}(x) - \frac{1}{2} g^{B}_{\mu\nu}(x) R_{B}(x) + \Lambda g^{B}_{\mu\nu}(x) \Big) \nn && + \frac{1}{g_{B}(x)} g^{B}_{\mu\mu'}(x) g^{B}_{\nu\nu'}(x) \partial_{\mu''} \partial_{\nu''} \Big\{g_{B}(x) \Big( g_{B}^{\mu'\nu'}(x) g_{B}^{\mu''\nu''}(x) - g_{B}^{\mu'\mu''}(x) g_{B}^{\nu'\nu''}(x) \Big) \Big\} \Big] \frac{\partial g_{B n}^{\mu\nu}(r,x_{\perp})}{\partial n} \Big|_{n = 1} \nn && - \frac{N}{2 \kappa} \int d^{D-2} x_{\perp} \int_{0}^{2 \pi} d \theta \int_{0}^{\infty} d r \sqrt{g_{B}(x)} \Big( R_{B}(x) - 2 \Lambda \Big)  \eqa
%
%\subsubsection{Invariance of entanglement entropy with respect to renormalization group transformations}
%
In other words, we have
\bqa && \mathcal{S}_{EE} = \mathcal{S}_{EE}^{(1)} . \eqa
As a result, we have to obtain
\bqa && \frac{N}{6 (D - 2) (4\pi)^{\frac{D}{2} - 1}} \frac{1}{\epsilon^{D-2}}
\Bigg( \int_{\Sigma_{B}} d^{D-2} x_{\perp} \sqrt{g^{B}_{\perp}(x)} -
\int_{\Sigma^{(1)}} d^{D-2} x_{\perp} \sqrt{g_{\perp}^{(1)}(x)} \Bigg)
\nn && =
\frac{N}{2 \kappa} \int d^{D-2} x_{\perp} \int_{0}^{2 \pi} d \theta
\int_{0}^{\infty} d r \sqrt{g_{B,n}(x)} \Big[ 2 \Big( R^{B}_{\mu\nu,n}(x) -
\frac{1}{2} g^{B}_{\mu\nu,n}(x) R_{B,n}(x) + \Lambda g^{B}_{\mu\nu,n}(x) \Big)
\nn &&\quad + \frac{1}{g_{B,n}(x)} g^{B}_{\mu\mu',n}(x) g^{B}_{\nu\nu',n}(x)
\partial_{\mu''} \partial_{\nu''} \Big\{g_{B,n}(x) \Big( g_{B,n}^{\mu'\nu'}(x)
g_{B,n}^{\mu''\nu''}(x) - g_{B,n}^{\mu'\mu''}(x) g_{B,n}^{\nu'\nu''}(x) \Big) \Big\}
\Big] \frac{\partial g_{B, n}^{\mu\nu}(r,x_{\perp})}{\partial n} \Big|_{n = 1}
%
%\nn &&\quad - \frac{N}{2 \kappa} \int d^{D-2} x_{\perp} \int_{0}^{2 \pi} d \theta \int_{0}^{\infty} d r \sqrt{g_{B}(x)} \Big( R_{B}(x) - 2 \Lambda \Big)
%
. \nn \eqa

%
%\subsubsection{$\sqrt{g^{(1)}}$}
%

Recalling the evolution equation of
\bqa && g^{(1)\mu\nu}(x) = g_{B}^{\mu\nu}(x) + g_{B}^{\mu\nu'}(x) (\partial_{\nu'} \partial_{\mu'} G_{xx'})_{x' \rightarrow x} g_{B}^{\mu'\nu}(x) , \eqa
%
%\bqa && g^{(1)}_{\mu\nu}(x) = g^{B}_{\mu\nu}(x) - (\partial_{\mu} \partial_{\nu} G_{xx'})_{x' \rightarrow x} \eqa
%
we obtain
\bqa && \sqrt{g^{(1)}(x)} = \sqrt{g_{B}(x)}\Big\{ 1 - \frac{1}{2} g_{B}^{\mu\nu}(x) (\partial_{\mu} \partial_{\nu} G_{xx'})_{x' \rightarrow x} \Big\} . \eqa
%
%\subsubsection{$\int_{\Sigma_{B}} d^{D-2} x_{\perp} \sqrt{g^{B}_{\perp}(x)} - \int_{\Sigma^{(1)}} d^{D-2} x_{\perp} \sqrt{g_{\perp}^{(1)}(x)}$}
%
As a result, we have
\begin{align}
  & \int_{\Sigma_{B}} d^{D-2} x_{\perp} \sqrt{g^{B}_{\perp}(x)} - \int_{\Sigma^{(1)}} d^{D-2} x_{\perp} \sqrt{g_{\perp}^{(1)}(x)}
    \nonumber \\
  &\quad
    = \frac{1}{2} \int_{\Sigma_{B}} d^{D-2} x_{\perp} \sqrt{g^{B}_{\perp}(x)} g_{B \perp}^{\mu\nu}(x) (\partial_{\mu} \partial_{\nu} G_{xx'})_{x' \rightarrow x} .
\end{align}

%
%\subsubsection{Final equation to be proven}
%

The final equation to be verified is
\bqa && \frac{N}{6 (D - 2) (4\pi)^{\frac{D}{2} - 1}} \frac{1}{2 \epsilon^{D-2}}
\int_{\Sigma_{B}} d^{D-2} x_{\perp} \sqrt{g^{B}_{\perp}(x)} g_{B
  \perp}^{\mu\nu}(x) (\partial_{\mu} \partial_{\nu} G_{xx'})_{x' \rightarrow x}
\nn && = \frac{N}{2 \kappa} \int d^{D-2} x_{\perp} \int_{0}^{2 \pi} d \theta
\int_{0}^{\infty} d r \sqrt{g_{B,n}(x)} \Big[ 2 \Big( R^{B}_{\mu\nu,n}(x) -
\frac{1}{2} g^{B}_{\mu\nu,n}(x) R_{B,n}(x) + \Lambda g^{B}_{\mu\nu,n}(x) \Big)
\nn &&\quad + \frac{1}{g_{B,n}(x)} g^{B}_{\mu\mu',n}(x) g^{B}_{\nu\nu',n}(x)
\partial_{\mu''} \partial_{\nu''} \Big\{g_{B,n}(x) \Big( g_{B,n}^{\mu'\nu'}(x)
g_{B,n}^{\mu''\nu''}(x) - g_{B,n}^{\mu'\mu''}(x) g_{B,n}^{\nu'\nu''}(x) \Big) \Big\}
\Big] \frac{\partial g_{B, n}^{\mu\nu}(r,x_{\perp})}{\partial n} \Big|_{n = 1}
%
%\nn && \quad - \frac{N}{2 \kappa} \int d^{D-2} x_{\perp} \int_{0}^{2 \pi} d \theta \int_{0}^{\infty} d r \sqrt{g_{B}(x)} \Big( R_{B}(x) - 2 \Lambda \Big)
%
. \nn \label{Entanglement_Entropy_Invariance_First_RG} \eqa
Here, the background metric is
\bqa && d s^{2} = d r^{2} + r^{2} d \theta^{2} + \delta_{ij} d x_{\perp}^{i} d x_{\perp}^{j} , ~~~~~ g^{B}_{rr} = 1 , ~~~ g^{B}_{\theta\theta} = r^{2} , ~~~ g_{ij}^{B\perp} = \delta_{ij}, ~~~~~ \sqrt{g_{B}} = r \eqa
with
\bqa && \frac{\partial g_{B n}^{\mu\nu}(r,x_{\perp})}{\partial n} \Big|_{n = 1} = 2 r^{2} \delta_{\mu \theta} \delta_{\nu \theta} \eqa
and the Green's function is \cite{Entanglement_Entropy_Calabrese_Cardy_I,Entanglement_Entropy_Calabrese_Cardy_II}
\begin{align}
  & G_{xx'}^{n} \equiv G_{n}(r,r',\theta,\theta',x_{\perp} - x_{\perp}')
    \nn
  & = \int \frac{d^{D-2} p_{\perp}}{(2\pi)^{D-2}} e^{i p_{\perp} (x_{\perp} - x_{\perp}')} \frac{1}{2 \pi n} \sum_{k = 0}^{\infty} d_{k} \int_{0}^{\infty} d \lambda \lambda \frac{J_{k/n}(\lambda r) J_{k/n}(\lambda r')}{p_{\perp}^{2} + \lambda^{2} + m^{2}/(e^{2 d z} - 1)} \cos \frac{k}{n} (\theta - \theta')
\end{align}
with
\bqa && d_{0} = 1 , ~~~~~ d_{k > 0} = 2 . \eqa
Inserting this Green's function with $n = 1$ into Eq. (\ref{Entanglement_Entropy_Invariance_First_RG}), the left-hand-side of this equation is written in terms of $\epsilon$ and $m$ with $\Sigma_{B}$. On the other hand, the right-hand-side of this equation is given by $\kappa$ and $\Lambda$ with a UV cutoff in the $r$ integral. As a result, renormalizing both coefficients of $\kappa$ and $\Lambda$ appropriately, one can verify this identity.


\end{document}